\documentclass[12pt]{article}
\usepackage{axodraw}
\usepackage{pstricks}
\usepackage{color}
\usepackage{amssymb,amsmath,bm,bbm}
\usepackage{epsf}
\usepackage{epsfig}
\usepackage{afterpage}
\usepackage{longtable}
\usepackage{cite}
\usepackage{latexsym, mathrsfs}
\usepackage{graphics}
\usepackage{url}
\usepackage{paralist}
\usepackage{bbold}
\newcommand{\no}{\nonumber}

\newcommand{\Br}{\mathrm{Br}}

\newcommand{\model}{\overline{331}}

\newcommand{\ord}{\mathcal{O}}

\newcommand{\Heff}{H_{eff}}

\newcommand{\tev}{\, {\rm TeV}}
\newcommand{\gev}{\, {\rm GeV}}
\newcommand{\mev}{\, {\rm MeV}}

\newcommand{\vcb}{|V_{cb}|}
\newcommand{\vtd}{|V_{td}|}
\newcommand{\vub}{|V_{ub}|}
\newcommand{\vts}{|V_{ts}|}
\newcommand{\vus}{|V_{us}|}

\def\epe{\varepsilon'/\varepsilon}
\newcommand{\beq}{\begin{equation}}
\newcommand{\eeq}{\end{equation}}
\newcommand{\be}{\begin{equation}}
\newcommand{\ee}{\end{equation}}
\newcommand{\bi}{\begin{itemize}}
\newcommand{\ei}{\end{itemize}}
\newcommand{\ba}{\begin{array}}
\newcommand{\ea}{\end{array}}
\newcommand{\beqa}{\begin{eqnarray}}
\newcommand{\eeqa}{\end{eqnarray}}
\newcommand{\bea}{\begin{eqnarray}}
\newcommand{\eea}{\end{eqnarray}}
\newcommand{\beqn}{\begin{eqnarray}}
\newcommand{\eeqn}{\end{eqnarray}}

\newcommand{\D}{\Delta}

\newcommand{\eps}{\epsilon}
\newcommand{\nn}{\nonumber}

\definecolor{red}{cmyk}{0,1,1,0.4}

\def\kpn{K^+\rightarrow\pi^+\nu\bar\nu}
\def\klpn{K_{L}\rightarrow\pi^0\nu\bar\nu}

\usepackage{fancyhdr}
\advance \headheight by 3.0truept       % for 12pt mandatory...
\begin{document}

\begin{flushright}
    {FLAVOUR(267104)-ERC-25}\\
    {BARI-TH/12-658}
\end{flushright}

\medskip

\begin{center}
{\LARGE\bf
\boldmath{The Anatomy of Quark Flavour Observables\\ in 331 Models in the
Flavour Precision Era
}}\\[0.8 cm]
{\bf Andrzej~J.~Buras$^{a,b}$, Fulvia~De~Fazio$^{c}$,
Jennifer Girrbach$^{a,b}$,  Maria~V.~Carlucci$^{a}$
 \\[0.5 cm]}
{\small
$^a$TUM Institute for Advanced Study, Lichtenbergstr. 2a, D-85747 Garching, Germany\\
$^b$Physik Department, Technische Universit\"at M\"unchen,
James-Franck-Stra{\ss}e, \\D-85747 Garching, Germany\\
$^c$Istituto Nazionale di Fisica Nucleare, Sezione di Bari, Via Orabona 4,
I-70126 Bari, Italy}
\end{center}

\vskip0.61cm

%{\em Version of \today}

\abstract{%
\noindent
The coming flavour precision era will allow to uncover various patterns
of flavour violation in different  New Physics (NP) scenarios that are
presently washed out by hadronic and experimental uncertainties. We illustrate
this by performing the anatomy of flavour violation in
the 331 models, based on the gauge group $SU(3)_C\times SU(3)_L\times U(1)_X$
that are among the simplest NP scenarios with new sources of flavour and CP violation. The latter originate dominantly through the flavour
violating interactions of ordinary quarks
and leptons with a new heavy $Z^\prime$ gauge boson.
After discussing first these models in some generality,
we present a detailed study of $\Delta F=2$ observables and of rare $K$ and $B$ meson decays in the specific ``$\beta=1/\sqrt{3}$'' model
(to be called $\overline{331}$ model) assuming significantly smaller
uncertainties in CKM and hadronic parameters than presently available.
The most prominent roles in our analysis play $\varepsilon_K$, $\Delta M_q$ $(q=d,s)$, the mixing induced CP asymmetries $S_{\psi K_S}$ and $S_{\psi\phi}$,
rare decays $B_{s,d}\to \mu^+\mu^-$ and in particular the CP-asymmetry
$S_{\mu^+\mu^-}^s$ in $B_{s}\to\mu^+\mu^-$.
As the phenomenology of $Z'$
contributions is governed only by $M_{Z'}$ and four new parameters
$\tilde s_{13}$, $\tilde s_{23}$, $\delta_1$ and $\delta_2$,
we identify a number of correlations between various observables that
differ from those known from CMFV models.
While, the $\Delta F=2$ observables allow still
for four oases in the new parameter space, we demonstrate how the inclusion
of $\Delta F=1$ observables can in the future identify the optimal
oasis for this model.
Favouring the inclusive value of $\vub$,
for $1\tev \le M_{Z'}\le 3\tev$, the  {$\overline{331}$}   model  is in
good agreement with the recent data for
$B^+\to\tau^+\nu_\tau$, it
removes
the $\varepsilon_K-S_{\psi K_S}$ tension present in the SM and
the $\varepsilon_K-\Delta M_{s,d}$ tension present in CMFV models.
Simultaneously, while differing from the SM, it is consistent with the present data on $\mathcal{B}(B_{s,d}\to\mu^+\mu^-)$ and $S_{\psi\phi}$.
We identify the triple correlation
$\mathcal{B}(B_{s}\to\mu^+\mu^-)-S_{\psi\phi}-S_{\mu^+\mu^-}^s$ that constitutes
an important test of this NP scenario.
}

\thispagestyle{empty}
\newpage
\setcounter{page}{1}

\section{Introduction}

The LHCb experiment  opened  this year a new era of precision flavour
physics \cite{Bediaga:2012py}.     This indirect route to shortest
distance scales with the goal to discover new physics (NP) is also followed
by ATLAS and CMS to be joined
later in this decade in particular
 by SuperKEKB, SuperB in Rome and  dedicated Kaon experiments NA62
at CERN, KOTO at J-Parc and ORKA at Fermilab.

While, as discussed below and reviewed in \cite{Buras:2012ts}, there are several anomalies
in the present data, none of them are conclusive and we should be prepared
that NP effects in flavour observables will be present  at most at the $50\%$
level of the Standard Model (SM) predictions
and not an order of magnitude as initially expected and
hoped for. In order to identify NP appearing at this level through flavour
violating processes, it will be crucial
to have both precise experiments and precise theory.

Indeed in indirect searches for NP through
particle-antiparticle mixings in $K$, $B_{s,d}$ and $D$
meson systems and rare decays of $K$, $B_{s,d}$ and $D$ mesons it is crucial
to know the background to NP: the predictions for various flavour observables within SM. If these predictions suffer from large uncertainties
also the room left for NP is rather uncertain and if a given NP model
contains many free parameters, the characteristic flavour violating features
of this model cannot be transparently seen. They are simply often washed out
by hadronic and parametric uncertainties even in the presence of accurate
data. Most prominent examples of this type are the mass differences $\Delta M_{d,s}$ and the parameter $\varepsilon_K$.

While the sophisticated model independent analyses of NP by
 professionals like UTfitters \cite{Bona:2007vi}, CKMfitters \cite{Lenz:2012az}, other bayesian \cite{Beaujean:2012uj} and frequentist analyses 
\cite{DescotesGenon:2012zf}  and analyses using 
sophisticated Markov-chain Monte Carlos, in particular improved versions
of the  classical Metropolis algorithm \cite{Barbieri:2011ci,Altmannshofer:2011gn,Altmannshofer:2012az,Botella:2012ju} should be appreciated, it is evident
from the plots presented in these papers that several correlations between
various observables are not clearly visible when all present
hadronic and  parametric uncertainties are taken into account.

In our view a complementary and a useful route to the identification of NP
is a simplified approach in which one investigates
how a given NP model would face the
future more precise data with more precise hadronic and parametric input \footnote{The recent lattice results reviewed in \cite{Davies:2012qf,Tarantino:2012mq} and
future experimental prospects \cite{Bediaga:2012py,Bona:2007qt} assure us
that the flavour precision era is just ahead of us.}.
In this manner some of the characteristic features
of the particular NP model are not washed out and one discovers patterns of flavour violations, in particular correlations between different observables,
that could distinguish between different NP  models. While this approach
has been already partially followed in the papers reviewed in
\cite{Buras:2012ts}, one recent study in this spirit
of a concrete model shows that such studies could give interesting results.

Indeed, by proceeding in this manner and analysing the dependence of
the allowed size of NP contributions in models with $U(2)^3$ flavour symmetry
on the value of $\vub$, it was possible to identify
the correlation $S_{\psi K_S}-S_{\psi\phi}-\vub$ \cite{Buras:2012sd} characteristic for this class of models. This correlation was washed out in
a much more sophisticated earlier analysis in \cite{Barbieri:2011ci} and
could not be identified by these authors. Other similar
examples in the literature could be given here.

While the more sophisticated analyses listed above are clearly
legimitate and possibly could appear as the only way to analyze NP
models at present, the example in \cite{Buras:2012sd} shows that it
is also useful to work in a different spirit, even if it is a bit futuristic,
and investigate how the pattern of flavour violation becomes visible when
the uncertainties in CKM parameters and hadronic uncertainies are
reduced. Such studies are not entirely new and can be found occasionally in various proposals, like the one in \cite{Bona:2007qt}, but here we will illustrate them
in a concrete model stressing the role of correlations between various
observables that become visible in this approach. Our strategy is explained
in concrete terms in Section~\ref{sec:7}.

One of the important questions in this context is whether the
frameworks of constrained Minimal Flavour Violation (CMFV) \cite{Buras:2000dm}
and the more general framework of MFV \cite{D'Ambrosio:2002ex} will be
capable of describing the future data. In models of this class,
when flavour blind phases (FBPs) are absent or set to zero, stringent
relations between various observables in $K$, $B_d^0$ and $B_s^0$ systems
 are present \cite{Buras:2003jf}. Consequently
the  departures from Standard Model (SM) expectations in this class of models in these three meson systems
are correlated with each other allowing very transparent tests of these
simple NP scenarios. However, generally these relations can be strongly
violated, implying often other correlations between observables characteristic
for a given NP scenario.
Such correlations being less
sensitive to the model parameters can often allow a transparent
distinction between various models proposed in the literature
\cite{Buras:2010wr,Buras:2012ts}. In this manner
the CMFV relations can be considered as {\it standard
candles of flavour physics} \cite{Buras:2012ts}.

Among the simplest extensions of the SM which go beyond the concept
of MFV are the so-called 331 models based on the gauge group $SU(3)_C\times SU(3)_L\times U(1)_X$  (331) \cite{Pisano:1991ee,Frampton:1992wt}.
Here the new sources of flavour
and CP violation originate dominantly through the flavour
violating interactions of ordinary quarks
and leptons with a new heavy $Z^\prime$ gauge boson. Also one-loop
contributions involving new charged gauge bosons and new heavy quarks
with exotic electric charges can be relevant in certain processes.

The difference between the 331 models and CMFV models  can be transparently
seen in the formulation of FCNC processes in terms of master
one-loop functions  \cite{Buras:2003jf}. This formulation can be used here because the 331 models
have the same operator structure as the SM and the CMFV models except that
the real and universal master functions in the latter models
become here complex quantities and the property of the universality of these
functions
is lost. Consequently the usual CMFV relations between $K$, $B_d$ and $B_s$
systems can be significantly violated. Explicitly, the new functions in the 331
models, similarly to the case of the
LHT model \cite{Blanke:2006eb}, will be denoted  as
follows ($i=K,d,s$):
\begin{equation}\label{eq31}
S_i\equiv|S_i|e^{i\theta_S^i},
\quad
X_i \equiv \left|X_i\right| e^{i\theta_X^{i}}, \quad
Y_i \equiv \left|Y_i\right| e^{i\theta_Y^i}, \quad
Z_i \equiv \left|Z_i\right| e^{i\theta_Z^i}\,,
\end{equation}
%\begin{equation} \label{eq32}
%D'_i \equiv \left|D_i'\right| e^{i\theta_{D'}^i}\,, \quad
%E'_i \equiv \left|E_i'\right| e^{i\theta_{E'}^i}\,.
%\end{equation}
Explicit expressions for these functions in specific 331 models
will be given in the course of
our presentation.

In 2007 two detailed analyses of flavour violation in the minimal
331 model to be defined below have
been presented \cite{Promberger:2007py,Promberger:2008xg}. During the last five years a number of changes both in
the experiment and theory took place so that a new analysis of this model   and other 331 models
is required. In particular:
\begin{itemize}
\item
The $\varepsilon_K-S_{\psi K_S}$ tension in the SM has been identified
 \cite{Lunghi:2008aa,Buras:2008nn,Buras:2009pj,Lunghi:2009sm,Lunghi:2010gv,UTfit-web}.
\item
The branching ratio for $B^+\to\tau^+\nu_\tau$ has been found to be roughly
by a factor of two larger than the SM prediction, even though the most recent data seem to  be consistent with the SM (see
Section~\ref{sec:5}).
\item
The new lattice input appears to imply some tension between
$\varepsilon_K$ and the mass differences $\Delta M_{s,d}$
within the CMFV models \cite{Buras:2012ts}.
\item
Most importantly the recent LHCb, ATLAS and CMS data imply SM-like values for
$\mathcal{B}(B_s\to\mu^+\mu^-)$ and for the mixing induced asymmetry $S_{\psi\phi}$ in the $B_s$ system,
whereas in 2007 still large departures from SM predictions for both
observables were possible.
\end{itemize}

The goal of the present paper is to make a fresh look at 331 models and
to confront their  flavour and CP violation structure with
recent very improved data on
$S_{\psi\phi}$ and rare decays $B_{s,d}\to\mu^+\mu^-$ from the LHCb, ATLAS and CMS
 taking into account
the most recent lattice input that has a definite impact on
various observables, in particular  CMFV relations.
Therefore in addition to  $S_{\psi\phi}$ and $B_{s,d}\to\mu^+\mu^-$
 we address simultaneously
the $\varepsilon_K-S_{\psi K_S}$ tension still present in the SM,
the mass differences  $\Delta M_{s,d}$, the rare decays $\kpn$ and
$\klpn$ and the tree-level decay $B^+\to\tau^+\nu_\tau$. We will also
consider
$K_L\to \mu^+\mu^-$,  $B\to X_{s,d}\nu\bar\nu$ and
$B\to X_s\gamma$,  while leaving the analysis of  $B\to X_s \ell^+\ell^-$
and $B\to K^*\ell^+\ell^-$ for the future.

We should announce already at this stage that in 2012
the minimal 331 model
analyzed in \cite{Promberger:2007py,Promberger:2008xg}, the so-called
``$\beta=\sqrt{3}$''-model, does not imply any longer an interesting phenomenology of processes considered by us: new physics (NP) effects
are simply very
small. Therefore, we will dominantly discuss non-minimal 331 models with
$\beta\not=\sqrt{3}$. While a significant part of our presentation will apply
to arbitrary $\beta$, we will present a detailed phenomenology of flavour
observables in a specific 331 model with $\beta=1/\sqrt{3}$ to be named
$\model$ model in what follows. This model,
while being rather economical, implies much more interesting phenomenology than
the minimal model. In particular, while consistent with the present data,
among rare decays, the largest deviations from SM expectations are found in
 $B_{s,d}\to\mu^+\mu^-$ branching ratios and the CP asymmetries
$S_{\mu^+\mu^-}^{s,d}$ in
 $B_{s,d}\to\mu^+\mu^-$, recently introduced in \cite{deBruyn:2012wj,deBruyn:2012wk}.

Our paper is organized as follows. In Section~\ref{sec:2} we recall in
some details
basic features of 331 models that are relevant for understanding of our analysis. In particular, we derive all Feynman rules for this class  of
models that we need for our phenomenology.
In Section~\ref{sec:3} we present the formulae for the effective
Hamiltonians governing particle-antiparticle mixings
$K^0-\bar K^0$ and $B_{d,s}-\bar B_{d,s}$ that  in addition to the
usual SM box diagrams receive tree-level contributions
from $Z^\prime$ exchanges and new box diagram contributions. We also give a compendium of formulae relevant
for numerical analysis of $\Delta F=2$ observables. In Section~\ref{sec:4}
the effective Hamiltonians for
$s\to d\nu\bar\nu$, $b\to q \nu\bar\nu$, $b\to q \ell^+\ell^-$ ($q=d,s$) and $b\to s\gamma$ transitions are given. In particular we present an improved QCD
analysis of
$Z'$ contribution to $B\to X_s\gamma$, demonstrating that it is negligibly
small compared to the SM contribution.
In Section \ref{sec:5} we calculate  the
most interesting rare decay
 branching ratios in {the} $K$ and $B$ meson systems, including {those for the processes} $\kpn$,
$\klpn$,  $B\to X_{s,d}\nu\bar\nu$, $B_{s,d}\to\mu^+\mu^-$, $B^+\to\tau^+\nu_\tau$ and  $K_L \to\mu^+\mu^-$.
In Section~\ref{sec:6} we recall the most
important CMFV relations with the goal to find out how these relations are violated
in the 331 models. In Section~\ref{sec:10} we emphasize stringent relations
between various observables in the 331 models that
originate in the dominance  of NP contributions by the tree-level
$Z'$ exchanges. In fact even without a detailed numerical analysis, it will
be possible to deduce from these relations the pattern of deviations from
the SM expectations predicted by the $\model$ model.
In Section~\ref{sec:7} we present first our strategy
for numerical analysis and summarize the recent LHCb data and new
lattice input. Subsequently, we demonstrate how already
$\Delta F=2$ observables allow to identify four allowed oases  in the
new parameter space. We demonstrate how the inclusion
of $\Delta F=1$ observables can in the future select the optimal
oasis for this model. In this context the branching ratios
$\mathcal{B}(B_{s,d}\to\mu^+\mu^-)$,
the CP-asymmetry $S_{\mu^+\mu^-}^s$ in $B_{s}\to\mu^+\mu^-$ and $S_{\psi\phi}$,
when considered simultaneously, play the prominent roles.
The correlations between these
observables in the $\model$ model should allow to monitor how this model
will face the data in the coming years.
We also analyze how these correlations differ from the ones present in
the CMFV models.
The highlights of our analysis
are listed in Section~\ref{sec:8}. There we also make a comparison of our
findings with those known from other simple models.
Few technicalities are collected in Appendices.

\section{Flavour Structure of the 331 Models \label{sec:2}}
\subsection{Preliminaries}
Among the many scenarios that have been proposed to extend the Standard Model (SM), one with rather appealing features is that based on the gauge group $SU(3)_C \times SU(3)_L \times U(1)_X$, originally developed in 
\cite{Pisano:1991ee,Frampton:1992wt}.
This group is spontaneously broken to the SM group $SU(2)_L \times U(1)_X$, subsequently broken down to $U(1)_Q$. Therefore, the 331 model has an extended Higgs sector \cite{Diaz:2003dk},
 with the first symmetry breaking occurring at a scale much larger than
the electroweak scale.

One of the nice features of this model is
that the requirement of anomaly cancelation, together with that of asymptotic freedom of QCD, constrains the number of generations to be equal to the number of colours, thus providing an explanation for the existence of 3 generations, long sought for in the SM.
Requirement of anomaly cancelation has also consequences on the transformation properties of fermions.
In fact, as a first consequence of the extension of the SM gauge group $SU(2)_L$ to $SU(3)_L$, one has that left-handed fermions transform as triplets (or antitriplets) under the action of $SU(3)_L$.
In order to have an anomaly-free   theory, the number of triplets should be equal to the number of antitriplets. A possible choice is to assume that the three lepton generations  transform as antitriplets, so that, taking into account the three colour possibilities for the quarks, the number of quark triplets should be equal to the number of antitriplets plus one. Hence  two quark generations should  transform as triplets, one as an antitriplet.
The common choice of having the third generation with different transformation properties might be at the origin of the large mass of the top quark with respect to the other quarks.
However, in contrast to the SM, where anomaly is canceled for each generation of fermions, in the 331 model the cancelation is fulfilled only when all the generations are considered.

As for the content of the (anti)triplets, the upper two components coincide with the known SM fermions. In the case of quarks, the third members of the generations correspond to new heavy quarks, while for leptons different choices are possible. The minimal version of the model  makes the choice that no new leptons are introduced, so that the third component of the lepton antitriplet is chosen as the conjugate of the charged lepton (in practice, all the three possible helicity states for the SM leptons are contained). In this minimal version, among the three new heavy quarks two have charge $-\displaystyle{4 \over 3}$ and one $\displaystyle{5 \over 3}$. Other variants are possible. In particular, in the lepton sector, another possibility could be that of having
 new heavy neutrinos as third components of the lepton antitriplets \cite{Hoang:1999yv}. Indeed in the present analysis we choose a variant of this latter kind that  has phenomenological implications that are more interesting than
found in the minimal model.

The extension of the SM gauge group $SU(2)_L$ to $SU(3)_L$ also implies the existence of 5 new gauge bosons.
These are a new neutral boson ($Z^\prime$) plus other four that might be charged depending on the variant of the model that one chooses. We denote them generically as  $V^{\pm Q_V}$ and  $Y^{\pm Q_Y}$.
As we shall see when describing the model in details, the new  bosons $V^{\pm Q_V}$ and $Y^{\pm Q_Y}$ couple two SM leptons, while this is not possible
for SM quarks, which can only be coupled to new heavy quarks by means
of these gauge bosons. Therefore the latter are assigned lepton number $L=\mp 2$, but carry no lepton generation number so that the lepton generation number
 can be violated and indeed processes like $\mu \to e \gamma$ or $\mu \to 3 e$ might proceed due to such new gauge boson mediation. Therefore, bounds on this model could come from the MEG experiment \cite{Adam:2011ch}
 the physics programme of which is devoted to the search for the lepton flavour violating process $\mu \to e \gamma$.
Detailed studies of the lepton sector of the model have been performed in \cite{Liu:1993gy,Diaz:2004fs}, in particular in \cite{Liu:1993gy} a mechanism to suppress large CP violation in this sector as well as large leptonic electric dipole moments was found.

The new neutral gauge boson $Z^\prime$ also introduces new features with respect to the SM as well as with respect to
other NP scenarios that also contain a new neutral $Z^\prime$, as for example grand unified theories (GUTs). In fact, in contrast to most of them, the $Z^\prime$ mass in some 331 models cannot be arbitrary, but is limited from above.
\footnote{Another similar case   is for example represented by heterotic string models with supergravity mediated supersymmetry
breaking \cite{Cvetic:1995rj}
 where also the $Z^\prime$ mass cannot be larger than around a TeV.} This occurs because the $U(1)_X$ gauge coupling $g_X$ and the $SU(3)_L$ one $g$ are related by:
\be
{g_X^2 \over g^2}={6 \sin^2 \theta_W \over 1-(1+\beta^2) \sin^2 \theta_W}
\ee
where $\theta_W$ is Weinberg angle and $\beta$ is a  parameter of the model
(see next subsection) that defines its variants. Therefore, considering for example the  minimal model with $\beta=\sqrt{3}$, one should require  $\sin^2 \theta_W(M_{Z^\prime})< \displaystyle{1 \over 4}$. Since we know that $\sin^2 \theta_W(M_{Z})\simeq 0.23$, the evolution up to $M_{Z^\prime}$ constrains  $M_{Z^\prime}$ below a few TeV \cite{Ng:1992st}.
On the other hand for $\beta=1/\sqrt{3}$, as dominatly considered in our
paper, this bound does not apply.

Also lower bounds on $M_{Z^\prime}$  were discussed in literature, stemming from study of $\mu$ decay within the model \cite{Ng:1992st,Liu:1993gy,Liu:1994rx}, from the analysis of $S$, $T$, $U$ parameters \cite{Hoang:1999yv} and from comparison with experimental results from CDF Collaboration for possible $Z^\prime$ decays to $e^+ \, e^-$ or $\mu^+ \mu^-$ \cite{Langacker:2000ju}. In principle those bounds are model dependent, since they depend also on the entries of the unitarity matrices that transform the quark gauge eigenstates into mass eigenstates. However, taking into account the results of the aforementioned analyses, we adopt as working range $1\,{\rm TeV} \le M_{Z^\prime}\le 3 \,{\rm TeV}$, which also shows that present experiments, mainly those at the LHC, have the possibility to test the model in the nearby future.
Along the same lines, a possible mixing angle between the SM $Z$ and the new $Z^\prime$ has been considered, with the result that $\theta_{Z-Z^\prime}<{\cal O}(10^{-3})$ \cite{Langacker:2000ju}. In what follows we will neglect this
mixing.

The main novelty with respect to the SM related to the $Z^\prime$ is the existence of tree level flavour changing neutral currents (FCNC). However, these arise only in the quark sector, since the universality of the coupling  of the $Z^\prime$ to leptons guarantees that no FCNC show up in this case. Furthermore, universality is also realized in the $Z^\prime$ couplings to right-handed quarks, so that the new FCNC are purely left-handed. The relevant couplings in this case are anyway small enough to make such tree level transition not larger than the
corresponding loop induced SM contribution.
Many processes induced by FCNC have been considered in the framework of the 331 model. Mixing of neutral mesons as well as a number of rare $K$ and $B_{d,s}$ decays have been considered in \cite{Liu:1994rx,Rodriguez:2004mw}. Some of the results have been updated in \cite{Promberger:2007py}, with particular attention to the neutral meson mixing and the study of possible new CP violating effects. The main result contained there is that the most sensitive quantity is the phase of $B_s$ mixing $\beta_s$, as determined from $B_s \to J/\psi \phi$. This is partly a consequence of the expectation that purely hadronic decays are potentially more sensitive to the $Z^\prime$ contributions since in addition to
being flavour violating, the flavour conserving couplings of $Z'$ to quarks is enhanced with respect to its coupling to leptons. We shall see this explicitly
below.
Finally, $b \to s \gamma$ was considered in \cite{Agrawal:1995vp} and reanalysed in \cite{Promberger:2008xg}. Notice that this mode is loop-induced also in the 331 model.

As a final remark, we mention that the 331 model  naturally possesses the Peccei-Quinn symmetry \cite{Peccei:1977hh}, solving the strong CP problem \cite{Pal:1994ba}.

\subsection{The Model}
\subsubsection{Gauge Bosons}
In this section we provide a description of the 331 model in general terms
and some of its variants parametrized by $\beta$.

  Leaving aside the case of $SU(3)_C$, that presents no differences with respect to the SM, the generators of the gauge group are: $T^a=\displaystyle{\lambda^a \over 2}$ for $SU(3)_L$ ($\lambda^a$ being the Gell-Mann matrices and $a=1,\dots 8$) plus $T^9$ which generates $U(1)_X$. $T^9$ can be conveniently defined as $T^9=\displaystyle{\mathbb{1} \over \sqrt{6}}$, the unit matrix being $3 \times 3$, when acting on triplets and just 1 when acting on singlets. With this definition, the generators satisfy the same relation as the eight generators of $SU(3)$: $Tr[T^aT^b]=\displaystyle{\delta^{ab} \over 2}$. In the case of antitriplets, transforming as $\bar 3$, the generators are $\bar T^a=-(T^a)^T$, $T$ meaning transpose.

The gauge bosons of the group $SU(3)_L$ are  denoted as $W^a$, $a=1,\dots 8$ and transform according to the adjoint representation of the group, while the gauge boson of $U(1)_X$ is denoted by $X$.
As a consequence, the covariant derivatives acting on the various fields read as follows:
\begin{itemize}
\item triplet $\psi_L$: $D_\mu \psi_L =\partial_\mu \psi_L -i\,g\,W^a_\mu T^a \, \psi_L -i g_X \,X \, X_\mu T^9 \psi_L$;
\item antitriplet $\bar \psi_L$: $D_\mu \bar \psi_L =\partial_\mu \bar \psi_L +i\,g\,W^a_\mu (T^a)^T \, \bar \psi_L -i g_X \,X \, X_\mu T^9 \bar \psi_L$;
\item singlet $\psi_R$: $D_\mu \psi_R =\partial_\mu \psi_R  -i g_X \,X \, X_\mu T^9 \psi_R$.
\end{itemize}
In the previous formulae, $g$, $g_X$ denote coupling constants and $X$ is the new quantum number corresponding to $U(1)_X$.

The matrix $W^a T^a$ can be rearranged as follows:
\be
W_\mu=W^a_\mu T^a= {1 \over 2} \,\left(\begin{array}{ccc}
W_\mu^3+{1 \over \sqrt{3}} W_\mu^8 & \sqrt{2} W_\mu^+ & \sqrt{2} Y_\mu^{Q_Y}  \\
\sqrt{2} W_\mu^-  & -W_\mu^3+{1 \over \sqrt{3}} W_\mu^8& \sqrt{2} V_\mu^{Q_V}   \\
\sqrt{2} Y_\mu^{-Q_Y}  & \sqrt{2} V_\mu^{-Q_V}  & -{2 \over \sqrt{3}} W_\mu^8
\end{array}\right) \label{WaTa}
\end{equation}
where the following definitions have been introduced:
\bea
W_\mu^\pm &=& {1 \over \sqrt{2}} (W_\mu^1 \mp i \, W_\mu^2),  \\
Y_\mu^{\pm Q_Y} &=& {1 \over \sqrt{2}} (W_\mu^4 \mp i \, W_\mu^5), \label{charged} \\
V_\mu^{\pm Q_V} &=& {1 \over \sqrt{2}} (W_\mu^6 \mp i \, W_\mu^7).
\eea
$W^3$ and $W^8$ correspond to neutral gauge bosons;
$W^\pm$  to the charged gauge bosons of the SM, while $V^{\pm Q_V}$ and $Y^{\pm Q_Y}$ might be charged or neutral. In order to assess this
precisely, we have to define the electric charge matrix.
In analogy with the SM, we introduce the following operator:
\bea
{\hat Q}_W &=& {\hat T}^3 + {{\hat Y} \over 2} \label{Qw} \\
{{\hat Y} \over 2}&=& \beta {\hat T}^8 +X  \mathbb{1} \,\,. \label{Y} \eea
The constant $\beta$ that has been just introduced plays a key role in the model: according to its value various versions of the model have been proposed, which significantly differ from each other. In particular, only  for some values of $\beta$ the gauge bosons turn out to have integer charges.
In the case of the minimal 331 model, the choice $\beta=\sqrt{3}$ is adopted.

Even though we shall recall the features of the minimal model, we shall instead consider the variant with $\beta=\displaystyle{1 \over \sqrt{3}}$,
which is phenomenologically more interesting than the $\beta=\sqrt{3}$ case.
Moreover, in contrast to the minimal model, it does not introduce
exotic electric charges of fermions in order to cancel the anomalies.  We also note that $|\beta|=1/\sqrt{3}$ is the smallest value of this parameter for
which the heavy gauge bosons have integer electric charges.

In (\ref{Qw},\ref{Y}),  the hat reminds us that we are dealing with operators and their action should be  distinguished from the simple
multiplication with the corresponding matrices.
Let us consider in turn the various cases:
\begin{itemize}
\item Triplets:
${\hat Q}_W \, \psi_L= Q_W^{triplet} \psi_L $ with
\be
Q_W^{triplet}=\left(\begin{array}{ccc}
{1 \over 2}+{\beta \over 2 \sqrt{3}}+X & 0 & 0  \\
0 & -{1 \over 2}+{\beta \over 2 \sqrt{3} }+X& 0  \\
0& 0&-{\beta \over  \sqrt{3}}+X
\end{array}\right) \label{Qw-triplet}
\end{equation}
\item Antitriplets:
${\hat Q}_W \, {\bar \psi}_L= Q_W^{antitriplet} {\bar \psi}_L $ with
\be
Q_W^{antitriplet}=\left(\begin{array}{ccc}
-{1 \over 2}-{\beta \over 2 \sqrt{3}}+X & 0 & 0  \\
0 & {1 \over 2}-{\beta \over 2 \sqrt{3} }+X& 0  \\
0& 0&{\beta \over  \sqrt{3}}+X
\end{array}\right) \label{Qw-antitriplet}
\end{equation}
\item Singlets
${\hat Q}_W \, \psi_R= Q_W^{singlet} \psi_R $ with $Q_W^{singlet}=X$.
\item Gauge bosons: ${\hat Q}_W \, W_\mu= [Q_W,W_\mu]=Q_W^{GB}W_\mu  $ so that the charges of the entries in~(\ref{WaTa}) can be read from
the matrix $Q_W^{GB}$ (GB standing for gauge bosons):
\be
Q_W^{GB}=\left(\begin{array}{ccc}
0 & 1 & {1 \over 2}+\beta { \sqrt{3} \over 2}  \\
-1 & 0 & -{1 \over 2}+\beta { \sqrt{3} \over 2}  \\
-{1 \over 2}-\beta { \sqrt{3} \over 2} & {1 \over 2}-\beta { \sqrt{3} \over 2} & 0
\end{array}\right) \label{Qw-GB}
\end{equation}
\end{itemize}
We can see that there are three neutral gauge bosons, two singly charged ones, that can be identified with the SM $W^\pm$ bosons, and other four  bosons. Choosing $\beta=\displaystyle{1 \over \sqrt{3}}$ these are two singly charged bosons $Y^{\pm }$ and two additional neutral bosons $V^0,\,{\bar V}^0$. For comparison, in the minimal model one finds instead two new singly charged bosons $V^\pm$ and two new doubly charged ones $Y^{\pm \pm}$.

After these general discussion we now consider the particle content of the
model, that is fermions, Higgs particles and gauge bosons. The main target in
this step is the derivation of the interaction Lagrangian in the mass eigenstate basis for all fields involved. This will allow to find directly the relevant
Feynman rules. We begin with fermions.
\subsubsection{Fermions}
 As already stated, left-handed fermions fit in (anti)triplets, while right-handed ones are singlets under the action of $SU(3)_L$.
In the case of the quarks, the first two generations  fill the two upper components of a triplet, while the third one fills those of an  antitriplet; the  third member of the quark (anti)triplet is a new heavy fermion.
Therefore we have:
\bea
\left(\begin{array}{c}
u  \\
 d   \\
D  \\
\end{array}\right)_L \hskip 2 cm
\left(\begin{array}{c}
c  \\
 s   \\
S  \\
\end{array}\right)_L \hskip 2 cm
\left(\begin{array}{c}
b  \\
 -t   \\
T  \\
\end{array}\right)_L .\hskip 2 cm
\label{quarksL}
\eea
 The corresponding right handed quarks are singlets. Notice that in the third generation of quarks a different order is chosen with respect to the SM. This stems from the requirement of the anomaly cancelation, as discussed at the beginning of  this section. The sign of the  fields in the antitriplet is fixed requiring that the SM couplings are reproduced.

 The  case of leptons is more involved. As already stated, left-handed leptons fit in antitriplets, and again the SM fields fill the upper two components of the antitriplets.
 The third member of the antitriplet is chosen in a different manner depending on the considered variant of the model.
 In the minimal version
 it is just the conjugate of the charged lepton in the same antitriplet, while in a generic version of the model it could be a new heavy lepton. In the latter case, one needs to introduce  the right handed charged leptons as singlets.
 Therefore, the antitriplets look as follows.
\bea
\left(\begin{array}{c}
e  \\
 -\nu_e   \\
E_e  \\
\end{array}\right)_L \hskip 2 cm
\left(\begin{array}{c}
\mu  \\
 -\nu_\mu   \\
E_\mu  \\
\end{array}\right)_L \hskip 2 cm
\left(\begin{array}{c}
\tau  \\
 -\nu_\tau   \\
E_\tau  \\
\end{array}\right)_L ,\hskip 2 cm
\,\, \eea
with $E_e=e^c,\,E_\mu=\mu^c,\,E_\tau=\tau^c$ only in the minimal version. Hence, in the minimal model one can avoid
the introduction of new leptons as well as right-handed leptons in separate multiplets. Right-handed neutrinos are not compatible
with the minimal model. In the case of $\beta = 1/\sqrt{3}$ we have $E_e=\nu_e^c,\,E_\mu=\nu_\mu^c,\,E_\tau=\nu_\tau^c$.

The electric charge of all fermions can be read from $Q_W^{triplet}$, $Q_W^{antitriplet}$ and $Q_W^{singlet}$. Furthermore, imposing that the SM
fermions have the usual charge,  the value of $X$ for such fields can be fixed. Due to gauge invariance, $X$ is the same for all the members
of a given (anti)triplet. Therefore, fixing for example the charge of the $u_L$ quark ${1 \over 2}+{\beta \over 2 \sqrt{3}}+X= {2 \over 3}$
allows to assign the  value  $X(u_L)={1 \over 6}-{\beta \over 2 \sqrt{3}}$ which in turn fixes the charge of the new heavy quark $D_L$  in
the same triplet to $Q_D=-{\beta \over  \sqrt{3}}+X={1 \over 6}-{\sqrt{3}\beta \over 2 }$, while correctly reproducing the $d_L$ quark
charge. Finally, for right-handed fields one has $Q_W=X$, $Q_W$ being of course the same as for left-handed fields. In Table \ref{QX} we
collect the electric charges of  all the fermions  and  their $X$  quantum numbers, both for arbitrary $\beta$ and for $\beta=\displaystyle{
1 \over \sqrt{3}}$. In the latter case we see that no exotic charges appear.

\begin{table*}[!tb]
\centering
\begin{tabular}{|c | c |c |c| c| }\hline
 & \multicolumn{2}{c|} {arbitrary $\beta$} & \multicolumn{2}{c|} {$\beta={ 1 \over \sqrt{3}}$} \\
\hline
fermions & $Q_W$ & $X$ & $Q_W$ & $X$ \\ \hline \hline
$u_L,\,c_L$ & ${2 \over 3}$ & ${1 \over 6}-{\beta \over 2 \sqrt{3}}$ & ${2 \over 3}$ & 0 \\ \hline
$u_R,\,c_R$ & ${2 \over 3}$ & ${2 \over 3}$ & ${2 \over 3}$ & ${2 \over 3}$ \\ \hline
$d_L,\,s_L$ & $-{1 \over 3}$ & ${1 \over 6}-{\beta \over 2 \sqrt{3}}$ & $-{1 \over 3}$ & 0 \\ \hline
$d_R,\,s_R$ & $-{1 \over 3}$ & $-{1 \over 3}$ & $-{1 \over 3}$ & $-{1 \over 3}$ \\ \hline
$D_L,\,S_L$ & ${1 \over 6}-{\sqrt{3}\beta \over 2 }$ & ${1 \over 6}-{\beta \over 2 \sqrt{3}}$ & $-{1 \over 3}$ & 0 \\ \hline
$D_R,\,S_R$ & ${1 \over 6}-{\sqrt{3}\beta \over 2 }$ & ${1 \over 6}-{\sqrt{3}\beta \over 2 }$ & $-{1 \over 3}$ &$-{1 \over 3}$  \\ \hline
\hline
$b_L$ & $-{1 \over 3}$ & ${1 \over 6}+{\beta \over 2 \sqrt{3}}$ & $-{1 \over 3}$ & ${1 \over 3}$  \\ \hline
$t_L$ & ${2 \over 3}$ & ${1 \over 6}+{\beta \over 2 \sqrt{3}}$ & ${2 \over 3}$ &  ${1 \over 3}$ \\ \hline
$b_R$ & $-{1 \over 3}$ &  $-{1 \over 3}$ &  $-{1 \over 3}$ &  $-{1 \over 3}$  \\ \hline
$t_R$ & ${2 \over 3}$ & ${2 \over 3}$ & ${2 \over 3}$ & ${2 \over 3}$ \\ \hline
$T_L$ & ${1 \over 6}+{\sqrt{3} \beta \over 2 }$ & ${1 \over 6}+{\beta \over 2 \sqrt{3}}$ & ${2 \over 3}$ & ${1 \over 3}$ \\ \hline
$T_R$ & ${1 \over 6}+{\sqrt{3} \beta \over 2 }$ &${1 \over 6}+{\sqrt{3} \beta \over 2 }$ & ${2 \over 3}$ & ${2 \over 3}$  \\ \hline \hline
$\ell_L$ & $-1$ & $-{1 \over 2}+{ \beta \over 2 \sqrt{3}}$ & $-1$ & $-{1 \over 3}$ \\ \hline
$\ell_R$ & $-1$ & $-1$ & $-1$ &$ -1$ \\ \hline
$\nu_{\ell \, L}$ & 0 &$-{1 \over 2}+{ \beta \over 2 \sqrt{3}}$ & 0 & $-{1 \over 3}$ \\ \hline
$\nu_{\ell \, R}$ & 0 &0 & 0 & 0 \\ \hline
$E_{\ell L}$ &  $-{1 \over 2}+{ \sqrt{3}\beta \over 2 }$ &$-{1 \over 2}+{ \beta \over 2 \sqrt{3}}$ & 0 & $-{1 \over 3}$ \\ \hline
$E_{\ell R}$ & $-{1 \over 2}+{ \sqrt{3}\beta \over 2 }$ & $-{1 \over 2}+{ \sqrt{3}\beta \over 2 }$ & 0 & 0 \\ \hline \hline
 \end{tabular}
\caption{\it Electric charges and $X$ quantum numbers of the fermions in the 331 model for arbitrary $\beta$ and for $\beta={ 1 \over
\sqrt{3}}$. In the case of leptons $\ell=e,\,\mu,\,\tau$. }\label{QX}~\\[-2mm]\hrule
\end{table*}

\subsubsection{Higgs sector}
Symmetry breaking  is accomplished in two steps. At first one has to recover the SM as a low energy effective theory deriving from the 331 model after the spontaneous symmetry breaking $SU(3)_L \times U(1)_X \to SU(2)_L \times U(1)_X$ at a given scale, that has to be properly identified with a vev of a Higgs field. Denoting with $<\phi_1>$ such a vev, in this step it should be required that

\be
{\hat T}^1<\phi_1>={\hat T}^2<\phi_1>={\hat T}^3<\phi_1>=(\beta {\hat T}^8+X  \mathbb{1})<\phi_1>=0 \,\,, \label{SSB1}
 \ee
 while all the other generators, when acting on the vacuum, give a non vanishing result.

In the second step, the familiar SM symmetry breaking $SU(2)_L \times U(1)_X \to U(1)_Q$ should be realized at  the electroweak (EW) scale, the corresponding Higgs vev being denoted by $<\phi_2>$. Now the only generator that leaves the vacuum invariant should be $Q_W$:
\be
{\hat Q}_W<\phi_2>=0 \,\,.
\label{SSB2}
\ee
Since the final goal is to give masses to the fermions and all the gauge bosons except the photon,
the procedure to be followed is still the building of a Lagrangian for the Higgs fields containing the coupling to gauge bosons through the
covariant derivative  and a Yukawa Lagrangian coupling such fields to fermions, in such a manner that the symmetries of the model are preserved. In
particular, the $X$ conservation should be guaranteed.
It turns out that invariant terms involving a fermion triplet and a singlet can be built  introducing a Higgs field that transforms either as a triplet or as a sextet. However, in the case of the sextet, requiring that the conditions (\ref{SSB1}) are satisfied constrains the parameter $\beta$ to specific values (different for quarks and leptons) that are not compatible with the scenario considered in this paper. Therefore, we are left with the choice of a Higgs triplet, that we denote as $\chi$. In this case, the first three conditions in (\ref{SSB1}) constrain  the vev of $\chi$ to the following structure:
\be
<\chi>={1 \over \sqrt{2}} \left(\begin{array}{c}
0  \\
 0   \\
u  \\
\end{array}\right) \,\,.
\label{vev-chi}
\ee
Furthermore, imposing the last condition in (\ref{SSB1}), i.e. $(\beta {\hat T}^8+X  \mathbb{1})<\chi>=0$, fixes the  quantum number $X$ for $\chi$ to the value $X_\chi=\displaystyle{\beta \over \sqrt{3}}$.

We now turn to the second step in the spontaneous symmetry breaking. Again, the Higgs field can transform under the action of $SU(3)_L$ either as a triplet or as a sextet. Let us start with the first possibility. Requiring that all the generators except $Q_W$ are broken implies that one can choose two possible Higgs triplets,  with the following vevs:
\be
<\rho>={1 \over \sqrt{2}} \left(\begin{array}{c}
0  \\
 v   \\
0  \\
\end{array}\right)\,\,, \hskip 2 cm
<\eta>={1 \over \sqrt{2}} \left(\begin{array}{c}
v^\prime  \\
 0   \\
0  \\
\end{array}\right)
\,\,.
\label{vev-eta}
\ee
The last condition: ${\hat Q}_W<\rho>={\hat Q}_W<\eta >=0$ finally fixes:
\be
X_\rho=\displaystyle{1 \over 2}-\displaystyle{\beta \over 2 \sqrt{3}}, \qquad
X_\eta=-\displaystyle{1 \over 2}-\displaystyle{\beta \over 2 \sqrt{3}}.
\ee

The case of the sextet can be carried out following the same steps and recalling that the action of the $SU(3)$ matrices $T^a$ on a generic sextet $\tilde S$ is defined as: ${\hat T}^a \tilde S=\tilde ST^a+(T^{a})^T\tilde S$.
The result is that there are four possible independent sextets.
A usual choice, that we adopt in this paper as well, is to keep only the following Higgs sextet:
\bea
\tilde S={1 \over 2} \left(\begin{array}{ccc}
0 & 0 & 0  \\
0 & 0 & w  \\
0 & w & 0  \\
\end{array}\right)\,; \hskip 0.5 cm X_{\tilde S}={1 \over 4}+{\beta \over 4 \sqrt{3}}\,.
 \eea
The  $X$ value reported above has been  fixed as done  in the case of the Higgs triplets.

Notice that, since the EW symmetry breaking should occur at a scale much lower than the scale at which $SU(3)_L$ is broken down to $SU(2)_L$, the following hierarchy among the vevs should hold: $u \gg v,\,v^\prime,\,w$.

The charge of the fields in the three Higgs triplets $\chi$, $\rho$, $\eta$ can be read from $Q^{triplet}$ by inserting the corresponding value of $X$. As for the Higgs sextet $\tilde S$, the electric charge matrix is:
\be
Q_W^{sextet}=\left(\begin{array}{ccc}
1+{\beta \over \sqrt{3}} +2X  & {\beta \over \sqrt{3}} +2X & {1 \over 2}-{\beta \over 2 \sqrt{3}} +2X  \\
{\beta \over \sqrt{3}} +2X & -1+{\beta \over \sqrt{3}} +2X & -{1 \over 2}-{\beta \over 2 \sqrt{3}} +2X  \\
{1 \over 2}-{\beta \over 2 \sqrt{3}} +2X & -{1 \over 2}-{\beta \over 2 \sqrt{3}} +2X & -{2\beta \over  \sqrt{3}} +2X
\end{array}\right)\,. \label{Qw-sextet}
\end{equation}
It turns out that the electric charges for  complete set of Higgs fields can be expressed in terms of the following $\beta$-dependent quantities
$Q_A={1 \over 2}+{\sqrt{3} \beta \over 2 }$, $Q_B=-{1 \over 2}+{\sqrt{3} \beta \over 2 }$ and $Q_C=-{3 \over 2}+{\sqrt{3} \beta \over 2 }$:
\bea
Q_\chi&=&\left(\begin{array}{ccc}
Q_A & 0 & 0  \\
0 &Q_B & 0  \\
0 &0 & 0  \\
\end{array}\right)\,,
 \hskip 0.2cm
 Q_\rho=\left(\begin{array}{ccc}
1 & 0 & 0  \\
0 & 0 & 0  \\
0 &0 & -Q_B  \\
\end{array}\right)\,,
\hskip 0.2cm
 Q_\eta=\left(\begin{array}{ccc}
0 & 0 & 0  \\
0 & -1 & 0  \\
0 &0 & -Q_A  \\
\end{array}\right)\,,  \\
Q_S&=&\left(\begin{array}{ccc}
Q_A & Q_B & 0  \\
Q_B & Q_C& -1  \\
0 & -1 & -Q_A  \\
\end{array}\right)\,.  \label{Higgs-sextet} \eea
For  $\beta=\displaystyle{1 \over \sqrt{3}}$ one has: $Q_A=1$, $Q_B=0$ and $Q_C=-1$ (for comparison, in the minimal model one finds $Q_A=2$, $Q_B=1$ and $Q_C=0$).
Since the sextet does not enter in the quark-Higgs interactions, we shall only concentrate on the three triplets and, on the basis of their electric charges, we adopt the following notation:
\be
\chi= \left(\begin{array}{c}
\chi_A  \\
 \chi_B   \\
\chi^0  \\
\end{array}\right)\,, \hskip 0.5 cm
\rho= \left(\begin{array}{c}
\rho^+  \\
 \rho^0   \\
\rho_B  \\
\end{array}\right)\,, \hskip 0.5 cm
\eta= \left(\begin{array}{c}
\eta^0  \\
 \eta^-   \\
\eta_A  \\
\end{array}\right) \label{higgs-charged} \,.\ee

The neutral fields $\chi^0$, $\rho^0$ and $\eta^0$ can be decomposed into their real and imaginary parts:
\bea
\chi^0 &=&{1 \over \sqrt{2}}(\xi_\chi +i \zeta_\chi)\,\,, \hskip 0.2cm  <\xi_\chi>=u\,\,,  \\
\rho^0 &=&{1 \over \sqrt{2}}(\xi_\rho +i \zeta_\rho)\,\,, \hskip 0.2cm  <\xi_\rho>=v \,\,, \\
\eta^0 &=&{1 \over \sqrt{2}}(\xi_\eta +i \zeta_\eta)\,\,, \hskip 0.2cm  <\xi_\eta>=v^\prime \,\,,
 \eea
 with the ratios of the vevs defining three  angles that govern the mixing among the fields in this rich Higgs sector:
\be
\sin^2 \beta_{v^\prime v}={v^{\prime 2} \over v^2+v^{\prime 2}}\,, \hskip 1 cm \sin^2 \beta_{vu}={v^2 \over u^2+v^2} \,,\hskip 1 cm \sin^2
\beta_{v^\prime u}={v^{\prime 2} \over u^2+v^{\prime 2}} \label{tan} \,\,.
\ee

Introducing the Higgs potential:
\bea
V_{Higgs}(\rho,\eta,\chi)&=& \mu_1^2 (\rho^\dagger \rho)+ \lambda_1 (\rho^\dagger \rho)^2+
 \mu_2^2 (\eta^\dagger \eta)+ \lambda_2 (\eta^\dagger \eta)^2+
 \mu_3^2 (\chi^\dagger \chi)+ \lambda_3 (\chi^\dagger \chi)^2  \nn \\
 &+& \lambda_{12}(\rho^\dagger \rho)(\eta^\dagger \eta)+
\lambda_{13}(\rho^\dagger \rho)(\chi^\dagger \chi)+
\lambda_{23}(\eta^\dagger \eta)(\chi^\dagger \chi)+  \nn \\
 &+& {\tilde \lambda}_{12}(\rho^\dagger \eta)(\eta^\dagger \rho)+
{\tilde \lambda}_{13}(\rho^\dagger \chi)(\chi^\dagger \rho)+
{\tilde \lambda}_{23}(\eta^\dagger \chi)(\chi^\dagger \eta )+ \nn  \\
&+& \sqrt{2} f_1 (\epsilon_{ijk}\rho^i \eta^j \chi^k+h.c.)
\eea
 and  minimizing it with respect to $u,\,v,\,v^\prime$ one can express $\mu^2_i$, $i=1,2,3$ in terms of the other parameters in
$V_{Higgs}(\rho,\eta,\chi)$.

 Higgs fields mix,   giving rise to the  mass eigenstates listed below.
\begin{itemize}
\item Singly charged states having $|L=0|$:
\be
\left(\begin{array}{c}
\phi_W^\pm  \\
H_0^\pm   \\
\end{array}\right)=\left(\begin{array}{cc}
\cos \beta_{v^\prime v} & -\sin \beta_{v^\prime v}  \\
\sin \beta_{v^\prime v} & \cos \beta_{v^\prime v}   \\
\end{array}\right)
\left(\begin{array}{c}
\rho^\pm  \\
\eta^\pm   \\
\end{array}\right) \label{Higgs-mix-1}
\ee
with $M_{\phi_W^\pm}=0$ and  $M_{H_0^\pm}^2={f_1 u  \over \sin \beta_{v^\prime v} \cos \beta_{v^\prime v}}$, where we have exploited the hierarchy  $u \gg v,\,v^\prime$.
\item
States with $L=2$ and charge $Q_B$:
\be
\left(\begin{array}{c}
\phi_V^B  \\
H_2^B   \\
\end{array}\right)=\left(\begin{array}{cc}
\cos \beta_{13} & -\sin \beta_{13}  \\
\sin \beta_{13} & \cos \beta_{13}   \\
\end{array}\right)
\left(\begin{array}{c}
\rho^B  \\
\chi^B   \\
\end{array}\right) \label{Higgs-mix-2}
\ee
with $M_{\phi_V^B}=0$ and  $M_{H_2^B}^2=f_1 u \tan \beta_{v^\prime v} +u^2 \displaystyle{{\tilde \lambda_{13}} \over 2}$ and $\beta_{13}=\displaystyle{\pi \over 2}-\beta_{vu}$.
\item States with charge $Q_A$:
\be
\left(\begin{array}{c}
\phi_Y^A  \\
H_2^A  \\
\end{array}\right)=\left(\begin{array}{cc}
\cos \beta_{23} & -\sin \beta_{23}  \\
\sin \beta_{23} & \cos \beta_{23}   \\
\end{array}\right)
\left(\begin{array}{c}
\eta^A \\
\chi^A \\
\end{array}\right) \label{Higgs-mix-3}
\ee
with $M_{\phi_Y^A}=0$ and  $M_{H_2^A}^2=f_1 u \cot \beta_{v^\prime v} +u^2 \displaystyle{{\tilde \lambda_{23}} \over 2}$ and $\beta_{13}=\displaystyle{\pi \over 2}-\beta_{v^\prime u}$.
\item Neutral pseudoscalar states:
\be
\left(\begin{array}{c}
\phi_Z  \\
\phi_{{\bar H}_0}  \\
\phi_{Z^\prime} \\
\end{array}\right)=\left(\begin{array}{ccc}
\cos \beta_{v^\prime v} & -\sin \beta_{v^\prime v} & 0  \\
\sin \beta_{v^\prime v} & \cos \beta_{v^\prime v} & 0  \\
0 & 0& -1
\end{array}\right)
\left(\begin{array}{c}
\xi_\rho  \\
\xi_\chi   \\
\xi_\eta  \\
\end{array}\right) \label{Higgs-mix-4}
\ee
\end{itemize}
with $M_{\phi_Z}^2=M_{\phi_{Z^\prime}}^2=0$ and $M^2_{\phi_{{\bar H}_0}}=\displaystyle{f_1 u \over 2 \sin \beta_{v^\prime v} \cos \beta_{v^\prime v}}$.
The real parts of the neutral Higgs fields gives raise to scalar states, with a complicated mixing pattern.
Quarks and gauge bosons couple to the various Higgs fields  defined by the above mixing.

\subsubsection{Masses of the gauge bosons.}
Having fixed the quantum numbers of the various Higgs fields, we can proceed to derive the masses of the gauge bosons.
These stem from the covariant derivative in the Higgs Lagrangian:
\be
L_{Higgs}=(D_\mu \chi)^\dagger (D^\mu \chi)+(D_\mu \rho)^\dagger (D^\mu \rho)+(D_\mu \eta)^\dagger (D^\mu \eta)+Tr[(D_\mu {\tilde S})^\dagger (D^\mu {\tilde S})]
\label{cov-der-H} \,\,.
\ee
The masses of the $W^\pm$, $V^{\pm Q_V}$, $Y^{\pm Q_Y}$ bosons read:
\bea
M^2_{W^\pm} &=& {g^2 \over 4}v_+^2 \,\,,\label{mW} \\
M^2_{V^{\pm Q_V}} &=& {g^2 \over 4} u^2 \left[ 1- {v_-^2 \over 2 u^2}+{v_+^2 \over 2 u^2} \right] \,\,,\label{mV} \\
M^2_{Y^{\pm Q_Y}} &=& {g^2 \over 4} u^2 \left[ 1+ {v_-^2 \over 2 u^2}+{v_+^2 \over 2 u^2} + 4 {w^2 \over u^2} \right] \label{mY} \,\,,\eea
where we have defined: $v_+^2=v^2+v^{\prime 2}+w^2$ and $v_-^2=v^{\prime 2}-v^2-w^2$.
The remaining neutral gauge bosons mix with each other.
In a first step, the two neutral gauge bosons $W^8_\mu$ and $X_\mu$ mix, giving rise to the two bosons $B_\mu$ and $Z^\prime_\mu$. The mixing angle is denoted by $\theta_{331}$ and is given by  $\sin \theta_{331}=\displaystyle{g \over \sqrt{g^2 + {g_X^2 \beta^2 \over 6}}}$. $B_\mu$ and $Z^\prime_\mu$ mix with $W_\mu^3$. However, taking into account the
hierarchy $u \gg v,\,v^\prime,\,w$, it turns out that the mixing involves only $B_\mu$ and $W^3_\mu$ finally giving a massless photon
$A_\mu$ and the $Z_\mu$ to be identified with the SM $Z$.
Introducing a new constant
\be
{1 \over g_Y^2}={6 \over g_X^2}+{\beta^2 \over g^2}
\ee
it turns out that the mixing angle that defines the rotation matrix providing the mass eigenstates $A,\,Z$ starting from $W^3,\,B$ is given by
\be
\sin \theta_W={g_Y \over \sqrt{g^2+g_Y^2}} \,\,.
\ee
Therefore, in this model  the following relation between the two mixing angles $\theta_{331}$ and $\theta_W$  holds:
\be
\cos \theta_{331}=\beta \tan \theta_W \,\,.
\ee
Thus the mixing of these neutral gauge bosons can be summarized as follows:
\begin{equation}
X_\mu\,,W^8_\mu\,,W^3_\mu\,\, \xrightarrow{\theta_{331}}\,\, Z^\prime_\mu\,,B_\mu\,,W^3_\mu\,\,\xrightarrow{\theta_W}\,\,
Z^\prime_\mu\,,Z_\mu\,,A_\mu\,.
\end{equation}
The  resulting masses for the neutral gauge bosons are:
\bea
M_A^2&=& 0 \,\,,\label{mA} \\
M_Z^2&=& {g^2 \over 4 \cos \theta_W}v_+^2\,\,, \label{mZ} \\
M^2_{Z^\prime} &=& {g^2 u^2 \cos^2 \theta_W \over 3[1-(1+\beta^2) \sin^2 \theta_W]} \left[ 1+{v_+^2 \over 4 u^2} \left(1+3 \beta^2 {\sin^4 \theta_W \over \cos^4 \theta_W} \right)+{\sqrt{3} \beta \over 2}{v_-^2 \over u^2} {\sin^2 \theta_W \over \cos^2 \theta_W} \right] \label{mZprime} \,\,.
\eea

Considering again that  $u \gg v,\,v^\prime,\,w$  one is lead to the final expressions for all the gauge boson masses:
\bea
M^2_{W^\pm} &=& {g^2 \over 4}v_+^2\,\,, \label{mWfin} \\
M^2_{V^{\pm Q_V}} &=& M^2_{Y^{\pm Q_Y}} ={g^2 \over 4} u^2 \,\,, \label{mVYfin} \\
M^2_A &=& 0\,\,, \label{mAfin} \\
M_Z^2&=& {g^2 \over 4 \cos \theta_W}v_+^2 \,\,,\label{mZfin} \\
M^2_{Z^\prime} &=& {g^2 u^2 \cos^2 \theta_W \over 3[1-(1+\beta^2) \sin^2 \theta_W]} \label{mZprimefin} \,\,.
\eea
Therefore,  as in the SM, one has that $\displaystyle{M_W^2 \over M_Z^2}=\cos \theta_W$, which allows to identifiy the angle $\theta_W$ with the Weinberg angle.

\subsubsection{Yukawa Lagrangian and  interaction Lagrangian terms in their final form.}

The next step is the definition of the quark mass eigenstates. The pattern is similar to the 2HDM or SUSY one, with up-type quarks getting mass from the coupling to the Higgs triplet $\rho$ and down-type quarks from the coupling to the Higgs triplet $\eta$. The new quarks $D,\,S,\,T$ get mass from the coupling to the Higgs triplet  $\chi$.
The most general Yukawa Lagrangian coupling left and right handed quarks to Higgs fields in a gauge invariant way is
\bea
L_{Yuk} = && \lambda^d_{i,a} {\bar Q}_i \rho d_{a,R}+\lambda^d_{3,a} {\bar Q}_3 \eta^\star d_{a,R} \nn \\
&+&\lambda^u_{i,a} {\bar Q}_i \eta u_{a,R}+\lambda^u_{3,a} {\bar Q}_3 \rho^\star u_{a,R} \nn \\
&+&\lambda^J_{i,j} {\bar Q}_i \chi J_{j,R}+\lambda^J_{3,3} {\bar Q}_3 \chi^\star T_{R} + h.c. \label{Yuk-gen} \eea
where $Q_i$, $i=1,2$, are the left-handed triplets of the first two quark generations while $Q_3$ is the corresponding one of the third generation;  $J_{1,\,2}=D,\,S$; $a=1,2,3$ with $d_{(1,2,3)R}=d_R,\,s_R,\,b_R$ and $u_{(1,2,3)R}=u_R,\,c_R,\,t_R$.

Upon rotation through unitary matrices one defines the mass eigenstates:
\be
\left(\begin{array}{c}
u^\prime_L  \\
c^\prime_L   \\
t^\prime_L  \\
\end{array}\right)=U_L^{-1}\left(\begin{array}{c}
u_L  \\
c_L   \\
t_L  \\
\end{array}\right)\,, \hskip 1 cm
\left(\begin{array}{c}
d^\prime_L  \\
s^\prime_L   \\
b^\prime_L  \\
\end{array}\right)=V_L^{-1}\left(\begin{array}{c}
d_L  \\
s_L   \\
b_L  \\
\end{array}\right)\,, \label{rotation}
\ee
where the rotation matrices are unitary
\be
U_L^\dagger U_L=U_L U_L^\dagger=V_L^\dagger V_L=V_L V_L^\dagger=1
\ee
and their matrix elements will be denoted as follows
\be
v_{ij} = (V_L)_{ij}, \qquad u_{ij} = (U_L)_{ij}.
\ee
 They will be crucial ingredients in the interaction Lagrangians  below. In the following we shall omit the prime in the notation for these
fields, understanding that we are considering the mass eigenstates. It can be noticed that, among the new quarks, $T$ cannot mix, since its
charge is different from that of the other new quarks.

As for the $D\,,S$ quarks, in principle they could also mix; however, as discussed in
\cite{Promberger:2007py,Liu:1994rx}, a convenient choice is  that of taking them simultaneously as gauge and mass eigenstates.
Indeed the mass matrix $M_J$ of $D,S,T$ is block-diagonal, since $T$ cannot mix. With two
block-diagonal rotation matrices $W_L$ and $W_R$, $M_J$ can be diagonalized. Then $W_R$ can be removed by a redefinition of right-handed
heavy quark fields and $W_L$ can be absorbed into the definition of $U_L$ and $V_L$. Below in Eq.~(\ref{VL-param}) we will choose a
parametrization for $V_L$ that is compatible with this redefinition as discussed in  \cite{Promberger:2007py}.

As in the SM, the rotation of the quark fields affects the charged currents. In complete analogy with the SM, it is possible to define a CKM matrix as
\be
V_{CKM}=U_L^\dagger V_L,
\ee
 which weights the transitions between up-type and down-type SM quarks belonging to different generations mediated by the $W^\pm$.

However, in the present model also neutral currents mediated by the $Z^\prime$ are affected by the quark mixing. This happens because the
couplings of the $Z^\prime$ to the third generation of quarks differ
from the ones to the first two generations resulting in tree-level FCNC
mediated by $Z'$. In order to see this explicitly, we look at $Z^\prime$
couplings to SM quarks
\begin{align}
 &\mathcal{L}^{Z^\prime}= J_\mu Z^{\prime\mu}\,,\\
&J_\mu = \bar u_L\gamma_\mu U_L^\dagger\begin{pmatrix}
                                        a& & \\ & a &\\ & & b
                                       \end{pmatrix}U_L u_L + \bar d_L\gamma_\mu V_L^\dagger\begin{pmatrix}
                                        a& & \\ & a &\\ & & b
                                       \end{pmatrix}V_L d_L\,,
\end{align}
where the couplings $a$ and $b$ depend on the chosen $\beta$. Separating the flavour conserving and flavour changing parts we find
\begin{align}\begin{split}
 J_\mu =&a \left(\bar u_L\gamma_\mu u_L + \bar d_L\gamma_\mu d_L\right) \\
&+ \bar u_L\gamma_\mu U_L^\dagger\begin{pmatrix}
                                        0& & \\ & 0 &\\ & & b-a
                                       \end{pmatrix}U_L u_L + \bar d_L\gamma_\mu V_L^\dagger\begin{pmatrix}
                                        0& & \\ & 0 &\\ & & b-a
                                       \end{pmatrix}V_L d_L\,,\end{split}
\end{align}
which shows  that the different treatment of the third
generation in 331 models generates FCNCs at tree level.

We are now ready to list the interaction Lagrangian densities (multiplied by $i$, in order to directly have the Feynman rules) that give the
couplings of fermions to
various gauge bosons. We list first the terms containing the currents generalizing the analogous SM ones and then those containing the
currents mediated by the new gauge bosons.
\begin{itemize}
\item charged current mediated by $W^\pm$
\bea
i\,L_{int}^W &=& i{g \over \sqrt{2}} \sum_{\ell=e,\mu,\tau} \left( {\bar \nu}_{\ell \, L} \gamma_\mu \ell_L \,W^{+ \mu}+
{\bar \ell}_L \gamma_\mu \nu_{\ell \,L} \,W^{- \mu} \right)
 \\
&+& i{g \over \sqrt{2}}\left( \left(
{\bar u}_L  \,\,
{\bar c}_L  \,\,
{\bar t}_L
\right) \gamma_\mu V_{CKM} \left(\begin{array}{c}
d_L  \\
s_L   \\
b_L  \\
\end{array}\right) W^{+ \mu} +h.c. \right)
\label{cc-current}
\eea

\item e.m. current
\be
i\,L_{int}^\gamma= \sum_f i \,Q e {\bar f} \gamma_\mu f \, A^\mu \,\,, \label{em-current}
\ee
$f$ is a generic lepton or quark field of charge $Q$ in units of the positron charge $e=g\,\sin \theta_W$.
\item
neutral current mediated by $Z$ boson
{\allowdisplaybreaks
\bea
i\,L_{int}^Z &=& { i g \over 2 c_W} Z^\mu \, \nn \\
&\Bigg\{  &\sum_{\ell=e,\mu,\tau} \Big[ {\bar \nu}_{\ell \,L} \gamma_\mu \nu_{\ell L}-(1 -2 s_W^2) {\bar \ell}_L \gamma_\mu \ell_L + 2 s_W^2 {\bar \ell}_R \gamma_\mu \ell_R +(1-\sqrt{3} \beta)s_W^2 {\bar E}_\ell \gamma_\mu E_\ell\Big] \nn
 \\
&+& \sum_{q_u=u,c,t} \Big[\left( 1-{4 \over 3}s_W^2 \right) {\bar q}_{u L} \gamma_\mu q_{u L} -{4 \over 3}s_W^2 {\bar q}_{uR} \gamma_\mu q_{u R}  \Big] \nn \\
&+& \sum_{q_d=d,s,b} \Big[\left( -1+{2 \over 3}s_W^2 \right) {\bar q}_{d L} \gamma_\mu q_{d L} +{2 \over 3} s_W^2{\bar q}_{d R} \gamma_\mu q_{d R}  \Big] \nn \\
&+& 2\left(-{1 \over 6}+{\sqrt{3} \beta \over 2}\right)s_W^2 \left[{\bar D} \gamma_\mu D+ {\bar S} \gamma_\mu S \right]-2\left({1 \over 6}+{\sqrt{3} \beta \over 2}\right) s_W^2 {\bar T} \gamma_\mu T \Bigg\}
\eea}%
where we adopted the notation $s_W=\sin \theta_W$, $c_W=\cos \theta_W$. As in the SM, for ordinary quarks  this current contains a piece proportional to the e.m. current plus a term in which only left-handed fermions are involved. In the case of the new quarks, only the contribution proportional to the e.m. current is present.
\item  current mediated by $V^{\pm Q_V}$
\bea
i\,L_{int}^V &=&i{g \over \sqrt{2}}
\sum_{\ell=e,\mu,\tau} \left[ {\bar \nu}_{\ell \, L} \gamma_\mu E_{\ell L} \,V^{-Q_V \mu}+
{\bar E}_{\ell L} \gamma_\mu \nu_{\ell \,L} \,V^{+Q_V \mu} \right]
 \nn \\
&+& i{g \over \sqrt{2}} \Big\{\sum_{i=1,2,3;\,j = 1,2}\left[ ({\bar q}_{dL})_i \gamma_\mu (Q_{DL})_j  v_{ij}^*  V^{+Q_V\mu} +(\bar
Q_{DL})_j \gamma_\mu (q_{dL})_i v_{ij} V^{-Q_V\mu}\right]
\nn \\
&+& \sum_{i=1,2,3} \left[ {\bar T}_L \gamma_\mu (q_{uL})_i u_{3i} V^{+Q_V \mu} + ({\bar q}_{uL})_i \gamma_\mu T_L
u_{3i}^* V^{-Q_V \mu} \right] \Big\}
\eea
 where $q_{ui}$ $i=1,2,3$ correspond to $u,c,t$; $q_{di}$ $i=1,2,3$ correspond to $d,s,b$; $Q_{Di}$ $i=1,2$ correspond to $D,S$. The elements of the matrices $U_L$ and $V_L$ have been denoted by $u_{ij}$ and $v_{ij}$ respectively.
\item  current mediated by $Y^{\pm Q_Y}$
\bea
i\,L_{int}^Y &=&-i{g \over \sqrt{2}}
\sum_{\ell=e,\mu,\tau} \left[
{\bar E}_{\ell L} \gamma_\mu \ell_L \,Y^{+ Q_Y \mu} + {\bar \ell}_L \gamma_\mu E_{\ell L}  \,Y^{- Q_Y\mu}
\right]
 \nn \\
&+& i{g \over \sqrt{2}} \Big\{ \sum_{i=1,2,3;\,j = 1,2} \left[({\bar q}_{uL})_i \gamma_\mu (Q_{DL})_j u_{ij}^* Y^{+Q_Y\mu} +({\bar
Q}_{DL})_i \gamma_\mu (q_{uL})_j u_{ij} Y^{-Q_Y\mu} \right]
\nn \\
&-& \sum_{i=1,2,3} \left[ {\bar T}_L \gamma_\mu (q_{dL})_i v_{3i} Y^{+Q_Y \mu} +
({\bar q}_{dL})_i \gamma_\mu T_L v_{3i}^*Y^{-Q_Y\mu}
\right] \Big\}
\eea

%%%%%%%%%%%%%%%%%%%%%%%%%%%%%%%%%%%%%%%%%%%%%%%%%%%%%%%%%%%%%%%%%%%%%%%%%%%%%%%%%%%%%%%%%%

\item neutral current mediated by $Z^\prime$
{\allowdisplaybreaks
\bea\label{ZprimeFR}
i\,L_{int}^{Z^\prime} &=&  i{g  Z^{\prime \mu}\over 2 \sqrt{3} c_W \sqrt{1-(1+\beta^2) s_W^2}} \nn\\
 && \Bigg\{ \sum_{\ell=e,\mu,\tau}\Big\{\left[1-(1+\sqrt{3} \beta) s_W^2 \right] \left({\bar \nu}_{\ell \, L} \gamma_\mu \nu_{\ell \, L}+ {\bar \ell}_L \gamma_\mu \ell_L \right)- 2 \sqrt{3} \beta s_W^2 {\bar \ell}_R \gamma_\mu \ell_R \nn
\\
&-&\left[ 2-\left(2-\sqrt{3} \beta(1-\sqrt{3} \beta) \right) s_W^2 \right] {\bar E}_{\ell L} \gamma_\mu E_{\ell L}- \sqrt{3} (1-\sqrt{3} \beta ) \beta s_W^2 {\bar E}_{\ell R} \gamma_\mu E_{\ell R}
\Big\}  \nn \\
&+&
\sum_{i,j=1,2,3} \Big\{ \big[-1+(1+{\beta \over \sqrt{3}})s_W^2\big]({\bar q}_{uL})_i \gamma_\mu (q_{uL})_j \delta_{ij}+ 2c_W^2 ({\bar q}_{uL})_i \gamma_\mu (q_{uL})_j u_{3i}^* u_{3j} \nn \\
&+& \big[-1+(1+{\beta \over \sqrt{3}})s_W^2\big]({\bar q}_{dL})_i \gamma_\mu (q_{dL})_j \delta_{ij}+ 2c_W^2 ({\bar q}_{dL})_i \gamma_\mu (q_{dL})_j v_{3i}^* v_{3j} \nn \\
&+&{4 \over \sqrt{3}}\beta s_W^2({\bar q}_{uR})_i \gamma_\mu (q_{uR})_j \delta_{ij}
-{2 \over \sqrt{3}}\beta s_W^2({\bar q}_{dR})_i \gamma_\mu (q_{dR})_j \delta_{ij}\Big\}
\nn \\
&+&\sum_{i=1,2} \Big\{ \big[ 2+{ -2 \sqrt{3}+\beta(1-3\sqrt{3} \beta) \over \sqrt{3}} s_W^2 \big]({\bar Q}_{DL})_i \gamma_\mu (Q_{DL})_i \nn \\
&+&{\beta (1-3\sqrt{3} \beta) \over \sqrt{3}} s_W^2  ({\bar Q}_{DR})_i \gamma_\mu (Q_{DR})_i \nn \\
 &+&\big[ -2+{ 2 \sqrt{3}+\beta(1+3\sqrt{3} \beta) \over \sqrt{3}} s_W^2 \big]{\bar T}_L \gamma_\mu T_L + {\beta(1+3\sqrt{3} \beta) \over \sqrt{3}} s_W^2 {\bar T}_R \gamma_\mu T_R \Big\}\Bigg\} \,\,.
\eea}%
\end{itemize}
It can be noticed that while the current  mediated by the SM $W^\pm$ bosons depends on the CKM matrix elements, the currents mediated by the new gauge bosons $V,\,Y$ and $Z^\prime$ depend on the elements of the two rotation matrices $U_L$ and $V_L$ introduced above. Due to the relation $U_L^\dagger V_L=V_{CKM}$ one can choose $V_L$ and $V_{CKM}$ as independent matrices. In the following we adopt the parametrization of the matrix $V_L$ reported below \cite{Promberger:2007py}\footnote{A similar parametrization had been proposed in \cite{Liu:1994rx} with the difference that the weak phases were neglected.}:
\be
V_L=\left(\begin{array}{ccc}
\tilde c_{12}\tilde c_{13} & \tilde s_{12}\tilde c_{23} e^{i \delta_3}-\tilde c_{12}\tilde s_{13}\tilde s_{23}e^{i(\delta_1
-\delta_2)} & \tilde c_{12}\tilde c_{23}\tilde s_{13} e^{i \delta_1}+\tilde s_{12}\tilde s_{23}e^{i(\delta_2+\delta_3)} \\
-\tilde c_{13}\tilde s_{12}e^{-i\delta_3} & \tilde c_{12}\tilde c_{23} +\tilde s_{12}\tilde
\tilde s_{13}\tilde s_{23}e^{i(\delta_1-\delta_2-\delta_3)} & -\tilde s_{12}\tilde s_{13}\tilde c_{23}e^{i(\delta_1 -\delta_3)}
-\tilde c_{12}\tilde s_{23} e^{i \delta_2} \\
-\tilde s_{13}e^{-i\delta_1} & -\tilde c_{13}\tilde s_{23}e^{-i\delta_2} & \tilde c_{13}\tilde c_{23}
\end{array}\right) \label{VL-param}
\end{equation}
One advantage of this parametrization is that the elements of the third line which will appear in the $Z'$ contributions to FCNC observables have a
small number of
parameters, e.g.  $\tilde s_{12}$, $\tilde c_{12}$ and the CP violating phase $\delta_3$ do not occur. As pointed out in
\cite{Promberger:2007py} this parametrization is compatible with the choice to treat $D$ and $S$ both as mass and
interaction eigentstates.

\boldmath
\section{$\Delta F=2$ Transitions}\label{sec:3}
\unboldmath
\subsection{Preliminaries}
In order to have transparent formulae for various observables we introduce
a compact notation for the couplings presented above. First of all we
denote the $Z'$ couplings to down-quarks as follows:
\be\label{eq:3.15}
 i\mathcal{L}_{L}(Z') = i\left[ \Delta_L^{sd}(Z')(\bar s \gamma^\mu P_L d)+\Delta_L^{bd}(Z')(\bar b \gamma^\mu P_L d)
+\Delta_L^{bs}(Z')(\bar b \gamma^\mu P_L s)\right] Z'_{\mu}
\ee
with the first upper index denoting outgoing quark and the second incoming
one. Consequently
\be
 \Delta_L^{ji}(Z')= (\Delta_L^{ij}(Z'))^*.
\ee
The expressions for $\Delta_L^{ij}(Z')$ in terms of the fundamental parameters
of the 331 models considered are collected in Appendix~\ref{app:Deltas}.
There the $Z'$ couplings to leptons, like $\Delta_L^{\nu\bar\nu}(Z')$ and
$\Delta_{L,R}^{\mu\bar\mu}(Z')$,
 are defined in analogy to (\ref{eq:3.15}).

\subsection{Standard Model Results}
We begin the presentation of
$\Delta F=2$ transitions by summarizing the known results within the SM.
SM contributions to the off-diagonal elements $M^i_{12}$ in the neutral $K$ and $B_{q}$ meson mass matrices are given as follows
\bea\label{eq:3.4}
\left(M_{12}^K\right)^*_\text{SM}&=&\frac{G_F^2}{12\pi^2}F_K^2\hat
B_K m_K M_{W}^2\left[
\lambda_c^{2}\eta_1S_0(x_c)+\lambda_t^{2}\eta_2S_0(x_t)+
2\lambda_c\lambda_t\eta_3S_0(x_c,x_t)
\right]
\,,\\
\left(M_{12}^q\right)^*_\text{SM}&=&{\frac{G_F^2}{12\pi^2}F_{B_d}^2\hat
B_{B_d}m_{B_d}M_{W}^2
\left[
\left(\lambda_t^{(q)}\right)^2\eta_B S_0(x_t)
\right]}\,,\label{eq:3.6}
\eea
where $x_i=m_i^2/M_W^2$ and
\be
\lambda^{(K)}_i=V_{is}^*V_{id},\qquad \lambda_t^{(q)}=V_{tb}^*V_{tq}
\ee
 with $V_{ij}$ being the elements of the CKM matrix. Here, $S_0(x_i)$ and
$S_0(x_c,x_t)$
are one-loop box functions for which explicit expressions are given e.\,g.~in \cite{Blanke:2006sb}. The
factors $\eta_i$ are QCD {corrections} evaluated at the NLO level in
\cite{Herrlich:1993yv,Herrlich:1995hh,Herrlich:1996vf,Buras:1990fn,Urban:1997gw}. For $\eta_1$ and $\eta_3$ also NNLO corrections
have been
calculated \cite{Brod:2010mj,Brod:2011ty}. Finally $\hat B_K$ and $\hat B_{B_q}$ are the well-known
non-perturbative factors.

In the SM only a single operator
\be
{Q}_1^\text{VLL}(K)=\left(\bar s\gamma_\mu P_L d\right)\left(\bar s\gamma^\mu P_L d\right)\, \qquad
{Q}_1^\text{VLL}(B_q)=\left(\bar b\gamma_\mu P_L q\right)\left(\bar b\gamma^\mu P_L q\right)\,
\ee
contributes to $M_{12}^K$ and $M_{12}^q\;(q=d,s)$, respectively. Moreover
flavour and CP violation is governed totally by the CKM matrix.

\begin{figure}[!tb]
\begin{center}
\includegraphics[width = 0.3\textwidth]{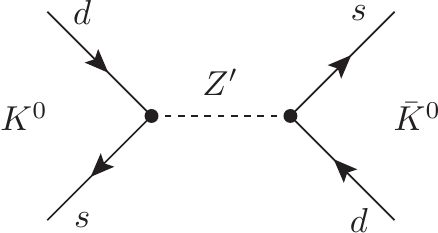}
    \caption{\it Tree-level flavour changing $Z'$ contribution to $K^0-\bar K^0$
mixing (the diagram rotated with $90^\circ$ also exists). }\label{fig:ZprimeKmix}~\\[-2mm]\hrule
\end{center}
\end{figure}

In the 331 model the operator structure remains unchanged, which does not
increase the hadronic uncertainties relatively to the SM. But there are new
flavour violating and CP-violating interactions originating dominantly in
tree-level $Z^\prime$ exchanges shown in Fig.~\ref{fig:ZprimeKmix}. Also box diagrams with
new charged and neutral gauge bosons and new quark exchanges
can contribute and we comment on their
importance soon.

The effect of these new contributions can be summarized by replacing the
flavour independent
$S_0(x_t)$ in the SM formulae by the functions $S_i$
($i=K,d,s$):
\begin{equation}\label{Seff}
S_i=S_0(x_t)+\Delta S_i(Z^\prime)+\Delta S_i({\rm Box})\equiv|S_i|e^{i\theta_S^i}.
\end{equation}
The important new property is the flavour dependence in these functions and
the fact that they carry new complex phases. In what follows we will
give the formulae for the new contributions in (\ref{Seff}).

\boldmath
\subsection{Tree Level $Z^\prime$ Contributions}
\unboldmath
We begin our discussion with the tree level $Z'$ exchanges contributing
to $\Delta F=2$ transitions in Fig.~\ref{fig:ZprimeKmix}.
Defining
\be\label{gsm}
g_{\text{SM}}^2=4\frac{G_F}{\sqrt 2}\frac{\alpha}{2\pi\sin^2\theta_W}\,
\ee
we find
\be\label{Zprime1}
\Delta S_q(Z^\prime)=
\left[\frac{\Delta_L^{bq}(Z^\prime)}{\lambda_t^{(q)}}\right]^2
\frac{4\tilde r}{M^2_{Z^\prime}g_{\text{SM}}^2}, \qquad
\Delta S_K(Z^\prime)=
\left[\frac{\Delta_L^{sd}(Z^\prime)}{\lambda_t^{(K)}}\right]^2
\frac{4\tilde r}{M^2_{Z^\prime}g_{\text{SM}}^2}
\ee
where
\be
\tilde r=\frac{C_1^{\rm VLL}(M_{Z^\prime})}{0.985}
             \eta_6^{6/21}\left[1+1.371\frac{\alpha_s^{(6)}(m_t)}{4\pi}(1-\eta_6)\right].
\ee
Here
\be\label{equ:WilsonZ}
C_1^\text{VLL}(\mu)=
1+\frac{\alpha_s}{4\pi}\left(-2\log\frac{M_{Z'}^2}{\mu^2}+\frac{11}{3}\right)
\ee
represents $\ord(\alpha_s)$ QCD corrections to $Z'$ tree-level exchange
\cite{Buras:2012fs}
and
the   two factors involving
\be
\eta_6=\frac{\alpha_s^{(6)}(M_{Z^\prime})}{\alpha_s^{(6)}(m_t)}
\ee
represent together NLO QCD renormalization group evolution from $m_t$ to
$M_{Z^\prime}$ as given in \cite{Buras:2001ra}. The renormalization
scheme dependence of this evolution is cancelled by the one of
$C_1^\text{VLL}$. We now explain the origin of $\tilde r$.

As the operator structure in 331 models
is unchanged with respect to the SM, the $\hat B_q$ and $\hat B_K$
parameters that have been factored out are the same as in the SM.
The same applies to the
renormalization group evolution in the SM and 331 models between
low-energy scales relevant for hadronic matrix elements and $\mu_t=\ord(m_t)$
entering the evaluation of $\eta_2$ and $\eta_B$ \cite{Buras:1990fn} that
have also been factored out. Therefore
the departure of $\tilde r$ from unity is governed
by the renormalization group effects between $\mu_t$ and
$\mu_{Z'}=\ord(M_{Z'})$, absent in the SM, and the difference in matching
conditions between full and effective theories, involving  tree diagrams in the 331 models but  box diagrams
in the SM. These are represented
by $C_1^{\rm VLL}(M_{Z^\prime})$ and the numerical factor $0.985$
\cite{Buras:1990fn}, respectively. The latter factor
describes flavour universal
 QCD correction to $S_0(x_t)$ in the SM and is usually included in
$\eta_B$ and $\eta_2$ \cite{Buras:1990fn}. The coefficient 1.371,
calculated in \cite{Buras:2001ra}, corresponds to the effective theory with six
flavours $f=6$. We assume tacitly that the new fermions $S,D,T$ are
heavier than $Z'$. Their inclusion into this formula would have a very
small impact on $\tilde r$. Note that $\tilde r$ is free from hadronic
uncertainties. It is also flavour universal, as all the flavour dependence
is already included in $\eta_2$, $\eta_B$, $\hat B_q$ and $\hat B_K$ that
have been factored out.

We find then
\be\label{rtilde}
\tilde r(M_{Z^\prime}=1\tev)=0.985, \quad \tilde r(M_{Z^\prime}=3\tev)=0.9534.
\ee

\boldmath
\subsection{New Box Contributions}
\unboldmath
In principle one would also have to include new box diagram contributions with
new charged and neutral gauge bosons and new quark exchanges. Also charged
Higgs particles
can contribute. However, without any calculation one can convince oneself
that these contributions are negligible. Indeed, compared to the SM contributions the new box contributions are suppressed automatically by a factor
$M_W^2/M^2_{V}\le 0.006$, where $M_V\ge 1\tev$ stands for a new gauge boson.
In models with new LR operators, like left-right symmetric models this
suppression could be compensated by enhanced hadronic matrix elements of new
operators and QCD renormalization group effects. But in 331 models only
SM operator is present and such enhancements are absent. Another enhancement
could be present through enhanced values of the elements of the mixing matrix
$V_L$. However, our analysis shows that in order to suppress sufficiently
tree-level $Z'$ contributions to $\Delta F=2$ processes, the hierarchical
structure of $V_L$ resembles the structure of the CKM matrix.

Box diagrams are also
suppressed with respect to $Z'$ tree-level contributions by a loop
factor like $1/(16\pi^2)$ and two additional vertices. Therefore
they cannot compete with the latter.

In summary it is safe to keep only $Z'$ contributions, which significantly
simplifies the analysis as new box diagrams depend generally on
new parameters, like masses of new gauge bosons and fermions and new
mixing parameters,
which are absent in $Z'$ contributions.

\subsection[Basic formulae for $\Delta F=2$ observables]
{\boldmath Basic formulae for $\Delta F=2$ observables}

Having the mixing amplitudes $M^i_{12}$ at hand we can calculate all relevant
$\Delta F=2$ observables. To this end we collect below those formulae that we used in our numerical analysis.

The $K_L-K_S$ mass difference is given by
\be
\Delta M_K=2\left[\Re\left(M_{12}^K\right)_\text{\rm SM}+\Re\left(M_{12}^K\right)_\text{NP}\right]\,,
\label{eq:3.34}
\ee
and the CP-violating parameter $\varepsilon_K$ by
\be
\varepsilon_K=\frac{\kappa_\eps e^{i\varphi_\eps}}{\sqrt{2}(\Delta M_K)_\text{exp}}\left[\Im\left(M_{12}^K\right)_\text{\rm SM}+\Im\left(M_{12}^K\right)_\text{NP}\right]\,,
\label{eq:3.35}
\ee
where $\varphi_\eps = (43.51\pm0.05)^\circ$ and $\kappa_\eps=0.94\pm0.02$ \cite{Buras:2008nn,Buras:2010pza} takes into account
that $\varphi_\eps\ne \tfrac{\pi}{4}$ and includes long distance effects in $\Im( \Gamma_{12})$ and $\Im (M_{12})$.

For the mass differences in the $B_{d,s}^0-\bar B_{d,s}^0$ systems we have
\be
\Delta M_q=2\left|\left(M_{12}^q\right)_\text{\rm SM}+\left(M_{12}^q\right)_\text{NP}\right|\qquad (q=d,s)\,.
\label{eq:3.36}
\ee
Let us then write \cite{Bona:2005eu}
\be
M_{12}^q=\left(M_{12}^q\right)_\text{\rm SM}+\left(M_{12}^q\right)_\text{NP}=\left(M_{12}^q\right)_\text{\rm SM}C_{B_q}e^{2i\varphi_{B_q}}\,,
\label{eq:3.37}
\ee
where
\be
\left(M_{12}^d\right)_\text{\rm SM}=\left|\left(M_{12}^d\right)_\text{\rm SM}\right|e^{2i\beta}\,,\qquad\beta\approx 22^\circ\,,
\label{eq:3.38}
\ee
\be
\left(M_{12}^s\right)_\text{\rm SM}=\left|\left(M_{12}^s\right)_\text{\rm SM}\right|e^{2i\beta_s}\,,\qquad\beta_s\simeq -1^\circ\,.
\label{eq:3.39}
\ee
Here the phases $\beta$ and $\beta_s$ are defined through
\be
V_{td}=|V_{td}|e^{-i\beta}\quad\textrm{and}\quad V_{ts}=-|V_{ts}|e^{-i\beta_s}\,.
\label{eq:3.40}
\ee
We find then
\be
\Delta M_q=(\Delta M_q)_\text{\rm SM}C_{B_q}\,,
\label{eq:3.41}
\ee
and
\be
S_{\psi K_S} = \sin(2\beta+2\varphi_{B_d})\,,
\qquad
S_{\psi\phi} =  \sin(2|\beta_s|-2\varphi_{B_s})\,,
\label{eq:3.42}
\ee
with the latter two observables being the coefficients of $\sin(\Delta M_d t)$ and $\sin(\Delta M_s t)$ in the
time dependent asymmetries in $B_d^0\to\psi K_S$ and $B_s^0\to\psi\phi$, respectively. Note that new phases are directly related to the phases of the functions
$S_q$:
\be
2\varphi_{B_q}=-\theta_S^q.
\ee

At this stage a few comments on the assumptions leading to expressions in
(\ref{eq:3.42}) are
in order. These simple formulae follow only if there are no weak phases in the decay amplitudes for
$B_d^0\to\psi K_S$ and $B_s^0\to\psi\phi$ as is the case in the SM and also in the LHT model, where due to
T-parity there are no new contributions to decay amplitudes at tree level so that these amplitudes are
dominated by SM contributions \cite{Blanke:2006sb}. Similarly in the model discussed in the present paper there are no new contributions
to decay amplitudes relevant for $S_{\psi K_S}$ and
$S_{\psi\phi}$ and the formulae given above apply.

\section{Effective Hamiltonians for \boldmath{$\Delta F=1$} Decays}
\label{sec:4}

\subsection{Preliminaries}
The goal of the present section is to give formulae for the effective
Hamiltonians relevant for rare $K$ and $B$ decays that in addition to
SM one-loop contributions include tree level contributions from the
$Z'$ gauge boson. While new penguin and box diagrams involving other
gauge bosons and heavy quarks can also contribute, they are subleading
with respect to $Z'$ tree-level contributions and we will neglect them in what
follows. Moreover, in contrast to $Z'$ contributions they involve additional
parameters and are more model dependent.

 \boldmath
\subsection{Effective Hamiltonian for $\bar s\to \bar d\nu\bar\nu$}\label{sec:sdnn}
\unboldmath
The effective Hamiltonian for $\bar s \rightarrow \bar d\nu\bar\nu$ transitions
 resulting from $Z$--penguin and box diagrams
is given in the SM as follows
\be\label{Heffpr}
\left[\Heff^{\nu\bar\nu}\right]^K_\text{SM}=g_{\text{SM}}^2\sum_{\ell=e,\mu,\tau}{\left[\lambda_c^{(K)}
{X_\text{NNL}^\ell(x_c)}+\lambda_t^{(K)} X(x_t)\right]}
(\bar s\gamma_\mu P_L d)(\bar\nu_\ell\gamma_\mu P_L \nu_\ell)+h.c.\,,
\ee
where $x_i=m_i^2/M_W^2$, $\lambda_i^{(K)} =V_{is}^*V_{id}^{}$ and $V_{ij}$ are the elements of the CKM matrix. {$X_\text{NNL}^\ell(x_c)$}
and $X(x_t)$ comprise
internal charm and top quark contributions, respectively. They are known to high accuracy including QCD corrections
\cite{Buchalla:1998ba,Buras:2005gr,Buras:2006gb} and electroweak corrections \cite{Brod:2008ss,Brod:2010hi}. For convenience we have
introduced $g_{\text{SM}}^2$ that is defined in (\ref{gsm}).

\begin{figure}[!tb]
\begin{center}
\includegraphics[width = 0.3\textwidth]{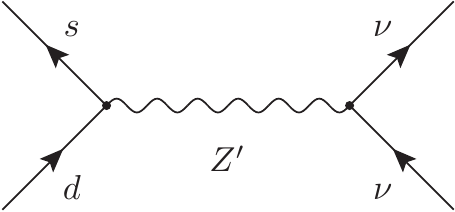}

\end{center}
\caption{\it Tree level contribution of $Z^\prime$ to the $s\to d\nu\bar\nu$ effective Hamiltonian.\label{KplusZ_H}}~\\[-2mm]\hrule
\end{figure}

 In the 331 model (\ref{Heffpr}) is modified by the tree-level diagram in
 Fig.\ \ref{KplusZ_H}.
A straightforward calculation of the diagram in Fig.\ \ref{KplusZ_H} results in a new contribution to $\left[\Heff^{\nu\bar\nu}\right]^K$.
Combining this contribution with the SM contribution in (\ref{Heffpr})
we find
\bea
\left[\Heff^{\nu\bar\nu}\right]^K &=&
g_{\text{SM}}^2\sum_{\ell=e,\mu,\tau}
{\left[\lambda_c^{(K)} {X_\text{NNL}^\ell(x_c)}+\lambda_t^{(K)}X(K)\right]}
(\bar s\gamma^\mu P_L d)(\bar\nu_\ell\gamma_\mu P_L\nu_\ell)\nn\\
&& {}
+h.c.\,.
\eea
where $X(K)$ is given as follows\footnote{In this section in order to increase
the transparency of basic formulae we use the notation
$X_K\equiv X(K)$ and $X_q\equiv X(B_q)$ and similarly for other functions.}
\be\label{XLK}
X(K)=X(x_t)+\frac{\Delta_L^{\nu\bar\nu}(Z')}{g^2_{\rm SM}M_{Z'}^2}
                                       \frac{\Delta_L^{sd}(Z')}{V_{ts}^* V_{td}}
\equiv \left|X(K)\right| e^{i\theta_X^K}.
\ee
Here
\be
X(x_t)=\eta_X~{\frac{x_t}{8}}\;\left[{\frac{x_t+2}{x_t-1}}
+ {\frac{3 x_t-6}{(x_t -1)^2}}\; \ln x_t\right], \qquad \eta_X=0.994
\ee
results within the SM from $Z$-penguin and box diagrams. $\eta_X$ is
QCD correction to these diagrams \cite{Buchalla:1998ba,Misiak:1999yg}
 when $m_t\equiv m_t(m_t)$.

The relevant couplings $\Delta^{ij}(Z')$ are defined in the Appendix~\ref{app:Deltas}.
As all NP contributions have been collected in the term proportional to $\lambda_t^{(K)}$, {$X_\text{NNL}^\ell(x_c)$} contains only the SM
contributions as stated above.

\boldmath
\subsection{Effective Hamiltonian for $b\to d\nu\bar\nu$ and $b\to s\nu\bar\nu$}
\label{sec:bqnn}\unboldmath

In order to generalize the result just obtained to the case of $b\to d\nu\bar\nu$ and $b\to s\nu\bar\nu$ transitions only two steps have to
be performed:
\begin{enumerate}
\item
All flavour indices have to be adjusted appropriately.
\item
The charm quark contribution can be safely neglected in $B$ physics.
\end{enumerate}
The effective Hamiltonian for $b\to q\nu\bar\nu$ ($q=d,s$) is then given as follows:
\bea
\left[\Heff^{\nu\bar\nu}\right]^{B_q} &=&
g_{\text{SM}}^2\sum_{\ell=e,\mu,\tau}
{\left[V_{tq}^\ast V_{tb} X(B_q)\right]}
(\bar q\gamma^\mu P_L b)(\bar\nu_\ell\gamma_\mu P_L\nu_\ell)\nn\\
&&
+h.c.\,,
\eea
with
\be\label{XLRB}
 X(B_q)=X(x_t)+\left[\frac{\Delta_{L}^{\nu\nu}(Z')}{g^2_{\rm SM}M_{Z'}^2}\right]
\frac{\Delta_{L}^{qb}(Z')}{ V_{tq}^\ast V_{tb}}\equiv \left|X(B_q)\right| e^{i\theta_X^q}
\ee
Again all relevant {$\Delta_{L}^{bq}$ entries} in the 331 model can be found
in Appendix~\ref{app:Deltas}.

Note that the functions $X(K)$ and $X(B_q)$ presented above
depend on the quark flavours involved, through the flavour indices in the $\Delta_{L}^{ij}(Z')$
{$(i,j=s,d,b)$} couplings and through the CKM elements that have been
factored out.
 While in principle $\Delta_{L}^{ij}(Z')$ could be aligned with the corresponding CKM factors, this is generally not the case and the
functions in question
become complex quantities that are flavour dependent.
 This should be contrasted {with} the case of the SM and CMFV models where $K$, $B_d$ and $B_s$ systems are governed by a \emph{flavour-universal} loop
function $X(x_t)$ and the only flavour dependence enters through
the CKM factors.
Consequently, as we will see below, certain SM-relations and more generally
CMFV-relations will be violated in the 331 model.

\begin{figure}[!tb]
\begin{center}
\includegraphics[width = 0.3\textwidth]{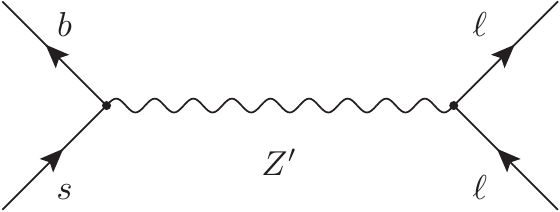}
\end{center}
\caption{\it Tree level contribution of $Z'$ to the $b\to s\ell^+\ell^-$ effective Hamiltonian.}\label{Bleptons}~\\[-2mm]\hrule
\end{figure}

\boldmath
\subsection{Effective Hamiltonian for $b\to d\ell^+\ell^-$ and  $b\to s\ell^+\ell^-$}\label{sec:bqll}
\unboldmath
The effective Hamiltonian for $b\to q\ell^+\ell^-$ ($q=d,s$) transitions in the
331 model is a simple generalization of the SM one.
For ($q=d,s$) we have
\be\label{eq:Heffqll}
\left[ \mathcal{H}_{\text{eff}}^{\ell\bar\ell}\right]^{B_q}
= \Heff(b\to q\gamma)
-\frac{g_{\rm SM}^2}{2}V_{tq}^\star V_{tb}\sum_{i = A,V} C_i(\mu)Q_i(\mu),
\end{equation}
where
\be\label{QAQV}
Q_A  = (\bar q\gamma_\mu P_L b)(\bar \ell\gamma^\mu\gamma_5\ell),\qquad
Q_V  = (\bar q\gamma_\mu P_L b)(\bar \ell\gamma^\mu\ell)
\ee
and $\Heff(b\to s\gamma)$ stands for the effective Hamiltonian for the
$b\to s\gamma$ transition that involves
the dipole operators.
An explicit formula
for the latter Hamiltonian  will be presented in the next subsection.

In the 331 model the second term in (\ref{eq:Heffqll}) receives contributions from the
usual SM penguin and box diagrams and the tree-level
$Z'$ contribution shown in Fig.\ \ref{Bleptons} that dominates
NP contributions. Usually such tree-level gauge
boson exchanges  generate primed operators obtained from the ones
in (\ref{QAQV}) by replacing $P_L$ with $P_R$. But in the 331 model similarly
to the SM right-handed FCNC's are
absent and primed operators are irrelevant.
We find then
for the  Wilson coefficients
\begin{align}
 C_V &=Y(x_t)-4\sin^2\theta_W Z(x_t)
-\frac{1}{g_{\text{SM}}^2}\frac{1}{M_{Z'}^2}
\frac{\Delta_L^{qb}(Z')\Delta_V^{\mu\bar\mu}(Z')}{V_{tq}^\star V_{tb}},\\
  C_A &= -Y(x_t) -\frac{1}{g_{\text{SM}}^2}\frac{1}{M_{Z'}^2}
\frac{\Delta_L^{qb}(Z')\Delta_A^{\mu\bar\mu}(Z')}{V_{tq}^\star V_{tb}},
 \end{align}
where we have defined
\begin{align}
\begin{split}
 &\Delta_V^{\mu\bar\mu}(Z')= \Delta_R^{\mu\bar\mu}(Z')+\Delta_L^{\mu\bar\mu}(Z'),\\
&\Delta_A^{\mu\bar\mu}(Z')= \Delta_R^{\mu\bar\mu}(Z')-\Delta_L^{\mu\bar\mu}(Z').
\end{split}
\end{align}
Here $Y(x_t)$ and $Z(x_t)$ are SM one-loop functions, analogous to $X(x_t)$, that
result from various penguin and box diagrams.
The relevant entries have been collected in Appendix \ref{app:Deltas}.
In particular,
\be
Y(x_t)=\eta_Y~\frac{x_t}{8}\left(\frac{x_t-4}{x_t-1} + \frac{3 x_t \log x_t}{(x_t-1)^2}\right), \qquad \eta_Y=1.012,
\ee
where $\eta_Y$ is QCD correction to these diagrams  \cite{Buchalla:1998ba,Misiak:1999yg} when $m_t\equiv m_t(m_t)$. All other
relevant entries have been collected in Appendix \ref{app:Deltas}.

Introducing
\be\label{YBq}
Y(B_q)=Y(x_t)+ \left[\frac{\Delta_{A}^{\mu\bar\mu}(Z')}{M_{Z'}^2g^2_{\rm SM}}\right]
\frac{\Delta_{L}^{qb}(Z')}{ V_{tq}^\ast V_{tb}}
\equiv \left|Y(B_q)\right| e^{i\theta_Y^q},
\ee
\be
Z(B_q)=Z(x_t)+\frac{1}{4\sin^2\theta_W}\left[\frac{2\Delta_{R}^{\mu\bar\mu}(Z')}{M_{Z'}^2g^2_{\rm SM}}\right]
\frac{\Delta_{L}^{qb}(Z')}{ V_{tq}^\ast V_{tb}}
\equiv \left|Z(B_q)\right| e^{i\theta_Z^q}\,,
\ee
the Wilson coefficients $C_V$ and $C_A$ can be cast in a SM-like form:
\begin{align}
 C_V &=V_{tq}^\star V_{tb} [Y(B_q)-4\sin^2\theta_W Z(B_q)], \\
  C_A &= -V_{tq}^\star V_{tb} Y(B_q).
 \end{align}

The effective Hamiltonian for  $s\to d\ell^+\ell^-$ transition can be obtained
directly from formulae given above by replacing
$q$ by $K$, appropriately changing the flavour indices and neglecting the
contributions of primed operators. Then the functions $Y(K)$ and $Z(K)$
enter:
\be\label{YK}
Y(K)=Y(x_t)+ \left[\frac{\Delta_{A}^{\mu\bar\mu}(Z')}{M_{Z'}^2g^2_{\rm SM}}\right]
\frac{\Delta_{L}^{sd}(Z')}{ V_{ts}^\ast V_{td}}
\equiv \left|Y(K)\right| e^{i\theta_Y^K},
\ee
\be
Z(K)=Z(x_t)+\frac{1}{4\sin^2\theta_W}\left[\frac{2\Delta_{R}^{\mu\bar\mu}(Z')}{M_{Z'}^2g^2_{\rm SM}}\right]
\frac{\Delta_{L}^{sd}(Z')}{ V_{ts}^\ast V_{td}}
\equiv \left|Z(K)\right| e^{i\theta_Z^K}\,.
\ee

As seen in the Appendix, in the $\model$ model the coupling $\Delta_V^{\mu\bar\mu}(Z')$ is strongly suppressed and by one order of
magnitude smaller
than $\Delta_A^{\mu\bar\mu}(Z')$. Consequently $Z'$ contribution to the
coefficient $C_V$ is very small and NP enters  $b\to s\ell^+\ell^-$ and
$b\to d\ell^+\ell^-$ transitions dominantly through the coefficient $C_A$, that
governs the decays $B_{s,d}\to\mu^+\mu^-$. In our numerical analysis we
will discuss only these decays, but we have checked that NP effects
in $b \to s \, d \ell^+ \ell^-$ transitions are within present experimental
and theoretical uncertainties.

In the case of the minimal 331 model,  $\Delta_V^{\mu\bar\mu}(Z')$ is by
a factor of three larger than  $\Delta_A^{\mu\bar\mu}(Z')$ but as the NP
contributions in this model are tiny, this change is irrelevant for all
practical purposes.

\boldmath
\subsection{Effective Hamiltonian for the $B\to X_s\gamma$ Decay}\label{sec:bsgamma}
\unboldmath
\subsubsection{Preliminaries}
Very detailed analyses of $B\to X_s\gamma$ decay in 331 models have been
presented in \cite{Agrawal:1995vp,Promberger:2008xg}, where further references
can be found. It has been found that the dominant NP contributions come from
the Higgs sector, while the gauge boson contributions are subleading. As
the Higgs contributions involve other set of new parameters than the ones involved in our analysis, there is no impact on our analysis from the constraints on
Higgs contributions to  $B\to X_s\gamma$ \cite{Promberger:2008xg}. Similarly
the subleading contributions involving the new gauge bosons $V^0,\,{\bar V}^0$ and $Y^\pm$ and the new heavy quarks involve new parameters and their contributions
can be suppressed if necessary without any impact on our analysis.

\begin{figure}[!tb]
\begin{center}
\includegraphics[width = 0.3\textwidth]{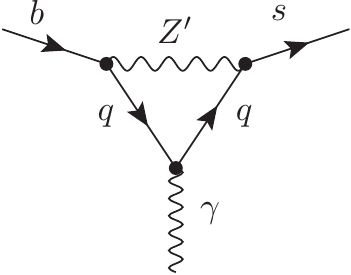}
\caption{\it New magnetic penguin diagram contributing to $B \to X_s \gamma$ with intermediate $Z^\prime$. $q$ denotes a SM quark.\label{Zprime-mag-pen}}~\\[-2mm]\hrule
\end{center}
\end{figure}

An exception is the $Z'$ contribution in Fig.~\ref{Zprime-mag-pen} which involves only SM quarks
and the parameters that enter our analysis. While on the basis of
\cite{Promberger:2008xg} we expect this contribution to be small, we
present here a new analysis which, as far as QCD corrections are concerned, goes
beyond the analysis of the latter paper.  Even if finally this contribution
will turn out to be negiligible in the 331 models, the formulae presented below
could be useful for other models.

Adopting the overall normalisation of the SM effective Hamiltonian
we have
{\begin{equation} \label{Heff_at_mu}
{\cal H}_{\rm eff}(b\to s\gamma) = - \frac{4 G_{\rm F}}{\sqrt{2}} V_{ts}^* V_{tb}
\left[  C_{7\gamma}(\mu_b) Q_{7\gamma} +  C_{8G}(\mu_b) Q_{8G} \right]\,,
\end{equation}}
where $\mu_b=\ord(m_b)$.
The dipole operators are defined as
\begin{equation}\label{O6B}
Q_{7\gamma}  =  \frac{e}{16\pi^2} m_b \bar{s}_\alpha \sigma^{\mu\nu}
P_R b_\alpha F_{\mu\nu}\,,\qquad
Q_{8G}     =  \frac{g_s}{16\pi^2} m_b \bar{s}_\alpha \sigma^{\mu\nu}
P_R T^a_{\alpha\beta} b_\beta G^a_{\mu\nu}\,.
\end{equation}
In writing (\ref{Heff_at_mu}) we have dropped
the primed operators that are obtained from (\ref{O6B}) by  replacing $P_{R}$
by $P_L$. In the SM the primed operators (RL) are suppressed by $m_s/m_b$
relative to the ones in (\ref{Heff_at_mu}). This is also the case of the
331 models. We have also suppressed
current-current operators which are important for the QCD analysis. We will
include these effects in the final formulae at the end of this subsection.

The coefficients $C_i(\mu_b)$ are calculated from their initial values at
high energy scales by means of renormalisation group methods. We distinguish
between SM quark contributions with the matching scale  $\mu_t=\ord(m_t)$
and the $Z'$ quark contributions with the matching scale
$\mu_{Z'}=\ord(M_{Z'})$.
While in the LO approximation the results depend on the choice of the
matching scale, the experience shows that taking as the matching scale
the largest mass in the diagram appears to be a very good choice at LO.
The choices made above follow this strategy.

We  decompose next the Wilson coefficients at the scale $\mu_b=\ord(m_b)$
as the sum of the SM contribution and the $Z'$ contributions:
\be
C_i(\mu_b)=C_i^{\rm SM}(\mu_b)
+\Delta C^{Z'}_i(\mu_b).
\label{cstart}\,
\ee

We recall that for the SM coefficients at $\mu_t=\ord(m_t)$
we have ($x_t=m_t^2/M_W^2$) without QCD corrections
\begin{equation}\label{c7}
C^{\rm SM}_{7\gamma} (\mu_t) = \frac{3 x_t^3-2 x_t^2}{4(x_t-1)^4}\ln x_t +
\frac{-8 x_t^3 - 5 x_t^2 + 7 x_t}{24(x_t-1)^3}\equiv C^{\rm SM}_{7\gamma}(x_t)\,,
\end{equation}
\begin{equation}\label{c8}
C^{\rm SM}_{8G}(\mu_t) = \frac{-3 x_t^2}{4(x_t-1)^4}\ln x_t +
\frac{-x_t^3 + 5 x_t^2 + 2 x_t}{8(x_t-1)^3}\,\equiv C^{\rm SM}_{8G}(x_t).
\end{equation}

\boldmath
\subsubsection{$Z'$ contribution without QCD Corrections}\label{eq:bsgamma-neutralBoson}
\unboldmath
A general analysis of neutral gauge boson contributions to $B\to X_s\gamma$
decay has been presented in \cite{Buras:2011zb}.
In addition to
SM-like LL contribution from $Z'$ we have a new LR one, where $L$ ($R$) stands for the $P_L$ ($P_R$) projector in the basic penguin diagram
involving the $s$($b$)-quark. In the 331 models the LR contributions to  $B\to X_s\gamma$
originate from flavour conserving $\bar b b Z'$ and $\bar s sZ'$ couplings
which in addition to left-handed component have also a right-handed component
as clearly seen in (\ref{ZprimeFR}).

In what follows
we present the results for a contribution of a fermion $f$ carrying electric
charge $-1/3$ and having the mass $m_f$. As only SM quarks with $m_f\ll M_{Z'}$
contribute, we can set $m_f=0$, whenever it is justified.

We first decompose the Wilson coefficients $\Delta C^\text{Z'}_i$ at the
$\mu_{Z'}$
scale as the sum of the SM-like LL contribution and a new LR one:
\be
\begin{aligned}
\Delta C^{Z'}_{7\gamma}(\mu_{Z'})&=\Delta^{LL} C^{Z'}_{7\gamma}(\mu_{Z'}) +\Delta^{LR}C^{Z'}_{7\gamma}(\mu_{Z'})\,,\\[2mm]
\Delta C^{Z'}_{8G}(\mu_{Z'})     &=\Delta^{LL} C^{Z'}_{8G}(\mu_{Z'}) +
\Delta^{LR}C^{Z'}_{8G}(\mu_{Z'})\,.
\end{aligned}
\label{eq:wilsonatmh}
\ee

Adapting the general formulae of \cite{Buras:2011zb}
to our notation and denoting by $f$ the down-quark exchanged in the diagram
we find
\begin{equation}
\begin{aligned}
\Delta^{LL}C^{Z'}_{7\gamma}(\mu_{Z'})
&=-\frac{2}{9}\,\frac{1}{g^2}\,\frac{M_W^2}{M_{Z'}^2}\,\sum_f\frac{\Delta_L^{fs*}(Z')\,\Delta_L^{fb}(Z')}{V_{ts}^*\,V_{tb}}
\approx
-\frac{2}{27}\,\frac{M_W^2}{ M_{Z'}^2}\,\frac{v^*_{32}v_{33}}{V_{ts}^*\,V_{tb}}
\\[2mm]
\Delta^{LL}C^{Z'}_{8 G}(\mu_{Z'}) &= - 3 \Delta^{LL}C^{Z'}_{7\gamma}(\mu_{Z'})\,,
\end{aligned}
\label{LLnew}
\end{equation}
where the summation is over SM down-quarks. As this
contribution is independent of quark masses the summation over $f$
can be performed implying a very simple result, in particular when
the d-quark contribution suppressed by additional mixing angles is
neglected.\\

For LR Wilson coefficients we find in the case of $\beta=1/\sqrt{3}$:
\begin{equation}
\begin{aligned}
\Delta^{LR}C^{Z'}_{7\gamma}(\mu_{Z'})&=\frac{2}{3}\,\frac{1}{g^2}\,\frac{M_W^2}{ M_{Z'}^2}\,\sum_f \frac{m_f}{m_b}\,
\frac{\Delta_L^{fs*}(Z')\,\Delta_R^{fb}(Z')}{V_{ts}^*\,V_{tb}}\approx
-\frac{2}{27}\,\frac{M_W^2}{ M_{Z'}^2}\,\frac{s_W^2}{1-\frac{4}{3}s_W^2}
\frac{v^*_{32}v_{33}}{V_{ts}^*\,V_{tb}}
,\\[2mm]
\Delta^{LR}C^{Z'}_{8G}(\mu_{Z'})&= -3\Delta^{LR}C^{Z'}_{7\gamma}(\mu_{Z'})\,,
\end{aligned}
\label{LRnew}
\end{equation}
where again the summation is over SM down-quarks. As LR contributions
arise from the chirality flip on the internal line, in this case
explict dependence on $m_f$ is present. However in accordance
with other approximations we should keep only the b-quark contribution,
which gives the final result. As
\be
C^{\rm SM}_{7\gamma} (x_t)=-0.193,  \quad   C^{\rm SM}_{8G}(x_t)=-0.096,
\ee
it is evident that without inclusion of QCD corrections, the $Z'$
contribution to $B\to X_s\gamma$ is totally negligible  for the values
of $M_{Z'}$ considered. We will now
demonstrate that this is also the case after the inclusion of QCD effects.

\subsubsection{Final Results including QCD corrections}\label{FinalWC}
In order to complete the analysis of $B\to X_s\gamma$ we have to include
QCD corrections which play a very important role in this decay.
In the SM these corrections are known at the NNLO level \cite{Misiak:2006zs}.
In the LR model
a complete LO analysis has been done by Cho and Misiak \cite{Cho:1993zb}
and after proper modification we can use their results in our model .
In this context the recent analyses \cite{Buras:2011zb,Blanke:2011ry}
turned out to be very useful.

We find then
\begin{equation}
\Delta C^{Z'}_{7\gamma}(\mu_b)=
\kappa_7(\mu_{Z'})~\Delta C^{Z'}_{7\gamma}(\mu_{Z'}) +
\kappa_8(\mu_{Z'})~\Delta C^{Z'}_{8G}(\mu_{Z'})+
\Delta^{\rm current}_Z(\mu_b)\,.
\label{eq:DeltaC7effA3}
\end{equation}
The last contribution in (\ref{eq:DeltaC7effA3}) results from the mixing of
new neutral current-current operators generated from the $Z'$ exchange
that mix with the dipole operators. The renormalization group analysis
of this contribution is very involved but fortunately the LO result is known
from  \cite{Buras:2011zb}. Therefore adapting the formulae (4.16), (4.17) and
(5.6) of this paper to our notation we find
\be\label{Zcurrent}
\Delta^{\rm current}_Z(\mu_b)=
\sum_{\substack{A=L,R\\f=u,c,t,d,s,b}}\!\!\!\!\! \kappa^{f}_{LA}~\Delta ^{LA} C_2^f(\mu_{Z'}) +\!\!\!\sum_{A=L,R}\!\!\!\!\hat{\kappa}^{d}_{LA}~\Delta ^{LA}
\hat{C}_2^d(\mu_{Z'}),
\ee
where
\begin{equation}
\Delta^{LA}C^f_2(\mu_{Z'})=-\frac{2}{g^2}\frac{M_W^2}{M_{Z'}^2}
\frac{\Delta_L^{sb*}(Z')\,\tilde\Delta_A^{ff}(Z')}{V_{ts}^*\,V_{tb}}\,,
\label{InitialConditionQnn}
\end{equation}
and
\begin{equation}
\Delta^{LA}\hat{C}_2^d(\mu_{Z'})=-\frac{2}{g^2}\frac{M_W^2}{M_{Z'}^2}
\frac{\Delta_L^{sd*}(Z')\,\Delta_A^{bd}(Z')}{V_{ts}^*\,V_{tb}}\,.
\label{InitialConditionQnnHat}
\end{equation}

The diagonal couplings $\tilde\Delta_A^{ff}(Z')$
introduce additional parameters.

Finally, $\kappa$'s are the NP magic numbers listed in Tab.~\ref{tab:Magic}
that is based on
\cite{Buras:2011zb} which used
$\alpha_s(M_Z=91.1876\,\text{GeV})=0.118$. They have been obtained for
$\mu_b=2.5\gev$ as used in the SM calculations.

 Using these formulae we find for $M_{Z'}=1\tev$
\begin{equation}
\Delta C^{Z'}_{7\gamma}(\mu_b)= \mathcal{O}(10^{-5})
\end{equation}
which is negligible when compared with the SM value of $-0.353$. Therefore
we will not consider the $B\to X_s\gamma$ further.

\begin{table}[h!]
\begin{center}
\begin{tabular}{|c||r|r|r|}
  \hline
  &&&\\[-4mm]
  $\mu_{Z'}$	 	& 1 TeV		& 5 TeV		& 10 TeV \\[1mm]
  \hline\hline
  $\kappa_7$		 	& 0.457		& 0.408		& 0.390	  \\			
  $\kappa_8$		 	& 0.125		& 0.129		& 0.130	 \\[1mm]		
  \hline
  &&&\\[-4mm]
  $\kappa_{LL}^{u,c}$	 	& 0.057		& 0.076		& 0.084	 \\			
  $\kappa_{LL}^{t}$	 	&-0.003		&-0.002		&-0.001	 \\			
  $\kappa_{LL}^{d}$	 	&-0.057		&-0.072		&-0.079	 \\			
  $\kappa_{LL}^{s,b}$	 	& 0.090		& 0.090		& 0.090	 \\			
  $\hat{\kappa}_{LL}^{d}$ 	& 0.147		& 0.163		& 0.168	 \\[1mm]
  \hline
  &&&\\[-4mm]                                                                            	
  $\kappa_{LR}^{u,c}$	 	& 0.128	 	& 0.173		& 0.193	 \\			
  $\kappa_{LR}^{t}$	 	& 0.012	 	& 0.023		& 0.028	 \\			
  $\kappa_{LR}^{d}$	 	&-0.025		&-0.036		&-0.041	 \\			
  $\kappa_{LR}^{s,b}$	 	&-0.092	 	&-0.106		&-0.111	 \\			
  $\hat{\kappa}_{LR}^{d}$	& 0.665		& 0.865		& 0.953	 \\[1mm]	
  \hline                                                                                         	
\end{tabular}
\caption{\it The NP magic numbers relevant for QCD calculations \cite{Buras:2011zb}.}
\label{tab:Magic}
\end{center}
\end{table}

 \section{Rare Decays}\label{sec:5}

\boldmath
\subsection{$K^+ \rightarrow \pi^+\nu\bar\nu$ and $K_L \rightarrow \pi^0\nu\bar\nu$}
\unboldmath

Having at hand the effective Hamiltonian for $\bar s\to \bar d\nu\bar\nu$ transitions
derived in Section \ref{sec:sdnn} it is now straightforward to obtain
explicit expressions for the branching ratios $\mathcal{B}(\kpn)$ and $\mathcal{B}(\klpn)$.
 Reviews of these two decays can be found in
\cite{Buras:2004uu,Isidori:2006yx,Smith:2006qg}.

The
branching ratios for the two
$K \to \pi \nu\bar \nu$ modes that follow from the Hamiltonian
in Section \ref{sec:sdnn}
 can be written generally as
\begin{gather} \label{eq:BRSMKp}
  \Br (K^+\to \pi^+ \nu\bar\nu) = \kappa_+ \left [ \left ( \frac{{\rm Im} X_{\rm eff} }{\lambda^5}
  \right )^2 + \left ( \frac{{\rm Re} X_{\rm eff} }{\lambda^5}
  - P_c(X)  \right )^2 \right ] \, , \\
\label{eq:BRSMKL} \Br( K_L \to \pi^0 \nu\bar\nu) = \kappa_L \left ( \frac{{\rm Im}
    X_{\rm eff} }{\lambda^5} \right )^2 \, ,
\end{gather}
where \cite{Mescia:2007kn}
\begin{equation}\label{kapp}
\kappa_+=(5.36\pm0.026)\cdot 10^{-11}\,, \quad \kappa_{\rm L}=(2.31\pm0.01)\cdot 10^{-10}
\ee
and \cite{Buras:2005gr,Buras:2006gb,Brod:2008ss,Isidori:2005xm,Mescia:2007kn}
\be
P_c(X)=0.42\pm0.03.
\end{equation}
The short distance contributions are described simply by
\be
X_{\rm eff} = V_{ts}^* V_{td} X(K) \equiv
V_{ts}^* V_{td} \left|X(K)\right| e^{i\theta_X^K},
\ee
where $X(K)$ is given in (\ref{XLK}).

These formulae
are in fact  very  general and apply to all extensions of the SM.
The correlation between the two branching ratios depends generally  on
two variables $\left|X(K)\right|$ and $\theta_X^K$ and measuring
these branching ratios one day will allow to determine them
 and compare them with model expectations.
 The numerical analysis of both decays is presented in Section \ref{sec:7}.

\subsection{\boldmath $B \to \{X_s,K, K^*\} \nu\bar \nu$}

Following the general analysis of \cite{Altmannshofer:2009ma},
the branching ratios of the $B \to \{X_s,K, K^*\}\nu\bar \nu$  in the 331
model can be simply obtained by removing the right-handed current contributions:
 \bea
 \mathcal{B}(B\to K \nu \bar \nu) &=&
 \mathcal{B}(B\to K \nu \bar \nu)_{\rm SM} \times \varrho^2~, \label{eq:BKnn}\\
 \mathcal{B}(B\to K^* \nu \bar \nu) &=&
 \mathcal{B}(B\to K^* \nu \bar \nu)_{\rm SM}\times\varrho^2~, \\
 \mathcal{B}(B\to X_s \nu \bar \nu) &=&
 \mathcal{B}(B\to X_s \nu \bar \nu)_{\rm SM} \times \varrho^2~,\label{eq:Xsnn}
 \eea
 where
 \be\label{etaepsilon}
 \varrho = \frac{ |X(B_s)|}{ X(x_t)}~
 \ee
with $X(B_q)$ defined in (\ref{XLRB}).

Evidently, the ratios of these branching ratios are equal to the corresponding
ratios in the SM and all three branching ratios depend on NP only through the
variable $\varrho$. It should also be emphasized that
the average of the $K^*$ longitudinal polarization fraction $F_L$,
also used in the studies of $B\to K^*\ell^+\ell^-$,  takes in the absence
of right-handed currents a fixed value
\be
\label{eq:epseta-FL}
 \langle F_L \rangle = 0.54 \,~.
\ee
Therefore, the measurement of $F_L$ constitutes an important test of 331
models.

We should remark that the expressions in Eqs.~(\ref{eq:BKnn})--(\ref{eq:Xsnn}),
as well as the SM results in (\ref{eq:BKnnSM}), refer only to the short-distance contributions
to these decays. The latter are obtained from the corresponding total rates
subtracting the reducible long-distance effects pointed out in~\cite{Kamenik:2009kc}.

The updated predictions for the SM branching  ratios
are~\cite{Bartsch:2009qp,Kamenik:2009kc,Altmannshofer:2009ma}
\bea
\mathcal{B}(B\to K \nu \bar \nu)_{\rm SM}   &=& (3.64 \pm 0.47)\times 10^{-6}~, \no \\
\mathcal{B}(B\to K^* \nu \bar \nu)_{\rm SM} &=& (7.2 \pm 1.1)\times 10^{-6}~, \no \\
\mathcal{B}(B\to X_s \nu \bar \nu)_{\rm SM} &=& (2.7 \pm 0.2)\times 10^{-5}~,
\label{eq:BKnnSM}
\eea
to be compared with the experimental bounds~\cite{Barate:2000rc,:2007zk,:2008fr}
\bea
\mathcal{B}(B\to K \nu \bar \nu)   &<&  1.4 \times 10^{-5}~, \no \\
\mathcal{B}(B\to K^* \nu \bar \nu) &<&  8.0 \times 10^{-5}~, \no \\
\mathcal{B}(B\to X_s \nu \bar \nu)  &<&  6.4 \times 10^{-4}~.
\label{eq:BKnn_exp}
\eea

\boldmath
\subsection{$B_{d,s} \to \mu^+ \mu^-$}
\unboldmath
We will next consider the two super stars of the LHCb, the
 decays $B_{d,s} \to \mu^+ \mu^-$, that suffer from helicity suppression in the
SM. This suppression cannot be removed through the tree level exchange of $Z'$
boson.
Assuming that the CKM parameters
have been determined independently
of NP and are universal we find
\be\label{GB/SM}
\frac{\mathcal{B}(B_q\to\mu^+\mu^-)}{\mathcal{B}(B_q\to\mu^+\mu^-)^{\rm SM}}
=\left|\frac{Y(B_q)}{Y(x_t)}\right|^2,
\ee
where $Y(B_q)$ is given in (\ref{YBq}).
The numerical results are given in Section~\ref{sec:7}.

The branching ratios $\mathcal{B}(B_q\to\mu^+\mu^-)$ are only sensitive to
the absolute value of $Y(B_q)$. However,
as pointed out recently in \cite{deBruyn:2012wj,deBruyn:2012wk}  in
the flavour precision era these decay could allow to get also some information
on the phase of $Y(B_s)$ and we want to investigate whether in
the $\model$ model this effect is significant.

First as stressed in 
\cite{DescotesGenon:2011pb,deBruyn:2012wj,deBruyn:2012wk} \footnote{We follow here presentation and notations of \cite{deBruyn:2012wj,deBruyn:2012wk}.},
when comparing
the theoretical branching ratio with experimental data quoted by LHCb, ATLAS and CMS,
a correction factor has to be included which takes care of $\Delta\Gamma_s$ effects
that influence the extraction of this branching ratio from the data:
\be
\label{Fleischer1}
\mathcal{B}(B_{s}\to\mu^+\mu^-)_{\rm th} =
r(y_s)~\mathcal{B}(B_{s}\to\mu^+\mu^-)_{\rm exp}, \quad r(0)=1.
\ee
Here
\be
r(y_s)\equiv\frac{1-y_s^2}{1+\mathcal{A}^\lambda_{\Delta\Gamma} y_s}
\approx 1 - \mathcal{A}^\lambda_{\Delta\Gamma} y_s
\ee
with
\be
y_s\equiv\tau_{B_s}\frac{\Delta\Gamma_s}{2}=0.088\pm0.014.
\ee
The quantity $\mathcal{A}^\lambda_{\Delta\Gamma}$ is discussed below.

As stressed in \cite{Buras:2012ts} it is a matter of choice whether the factor $r(y_s)$
is included in the experimental branching ratio or in the theoretical calculation,
provided $r(y_s)$ is not significantly affected by NP.
Once it is measured, its inclusion in the experimental value, as advocated in
\cite{deBruyn:2012wj}, should be favoured as it would have no impact on the theoretical
calculations of branching ratios that do not depend on $\Delta\Gamma_s$. As in the SM
and CMFV  $\mathcal{A}^\lambda_{\Delta\Gamma}=1$ \cite{deBruyn:2012wk} and
the factor $r(y_s)$ is universal, it is also a good idea to include this factor
in experimental branching ratio. In this manner various CMFV relations remain intact.
If this is done the experimental upper bound in (\ref{LHCb2}) is reduced by $9\%$
implying a more stringent upper bound of $3.8\times 10^{-9}$.
For the latest discussions see \cite{deBruyn:2012wj,deBruyn:2012wk} and \cite{Buras:2012ru,Fleischer:2012fy}.

 If a given model predicts  $\mathcal{A}^\lambda_{\Delta\Gamma}$ significantly different from unity and
the dependence of $r(y_s)$ on
model parameters is large  one may include this factor in
the theoretical branching ratio:
\be
\label{Fleischer2}
\mathcal{B}(B_{s}\to\mu^+\mu^-)_{\rm corr}=\frac{\mathcal{B}(B_{s}\to\mu^+\mu^-)_{\rm th}}{r(y_s)}
\ee
with the result for the SM \cite{deBruyn:2012wj,deBruyn:2012wk}
\be
\label{FleischerSM}
\mathcal{B}(B_{s}\to\mu^+\mu^-)^{\rm SM}_{\rm corr}= (3.5\pm0.2)\cdot 10^{-9}.
\ee
It is this branching that should be compared in such a case
with the results of LHCb, ATLAS and CMS. However, as we will see below in
the 331 models considered by us  $\mathcal{A}^\lambda_{\Delta\Gamma}$ is very
close to unity and it is more convenient to include this effect in the
experimental branching ratio.

What is interesting is that in addition to
$\mathcal{A}^\lambda_{\Delta\Gamma}$ it is possible to define a CP-asymmetry
$S_{\mu^+\mu^-}^s$ \cite{deBruyn:2012wk} analogous to $S_{\psi K_S}$ and
$S_{\psi\phi}$. Both observables can be measured one day providing
additional tests of NP models.

The authors of \cite{deBruyn:2012wk,Fleischer:2012fy} provide general expressions for
$\mathcal{A}^\lambda_{\Delta\Gamma}$ and
$S_{\mu^+\mu^-}^s$ as functions of Wilson coefficients involved. Using
these formulae we find in the 331 models very simple formulae that
reflect the fact that $Z'$ and not scalar operators dominate NP contributions:
\be
\mathcal{A}^\lambda_{\Delta\Gamma}=\cos (2\theta^{B_s}_Y-2\varphi_{B_s}), \quad
S_{\mu^+\mu^-}^s=\sin (2\theta^{B_s}_Y-2\varphi_{B_s})
\ee
Both $\mathcal{A}^\lambda_{\Delta\Gamma}$ and
$S_{\mu^+\mu^-}^s$ are theoretically clean observables.

In the SM and CMFV models
\be
\mathcal{A}^\lambda_{\Delta\Gamma}=1, \quad S_{\mu^+\mu^-}^s=0,
\quad r(y_s)=0.912\pm0.014
\ee
independently of NP parameters.  As we will see in the next section,
the first and consequently the third quantity above is valid rather
accurately in the models considered here. But
for $\model$ the CP-asymmetry  $S_{\mu^+\mu^-}^s$ will definitely be
different  from zero.
Moreover, it will be of interest to investigate how $S_{\mu^+\mu^-}^s$ is
correlated with $S_{\psi K_S}$ and
$S_{\psi\phi}$.

{While $\Delta\Gamma_d$ is very small and $y_d$ can be set to zero,
in the case of $B_d\to\mu^+\mu^-$ one can still consider the CP asymmetry
$S_{\mu^+\mu^-}^d$ \cite{Fleischer:2012fy}, for which we simply find
\be
S_{\mu^+\mu^-}^d=\sin(2\theta^{B_d}_Y-2\varphi_{B_d})
\ee
Even if this asymmetry, similarly to $S_{\mu^+\mu^-}^s$, is experimentally challenging, it will be interesting to find its size in the $\model$ model.}

\boldmath
\subsection{$K_L\to\mu^+\mu^-$}\label{sec:KLmumu}
\unboldmath
The discussion of the NP contributions to this decay is analogous
to $B_{d,s}\to \mu^+\mu^-$. Again only the SM operator $(V-A)\otimes (V-A)$
contributes and the real function $Y(x_t)$ is replaced by the complex
function in Eq.~(\ref{YK}).
Only the so-called short distance (SD)
part to a dispersive contribution
to $K_L\to\mu^+\mu^-$ can be reliably calculated. We have then
following \cite{Buras:2004ub}
($\lambda=0.226$)
\be
\mathcal{B}(K_L\to\mu^+\mu^-)_{\rm SD} =
 2.08\cdot 10^{-9} \left[\bar P_c\left(Y_K\right)+
A^2 R_t\left|Y(K)\right|\cos\bar\beta_{Y}^K\right]^2\,,
\ee
where $R_t $ is given in (\ref{eq:Rt_beta}) and
\be
A=\frac{\vcb}{\lambda^2},\qquad
\bar\beta_{Y}^K \equiv \beta-\beta_s-\theta^K_Y\,,
\qquad
\bar P_c\left(Y_K\right) \equiv \left(1-\frac{\lambda^2}{2}\right)P_c\left(Y_K\right)\,,
\ee
with $P_c\left(Y_K\right)=0.113\pm 0.017$
\cite{Gorbahn:2006bm}. Here
$\beta$ and $\beta_s$ are the phases of $V_{td}$ and $V_{ts}$ defined in
(\ref{eq:3.40}).

The extraction of the short distance
part from the data is subject to considerable uncertainties. The most recent
estimate gives \cite{Isidori:2003ts}
\be\label{eq:KLmm-bound}
\mathcal{B}(K_L\to\mu^+\mu^-)_{\rm SD} \le 2.5 \cdot 10^{-9}\,,
\ee
to be compared with $(0.8\pm0.1)\cdot 10^{-9}$ in the SM
\cite{Gorbahn:2006bm}.
The numerical results are discussed in Section~\ref{sec:7}.

\boldmath
\subsection{$B^+\to \tau^+\nu$}\label{btaunu}
\unboldmath
\subsubsection{Standard Model Results}
We now look at the tree-level decay $B^+ \to \tau^+ \nu$ which in the SM is
mediated by the $W^\pm$  exchange with the resulting branching ratio  given by
\begin{equation} \label{eq:Btaunu}
\mathcal{B}(B^+ \to \tau^+ \nu)_{\rm SM} = \frac{G_F^2 m_{B^+} m_\tau^2}{8\pi} \left(1-\frac{m_\tau^2}{m^2_{B^+}} \right)^2 F_{B^+}^2
|V_{ub}|^2 \tau_{B^+}~.
\end{equation}
Evidently this result is subject to significant parametric uncertainties induced in (\ref{eq:Btaunu}) by $F_{B^+}$ and $V_{ub}$. However, it is expected
that these uncertainties will be eliminated in this decade and
a precise prediction will be possible. Anticipating this we will present
the results for fixed values of these parameters.

In the literature
in order to find the SM prediction for this branching ratio one eliminates
these uncertainties by using $\Delta M_d$,  $\Delta M_d/\Delta M_s$ and
$S_{\psi K_S}$ \cite{Bona:2009cj,Altmannshofer:2009ne} and taking experimental
values for these three quantities. This method has a weak point as
 the experimental
values of $\Delta M_{d,s}$ used
in this strategy may not be the one corresponding to the true value of
the SM. However, proceeding in this manner one finds
\cite{Altmannshofer:2009ne}
\begin{equation}\label{eq:BtaunuSM1}
\mathcal{B}(B^+ \to \tau^+ \nu)_{\rm SM}= (0.80 \pm 0.12)\times 10^{-4},
\end{equation}
with a similar result obtained
by the UTfit collaboration
\cite{Bona:2009cj} and CKM-fitters \cite{Lenz:2010gu}.

Now, the experimental world avarage based
on older results  by BaBar \cite{Aubert:2007xj} and Belle
\cite{Ikado:2006un} was \cite{Nakamura:2010zzi}	
\begin{equation} \label{eq:Btaunu_exp1}
\mathcal{B}(B^+ \to \tau^+ \nu)_{\rm exp} = (1.65 \pm 0.34) \times 10^{-4}~,
\end{equation}
which is roughly by a factor of 2 higher than the SM value.
Very recently, new results have been provided by
BaBar Collaboration \cite{Lees:2012ju}:
\begin{equation} \label{eq:Btaunu_exp2}
\mathcal{B}(B^+ \to \tau^+ \nu)_{\rm exp} = (1.79 \pm 0.48) \times 10^{-4}~,\qquad {\rm BaBar}
\end{equation}
and by Belle Collaboration \cite{BelleICHEP}:
\begin{equation} \label{eq:Btaunu_exp3}
\mathcal{B}(B^+ \to \tau^+ \nu)_{\rm exp} = (0.72 \pm^{0.27}_{0.25} \pm^{0.46}_{0.51}) \times 10^{-4}~,\qquad {\rm Belle}
\end{equation}
implying a new world average provided by the UTfit collaboration \cite{Tarantino:2012mq}:
\begin{equation} \label{eq:world}
\mathcal{B}(B^+ \to \tau^+ \nu)_{\rm exp} = (0.99 \pm 0.25) \times 10^{-4}~,\qquad {\rm WA},
\end{equation}
which is consistent with the SM.
As we will see below the central value in (\ref{eq:world}) for central lattice
input corresponds to $\vub=0.0040$.

The  full clarification of the situation
will be provided by the
data from Super-B machines at KEK and Rome. In the meantime hopefully
improved values for $F_{B^+}$ from lattice
and $\vub$ from tree level decays will allow us to make a precise prediction
for this decay without using the experimental value for $\Delta M_d$.

\subsubsection{Result in the 331 Model}
In the 331 model there are no new tree-level contributions to
$B^+ \to \tau^+ \nu_\tau$. This means that the model favours a
value of $\vub$ corresponding to the inclusive determination of this CKM element. We will return to this point in
our numerical analysis.

\section{Breakdown of CMFV Relation in the 331 Model \label{sec:6}}
The presence of new sources of flavour and CP violation in the 331 model
modifies the usual CMFV relations \cite{Buras:2003jf}.
Therefore it is useful to
generalize the most interesting among these relations to include possible
breakdown of them as follows \cite{Buras:2012ts}:
\be\label{dmsdmd}
\frac{\Delta M_d}{\Delta M_s}=
\frac{m_{B_d}}{m_{B_s}}
\frac{\hat B_{d}}{\hat B_{s}}\frac{F^2_{B_d}}{F^2_{B_s}}
\left|\frac{V_{td}}{V_{ts}}\right|^2r(\Delta M)
\end{equation}
\begin{equation}\label{bxnn}
\frac{\mathcal{B}(B\to X_d\nu\bar\nu)}{\mathcal{B}(B\to X_s\nu\bar\nu)}=
\left|\frac{V_{td}}{V_{ts}}\right|^2r(\nu\bar\nu)
\end{equation}
\begin{equation}\label{bmumu}
\frac{\mathcal{B}(B_d\to\mu^+\mu^-)}{\mathcal{B}(B_s\to\mu^+\mu^-)}=
\frac{\tau({B_d})}{\tau({B_s})}\frac{m_{B_d}}{m_{B_s}}
\frac{F^2_{B_d}}{F^2_{B_s}}
\left|\frac{V_{td}}{V_{ts}}\right|^2 r(\mu^+\mu^-).
\end{equation}
The quantities $r(\Delta M)$, $r(\nu\bar\nu)$
and $r(\mu^+\mu^-)$ are all equal unity in models with CMFV. In the 331
model they can be entirely expressed in terms of the master functions
(\ref{eq31}):
\be\label{r1}
 r(\Delta M)=\left|\frac{S_d}{S_s}\right|, \quad
 r(\nu\bar\nu)=\left|\frac{X(B_d)}{X(B_s)}\right|^2, \quad
 r(\mu^+\mu^-)=\left|\frac{Y(B_d)}{Y(B_s)}\right|^2 ~.
\ee

Eliminating $|V_{td}/V_{ts}|$ from the three relations above allows
to obtain three relations between observables that are universal within the
CMFV models. In particular
from (\ref{dmsdmd}) and (\ref{bmumu}) one finds \cite{Buras:2003td}
\be\label{R1}
\frac{\mathcal{B}(B_{s}\to\mu^+\mu^-)}{\mathcal{B}(B_{d}\to\mu^+\mu^-)}
=\frac{\hat B_{d}}{\hat B_{s}}
\frac{\tau( B_{s})}{\tau( B_{d})}
\frac{\Delta M_{s}}{\Delta M_{d}}r, \quad r=\frac{r(\Delta M)}{r(\mu^+\mu^-)}
\ee
that does not
involve $F_{B_q}$ and consequently contains
smaller hadronic uncertainties than the formulae considered
above. It involves
only measurable quantities except for the ratio $\hat B_{s}/\hat B_{d}$
that is now known already from lattice calculations
with respectable precision \cite{Shigemitsu:2009jy,Laiho:2009eu}:
$1.05\pm0.07$.   A review of other CMFV relations is
given in \cite{Buras:2003jf}.

We also recall that within CMFV models there is a correlation between
$\varepsilon_K$, $\Delta M_s$ and $\Delta M_d$ as all these quantities
are described
by the same universal function $S_0(x_t)$. As a result, the enhancement of
one of these observables implies uniquely the enhancements of other
two. As we will see below also this correlation is broken in the
331 models.

\section{Correlations within 331 Model}\label{sec:10}
\subsection{Classification}
The CMFV correlations between various observables are violated in the
331 model, in particular the correlations between the $K$, $B_d$ and
$B_s$ systems are less stringent. Yet
we would like to emphasize that the fact that NP effects
originate dominantly from tree-level $Z'$ exchanges with only left-handed quark
couplings involved and leptonic $Z'$ couplings are fully described by SM parameters implies still very stringent correlations between departures from SM
and CMFV in three classes of observables. In particular there are interesting
correlations between $\Delta F=2$ and $\Delta F=1$ observables in each class
which imply important tests for this NP scenario.

Indeed in the three meson systems,  $K$, $B_d$ and
$B_s$,  NP effects are governed by
\be\label{Deltas}
\Delta^{sd}_L(Z'), \qquad \Delta^{bd}_L(Z') \qquad \Delta^{bs}_L(Z'),
\ee
respectively and by the value of the $Z'$ mass which is obviously common
to these three systems. As seen in Appendix~\ref{app:Deltas} the three
$\Delta^{ij}(Z')$ depend only on four new parameters:
\be\label{newpar}
\tilde s_{13}, \quad \tilde s_{23}, \quad \delta_1, \quad \delta_2
\ee
with $\tilde s_{13}$ and $\tilde s_{23}$ being positive definite and $\delta_i$ in
the range $[0,2\pi]$.
Therefore for fixed  $M_{Z'}$, the $Z'$ contributions to all processes
analyzed by us depend on only these parameters implying very strong
correlations between NP effects to various observables. The relations
discussed below make it clear why the correlations in question
are present.

As we will see in the next section  $\tilde s_{13}$ and $\tilde s_{23}$
are small numbers implying that  $\tilde c_{13}$ and $\tilde c_{23}$ are
close to unity. Therefore there is a very clear structure in the
dependence of various observables on the parameters in (\ref{newpar}).
In view of this structure we expect stringent direct correlations between
observables in the following three classes:

{\bf Class A:}

\be\label{ClassA}
\Delta M_s, \quad S_{\psi \phi}, \quad B_s\to\mu^+\mu^-, \quad S_{\mu^+\mu^-}^s,
\quad \mathcal{A}^\lambda_{\Delta\Gamma},\quad B\to X_s \nu \bar \nu,
\ee
which depend only on $\tilde s_{23}$ and $\delta_2$.

{\bf Class B:}

\be\label{ClassB}
\Delta M_d, \quad S_{\psi K_S}, \quad  B_d\to\mu^+\mu^-, \quad  S_{\mu^+\mu^-}^d
\ee
which depend only on $\tilde s_{13}$ and $\delta_1$.

 {\bf Class C:}

\be\label{ClassC}
\varepsilon_K, \quad \kpn, \quad \klpn, \quad K_L\to\mu^+\mu^-, \quad
K_L\to \pi^0 \ell^+\ell^-,\quad  \epe,
\ee
which depend on $\tilde s_{13}$, $\tilde s_{23}$ and $\delta_2-\delta_1$.

While the processes and related observables in classes A and B are at first sight uncorrelated
from each other, the fact that the processes in class C depend on
all four parameters implies indirect correlation between all classes
when experimental constraints are taken into account.

\subsection{Master Relations}
In view of a very simple structure of $Z'$ contributions, it is possible to
derive the relations between the modifications of the master functions $S_i$,
$X_i$ and $Y_i$ for arbitrary $\beta\not=\sqrt{3}$. We will first list
these general expressions and apply them to the case $\beta=1/\sqrt{3}$.
Subsequently we will give them for $\beta=\sqrt{3}$. Finally we will
discuss the implications for phenomenology performed in the next section,
where we will see the implied
correlations between various observables in explicit terms.  In
what follows we will denote the modifications of master functions $F_i$
coming from $Z'$ contributions simply by $\Delta F_i$.

We first find the relation between the $Z'$ effects in $\Delta F=2$
master functions
\be\label{master1}
\frac{\Delta S_K}{\Delta S_d\Delta S_s^*}=\frac{M^2_{Z'}g^2_{\rm SM}}{4 \tilde r}\left[\frac{\Delta_L^{sd}(Z')}{\Delta_L^{bd}(Z')\Delta_L^{bs*}(Z')}\right]^2=
\frac{3.68}{\tilde r}\left(\frac{M_{Z'}}{3\tev}\right)^2(1-(1+\beta^2)s_W^2).
\ee
Here and in following equations we set $|V_{tb}|=1$ and
$\tilde c_{13}=\tilde c_{23}=1$ if necessary. $\tilde r$ is given in (\ref{rtilde}).

Next relations between the $Z'$ contributions to $\Delta F=1$ functions $X_i$
and $S_i$ functions are given by
\be\label{master2}
\frac{\Delta X_q}{\sqrt{\Delta S_q^*}}=a_q\frac{\Delta_L^{\nu\bar\nu}(Z')}{2\sqrt{\tilde r}g_{\rm
SM}M_{Z'}}=-a_q\frac{0.085}{\sqrt{\tilde r}}\left(\frac{3\tev}{M_{Z'}}\right)
\frac{\left[1-(1+\sqrt{3}\beta)s_W^2\right]}{\sqrt{1-(1+\beta^2)s_W^2}},
\ee
where
\be
a_d=1, \qquad a_s=-1.
\ee
Moreover,
\be\label{master2a}
\frac{\Delta X_K}{\sqrt{\Delta S_K}}=\frac{\Delta X_s}{\sqrt{\Delta S_s^*}},
\ee
where with our definitions there is no
complex conjugate  on the l.h.s of this equation.

Finally, there is the relation between $Z'$ contributions to $X_i$ and $Y_i$:
\be\label{master3}
\frac{\Delta Y_i}{\Delta X_i}= -\frac{\left[1-(1-\sqrt{3}\beta)s_W^2\right]}{1-(1+\sqrt{3}\beta)s_W^2}.
\ee

Applying these formulae for the case $\beta=1/\sqrt{3}$ we find
\be\label{master1R}
\frac{\Delta S_K}{\Delta S_d\Delta S_s^*}=
\frac{2.55}{\tilde r}\left(\frac{M_{Z'}}{3\tev}\right)^2 \,,
\ee
\be\label{master2R}
\frac{\Delta X_q}{\sqrt{\Delta S_q^*}}=-a_q\frac{0.055}{\sqrt{\tilde r}}
\left(\frac{3\tev}{M_{Z'}}\right)\,,
\ee
\be\label{master3R}
\frac{\Delta Y_i}{\Delta X_i}= -1.86 \,,
\ee
\begin{align}\label{rel4}
 \Delta Y_i &  = \Delta Z_i \,.
\end{align}

For the $\sqrt{3}$ case we find

\be\label{master1RB}
\frac{\Delta S_K}{\Delta S_d\Delta S_s^*}=
\frac{0.27}{\tilde r}\left(\frac{M_{Z'}}{3\tev}\right)^2 \,,
\ee
\be\label{master2RB}
\frac{\Delta X_q}{\sqrt{\Delta S_q^*}}=-a_q\frac{0.023}{\sqrt{\tilde r}}
\left(\frac{3\tev}{M_{Z'}}\right)\,,
\ee
\be\label{master3RB}
\frac{\Delta Y_i}{\Delta X_i}= 1\,,
\ee

 \begin{equation}\label{rel3}
 \Delta Y_i = s_W^2 \Delta Z_i=\Delta X_i\,,\qquad i = K,d,s\,.
\end{equation}

\subsection{Implications}
While our statements to be  made below are more
general, in giving numerical examples we will concentrate on the $\model$ case. The master
relations listed above have a number of implications which one can see even
before a detailed numerical analysis is performed:
\begin{itemize}
\item
The relation (\ref{master1R}) implies that after the experimental
constraints on
$\Delta M_{s,d}$ have been taken into account, the effects in $\varepsilon_K$
must be smaller than in $\Delta M_{s,d}$ for values $M_{Z'}\le 3\tev$.
Indeed, NP contributions in $\Delta M_{s,d}$ are constrained to
be at most of $10\%$ of the SM values and therefore
the absolute values of $\Delta S_q$ can be
at most $0.25$. As $r_i\approx 1$,  we find then $\Delta S_K\le 0.16$, which
implies a correction of at most $6\%$.  However, as we will discuss
later on for much larger $M_{Z'}$ these effects will increase significantly.
\item
The relation (\ref{master2R}) implies that NP effects in the processes
with $\nu\bar\nu$ in the final state are rather small for $M_{Z'}=3\tev$.
Requiring again that $\Delta S_q\le 0.25$ and taking into account
that $X(x_t)\approx 1.5$, we find that for  $M_{Z'}=3\tev$
only effects of at most $5\%$ at the level of branching ratios are expected for
$b\to s\nu\bar\nu$ transitions. For $\kpn$ and $\klpn$ these effects are even
smaller. For  $M_{Z'}=1\tev$ the relevant mixing parameters $\tilde s_{13}$
and $\tilde s_{23}$ have to
be decreased by roughly a factor of 3 to satisfy $\Delta M_{s,d}$
constraint. The inspection of the dependence on mixing
angles shows then that this decrease of the latter is compensated
approximately by the decrease of $M_{Z'}$ in the case of $\kpn$ and $\klpn$.
However, it is overcompensated in the case of $b\to s \nu\bar\nu$ transitions,
so that the modifications of the  branching ratios in this case
by $15\%$ are possible.
\item
The relation (\ref{master3R}) implies that NP effects in decays with
$\mu^+\mu^-$ in the final state can be larger than in the case of
$\nu\bar\nu$. This is not only because of the factor $-1.9$ in this relation
but also because $Y(x_t)$ is smaller than $X(x_t)$: $Y(x_t)\approx 1$. Thus for  $M_{Z'}=3\tev$  effects of $10-15\%$ at the level of the branching ratio are still allowed
in $B_{s,d}\to\mu^+\mu^-$ and these effects  are expected to be increased up to
$30\%$ for $M_{Z'}=1\tev$. Interestingly the NP effects
in $B_{s,d}\to\mu^+\mu^-$ and $B\to X_s\nu\bar\nu$ are anti-correlated. The
suppression of $B_{s,d}\to\mu^+\mu^-$ implies enhancement of  $B\to X_s\nu\bar\nu$ and vice versa. Similar comments apply to $\kpn$ and $K_L\to\mu^+\mu^-$.
Unfortunately, due to small NP effects in decays with $\nu\bar\nu$ in the
final states,
these anti-correlations will be difficult to test.
\end{itemize}

Finally let us note that once the constraints on $\Delta S_q$ are taken
into account NP effects for $\beta=\sqrt{3}$ both in $\varepsilon_K$ and
in all rare decays are so  small that it will be difficult to
distinguish this model from the SM on the basis of flavour violation in
meson decays. Therefore, we will not consider this case further.

\section{Numerical Analysis \label{sec:7}}
\subsection{Recent Data and Anomalies}
Our prime motivation for a new analysis of the 331 models are
the most recent data on $S_{\psi\phi}$ and
$B_{s,d}\to\mu^+\mu^-$ decays from LHCb that read \cite{Clarke:1429149,Aaij:2012ac,LHCbBsmumu}
\be\label{LHCb1}
S_{\psi\phi}=0.002\pm 0.087, \quad S^{\rm SM}_{\psi\phi}=0.035\pm 0.002,
\ee
\be\label{LHCb2}
\mathcal{B}(B_{s}\to\mu^+\mu^-) = (3.2^{+1.5}_{-1.2}) \times 10^{-9}, \quad
\mathcal{B}(B_{s}\to\mu^+\mu^-)^{\rm SM}=(3.23\pm0.27)\times 10^{-9},
\ee
\be\label{LHCb3}
\mathcal{B}(B_{d}\to\mu^+\mu^-) \le 9.4\times 10^{-10}, \quad
\mathcal{B}(B_{d}\to\mu^+\mu^-)^{\rm SM}=(1.07\pm0.10)\times 10^{-10}.
\ee
We have also shown the most recent direct SM predictions  for
$\mathcal{B}(B_{q}\to\mu^+\mu^-)$ with new lattice input \cite{Buras:2012ru}.
Almost identical results are obtained
 by applying  the CMFV
relations between $\mathcal{B}(B_{q}\to\mu^+\mu^-)$ and
$\Delta M_q$  within the SM \cite{Buras:2010wr,Buras:2012ts}.  In quoting these results
we did not include the correction $r(y_s)$ but it has to be taken into
account either in the theory or experiment when the data improve.

While the SM still survived another test,
from the present perspective $S_{\psi\phi}$ could still be found
in the $2\sigma$ range
\be\label{Spsiphirange}
-0.18\le S_{\psi\phi}\le 0.18
\ee
and finding it to be negative would be a clear signal of NP. Moreover finding
it above $0.1$ would also be a signal of NP but not as pronounced as a
negative value.

Concerning $B_{s}\to\mu^+\mu^-$ its branching ratio can still be enhanced
by  a factor of 2 and finding it $(5-6)\cdot 10^{-9}$ would be a clear signal of NP at work.  Also finding it well below the
SM value is still possible. Similarly we should note that the upper bound on
$\mathcal{B}(B_{d}\to\mu^+\mu^-)$ is still one order of magnitude above the
SM value. It could turn out after all that it is $B_{d}\to\mu^+\mu^-$
and not  $B_{s}\to\mu^+\mu^-$ that will most clearly signal NP in these
decays.

Another motivation for our analysis is a number of anomalies seen from
the point of view of the SM that could become more relevant as the data
and the lattice input improve.
In this context it should be emphasized that because of  some visible
$\varepsilon_K-S_{\psi K_S}$ tension \cite{Lunghi:2008aa,Buras:2008nn,Buras:2009pj,Lunghi:2009sm,Lunghi:2010gv,UTfit-web} within
the SM the pattern of deviation from SM expectations depends on
whether $\varepsilon_K$ or $S_{\psi K_S}$ is used as a basic observable to
fit the CKM parameters. As both observables can receive important
contributions from NP, none of them is optimal for this goal. The solution
to this problem will be  solved one day by precise measurements of the CKM parameters
with the help
of tree-level decays. Unfortunately, the tension between the inclusive
and exclusive determinations of $\vub$ \footnote{For a recent review see
\cite{Ricciardi:2012pf}.} and the poor knowledge of  the
phase $\gamma$
from tree-level decays preclude this solution at present.

 In view of this,  it was useful already for some time \cite{Buras:2012ts}
to set $\gamma\approx 70^\circ$ in the ballpark of tree-level determinations and consider two scenarios for $\vub$:
 \begin{itemize}
\item
{\bf Exclusive (small) $\vub$ Scenario 1:}
$|\varepsilon_K|$ is smaller than its experimental determination,
while $S_{\psi K_S}$ is close to the central experimental value.
\item
{\bf Inclusive (large) $\vub$ Scenario 2:}
$|\varepsilon_K|$ is consistent with its experimental determination,
while $S_{\psi K_S}$ is significantly higher than its  experimental value.
\end{itemize}

Thus dependently which scenario is considered we will ask the $\model$ model
to provide
{\it constructive} NP contributions to $|\varepsilon_K|$ (Scenario 1)
or {\it destructive} NP contributions to  $S_{\psi K_S}$ (Scenario 2)
without spoiling
the agreement with the data
for $S_{\psi K_S}$ (Scenario 1) and for $|\varepsilon_K|$ (Scenario 2).

 This strategy turned out to be useful in the analyses
of other models \cite{Buras:2012ts} and helped
to identify the correlation $S_{\psi K_S}-S_{\psi\phi}-\vub$  in
the context of models with $U(2)^3$  flavour symmetry \cite{Buras:2012sd}.
Yet, one should emphasize the following difference between these two scenarios.
In Scenario 1, the central value of $|\varepsilon_K|$ is visibly smaller than
the very precise data  but the still  significant parametric uncertainty
due to $\vcb^4$ dependence in $|\varepsilon_K|$ and a large uncertainty
in the charm contribution found at the NNLO level in \cite{Brod:2011ty}
does not make  this anomaly as pronounced as this is the case of
Scenario 2, where large $\vub$ implies definitely a value of $S_{\psi K_S}$
that is by $2-3\sigma$ above the data.

While  models with many new parameters can face successfully both scenarios
removing the deviations from the data for certain range of their parameters
it is not obvious from the beginning whether with improved data and
new lattice input the $\model$ model will be successful in this respect, in
particular when we will request from the $\model$ model to remove simultaneously
other anomalies that we will list below.
In fact in  simpler models without new sources of flavour violation
only one scenario for $\vub$ can be admitted as only in that scenario  a given
model has a chance to fit $\varepsilon_K$ and $S_{\psi K_S}$ simultaneously.
For instance as reviewed in \cite{Buras:2012ts} models with CMFV
select Scenario 1, while the 2HDM with MFV and flavour blind phases,
${\rm 2HDM_{\overline{MFV}}}$, selects  Scenario 2 for
$\vub$.

% \begin{table}[!tb]
% \centering
% \begin{tabular}{|c||c|c|c|}
% \hline
%  & Scenario 1: & Scenario 2:   & Experiment\\
% \hline
% \hline
%   \parbox[0pt][1.6em][c]{0cm}{} $|\varepsilon_K|$ & $1.87(26)  \cdot 10^{-3}$  & $2.28(32)\cdot 10^{-3}$ &$ 2.228(11)\times 10^{-3}$ \\
%
%  \parbox[0pt][1.6em][c]{0cm}{}$\mathcal{B}(B^+\to \tau^+\nu_\tau)$&  $0.74(14) \cdot 10^{-4}$&$1.19(20)\cdot 10^{-4}$ & $1.73(35) \times
% 10^{-4}$\\
%  \parbox[0pt][1.6em][c]{0cm}{}$(\sin2\beta)_\text{true}$ & 0.676(25) &0.812(23)  & $0.679(20)$\\
%  \parbox[0pt][1.6em][c]{0cm}{}$\Delta M_s\, [\text{ps}^{-1}]$ &19.0(21)&  19.1(21) &$17.77(12)$ \\
%  \parbox[0pt][1.6em][c]{0cm}{} $\Delta M_d\, [\text{ps}^{-1}]$ &0.55(6) &0.56(6)   &  $0.507(4)$\\
%  \parbox[0pt][1.6em][c]{0cm}{}$\mathcal{B}(B\to X_s\gamma)$ &$3.15(23) \cdot 10^{-4}$&$3.15(23)\cdot 10^{-4}$ & $3.55(26) \times
% 10^{-4}$\\
% \hline
% \end{tabular}
% \caption{\it SM prediction for various observables for  $|V_{ub}|=3.4\cdot 10^{-3}$ and $|V_{ub}|=4.3\cdot 10^{-3}$ and $\gamma =
% 68^\circ$ compared to experiment.
% }\label{tab:SMpred}~\\[-2mm]\hrule
%
%\end{table}
\begin{table}[!tb]
\centering
\begin{tabular}{|c||c|c|c|}
\hline
 & Scenario 1: & Scenario 2:   & Experiment\\
\hline
\hline
  \parbox[0pt][1.6em][c]{0cm}{} $|\varepsilon_K|$ & $1.72(26)  \cdot 10^{-3}$  & $2.15(32)\cdot 10^{-3}$ &$ 2.228(11)\cdot
10^{-3}$ \\

 \parbox[0pt][1.6em][c]{0cm}{}$\mathcal{B}(B^+\to \tau^+\nu_\tau)$&  $0.62(14) \cdot 10^{-4}$&$1.02(20)\cdot 10^{-4}$ & $0.99(25)
\cdot
10^{-4}$\\
 \parbox[0pt][1.6em][c]{0cm}{}$(\sin2\beta)_\text{true}$ & 0.623(25) &0.770(23)  & $0.679(20)$\\
 \parbox[0pt][1.6em][c]{0cm}{}$\Delta M_s\, [\text{ps}^{-1}]$ &19.0(21)&  19.0(21) &$17.77(12)$ \\
 \parbox[0pt][1.6em][c]{0cm}{} $\Delta M_d\, [\text{ps}^{-1}]$ &0.56(6) &0.56(6)   &  $0.507(4)$\\
\hline
\end{tabular}
\caption{\it SM prediction for various observables for  $|V_{ub}|=3.1\cdot 10^{-3}$ and $|V_{ub}|=4.0\cdot 10^{-3}$ and $\gamma =
68^\circ$ compared to experiment.
}\label{tab:SMpred}~\\[-2mm]\hrule
\end{table}

Now the tensions within the SM discussed above constitute only a subset
of visible deviations of its predictions from the data.
In Table~\ref{tab:SMpred} we illustrate the SM predictions for some of
these observables in both scenarios setting $\gamma=68^\circ$. We only
want to emphasize two points related to this table.

First, the SM branching ratio for $B^+\to\tau^+\nu_\tau$ in Scenario 1 appears
to be  visibly
below the data, although the latter are not very
precise. Moreover, as discussed previously the most recent Belle result
\cite{BelleICHEP}
in (\ref{eq:Btaunu_exp3})  and the world average in (\ref{eq:world}) are fully consistent with the SM. In
 Scenario 2 there is good agreement with the data. Until the experimental
number will be clarified from the present perspective
in Scenario 1 for $\vub$  models
providing an {\it enhancement} of this branching ratio should be favoured.

What is also
striking in this table is that with the new lattice input in Table~\ref{tab:input}
the predicted central values of $\Delta M_s$  and $\Delta M_d$, although
slightly above the data,  are both in a good agreement with the latter
when hadronic uncertainties are taken into account. In particular
the central value of the ratio $\Delta M_s/\Delta M_d$ is
 very close to  the data.
These results depend strongly on the lattice input and in the case
of $\Delta M_d$ on the value of $\gamma$. Therefore to get a better insight
both lattice input and the tree level determination of $\gamma$
have to improve.
From
the present perspective, models providing  $10\%$ {\it suppression}
of {\it both} $\Delta M_s$ and $\Delta M_d$ with respect to SM values
appear to be slightly favoured. As pointed out in
\cite{Blanke:2006yh} this is not possible within the models with CMFV and as
 demonstrated in \cite{Buras:2012ts} these models do not offer
a good simultaneous discription of $\Delta M_{s,d}$ and $\varepsilon_K$.
It is then interesting to see whether the $\model$ model is performing better
in this respect.

\subsection{Strategy}
 As already advertised at the beginning of our paper,
it is not our goal to present a full-fledged numerical analysis of the
$\model$ model including present theoretical, parametric and experimental
uncertainties as this would only wash out various correlations between various
observables that we would like to emphasize. Recent sophisticated Monte-Carlo
analyses in the context of other models and model independent studies
\cite{Barbieri:2011ci,Altmannshofer:2011gn,Altmannshofer:2012ir,Botella:2012ju}
that in particular used some kind
of an average between inclusive and exclusive values of $\vub$, while
certainly interesting, do not allow often for a transparent presentation
of such correlations that are seen in our simplified approach.

In view of the comments above our strategy in our numerical analysis should
begin with finding out whether it is the inclusive or exclusive value
of $\vub$ which is favoured by the $\model$ model.
 As our discussion of previous section indicates, the imposition
of the experimental constraints on $\Delta M_{s,d}$  for
$1\tev\le M_{Z'}\le 3\tev$   implies very
small NP effects in $\varepsilon_K$. Consequently the $\model$ model
favours a value of $\vub$ that is closer to inclusive determinations.
Then not only $\varepsilon_K$ is in good agreement with the data, but
also the branching ratio for $B^+\to\tau^+\nu_\tau$ is closer to the
experimental value than in the SM. Therefore we will  perform our
numerical analysis
within Scenario 2 setting the CKM parameters at the following values

\noindent  {\bf Scenario 2:}
\be
\vus=0.2252, \qquad \vcb=0.0406, \qquad \gamma=68^\circ, \qquad
\vub= 4.0\cdot 10^{-3}\,.
\ee
The values for
 $|V_{us}|$ and  $|V_{cb}|$ correspond to their central values measured
 in tree level
decays and the value of $\vub$ is in the ballpark of inclusive determinations.
We checked that varying  $\vub$ within $\pm 5\%$ has only minor impact
on our results.
Other inputs are collected in
Table~\ref{tab:input}.

\begin{table}[!tb]
\center{\begin{tabular}{|l|l|}
\hline
$G_F = 1.16637(1)\times 10^{-5}\gev^{-2}$\hfill\cite{Nakamura:2010zzi} 	&  $m_{B_d}= 5279.5(3)\mev$\hfill\cite{Nakamura:2010zzi}\\
$M_W = 80.385(15) \gev$\hfill\cite{Nakamura:2010zzi}  								&	$m_{B_s} =
5366.3(6)\mev$\hfill\cite{Nakamura:2010zzi}\\
$\sin^2\theta_W = 0.23116(13)$\hfill\cite{Nakamura:2010zzi} 				& 	$F_{B_d} =
(190.6\pm4.7)\mev$\hfill\cite{Laiho:2009eu}\\
$\alpha(M_Z) = 1/127.9$\hfill\cite{Nakamura:2010zzi}									& 	$F_{B_s} =
(227.6\pm5.0)\mev$\hfill\cite{Laiho:2009eu}\\
$\alpha_s(M_Z)= 0.1184(7) $\hfill\cite{Nakamura:2010zzi}								&  $\hat B_{B_d} =
1.26(11)$\hfill\cite{Laiho:2009eu}\\\cline{1-1}
$m_u(2\gev)=(2.1\pm0.1)\mev $ 	\hfill\cite{Laiho:2009eu}						&  $\hat B_{B_s} =
1.33(6)$\hfill\cite{Laiho:2009eu}\\
$m_d(2\gev)=(4.73\pm0.12)\mev$	\hfill\cite{Laiho:2009eu}							& $\hat B_{B_s}/\hat B_{B_d}
= 1.05(7)$ \hfill \cite{Laiho:2009eu} \\
$m_s(2\gev)=(93.4\pm1.1) \mev$	\hfill\cite{Laiho:2009eu}				&
$F_{B_d} \sqrt{\hat
B_{B_d}} = 226(15)\mev$\hfill\cite{Laiho:2009eu} \\
$m_c(m_c) = (1.279\pm 0.013) \gev$ \hfill\cite{Chetyrkin:2009fv}					&
$F_{B_s} \sqrt{\hat B_{B_s}} =
279(15)\mev$\hfill\cite{Laiho:2009eu} \\
$m_b(m_b)=4.19^{+0.18}_{-0.06}\gev$\hfill\cite{Nakamura:2010zzi} 			& $\xi =
1.237(32)$\hfill\cite{Laiho:2009eu}
\\
$m_t(m_t) = 163(1)\gev$\hfill\cite{Laiho:2009eu,Allison:2008xk} &  $\eta_B=0.55(1)$\hfill\cite{Buras:1990fn,Urban:1997gw}  \\
$M_t=172.9\pm0.6\pm0.9 \gev$\hfill\cite{Nakamura:2010zzi} 						&  $\Delta M_d = 0.507(4)
\,\text{ps}^{-1}$\hfill\cite{Nakamura:2010zzi}\\\cline{1-1}
$m_K= 497.614(24)\mev$	\hfill\cite{Nakamura:2010zzi}								&  $\Delta M_s = 17.77(12)
\,\text{ps}^{-1}$\hfill\cite{Nakamura:2010zzi}\\	
$F_K = 156.1(11)\mev$\hfill\cite{Laiho:2009eu}												&
$S_{\psi K_S}= 0.679(20)$\hfill\cite{Nakamura:2010zzi}\\
$\hat B_K= 0.764(10)$\hfill\cite{Laiho:2009eu}												&
$S_{\psi\phi}= 0.002\pm 0.087$\hfill\cite{Clarke:1429149}\\
$\kappa_\epsilon=0.94(2)$\hfill\cite{Buras:2008nn,Buras:2010pza}										&
\\	
$\eta_1=1.87(76)$\hfill\cite{Brod:2011ty}												
	& $\tau(B_s)= 1.471(25)\,\text{ps}$\hfill\cite{Asner:2010qj}\\		
$\eta_2=0.5765(65)$\hfill\cite{Buras:1990fn}												
& $\tau(B_d)= 1.518(7) \,\text{ps}$\hfill\cite{Asner:2010qj}\\
$\eta_3= 0.496(47)$\hfill\cite{Brod:2010mj}												
& \\\cline{2-2}
$\Delta M_K= 0.5292(9)\times 10^{-2} \,\text{ps}^{-1}$\hfill\cite{Nakamura:2010zzi}	&
$|V_{us}|=0.2252(9)$\hfill\cite{Nakamura:2010zzi}\\
$|\eps_K|= 2.228(11)\times 10^{-3}$\hfill\cite{Nakamura:2010zzi}					& $|V_{cb}|=(40.6\pm1.3)\times
10^{-3}$\hfill\cite{Nakamura:2010zzi}\\\cline{1-1}
  $\mathcal{B}(B\to X_s\gamma)=(3.55\pm0.24\pm0.09) \times10^{-4}$\hfill\cite{Nakamura:2010zzi}                                                                &
$|V^\text{incl.}_{ub}|=(4.27\pm0.38)\times10^{-3}$\hfill\cite{Nakamura:2010zzi}\\
$\mathcal{B}(B^+\to\tau^+\nu)=(0.99\pm0.25)\times10^{-4}$\hfill\cite{Tarantino:2012mq}	&
$|V^\text{excl.}_{ub}|=(3.38\pm0.36)\times10^{-3}$\hfill\cite{Nakamura:2010zzi}\\\cline{1-1}					
$\tau_{B^\pm}=(1641\pm8)\times10^{-3}\,\text{ps}$\hfill\cite{Nakamura:2010zzi}
														&
\\

\hline
\end{tabular}  }
\caption {\textit{Values of the experimental and theoretical
    quantities used as input parameters.}}
\label{tab:input}~\\[-2mm]\hrule
\end{table}

Having fixed the four parameters of the CKM matrix, the {``true''} values
of the angle $\beta$  and
of the element $\vtd$
are  obtained from the unitarity of the CKM matrix:
\begin{equation} \label{eq:Rt_beta}
\vtd=\vus \vcb R_t,\quad
R_t=\sqrt{1+R_b^2-2 R_b\cos\gamma} ~,\quad
\cot\beta=\frac{1-R_b\cos\gamma}{R_b\sin\gamma}~,
\end{equation}
where
\be
 R_b=\left(1-\frac{\lambda^2}{2}\right)\frac{1}{\lambda}\frac{|V_{ub}|}{\vcb}.
\ee
We find then
\be\label{CKMinput}
\vts=
0.0399 \,, \qquad
\vtd= 8.51 \cdot 10^{-3}\quad\text{(S2)}\,.
\ee

Concerning the direct lower bound  on  $M_{Z'}$ from collider experiments, the
most stringent bounds are provided by CMS experiment \cite{Chatrchyan:2012it}.
The precise value depends
on the model considered. While for the so-called sequential $Z'$ the lower
bound for $M_{Z'}$ is in the ballpark of $2.5\tev$, in other models values
as low as $1\tev$ are still possible. In order to be on the safe side
we will choose as our minimal value $M_{Z'}=1\tev$ and will
also provide values for $M_{Z'}=3\tev$.
With the help of
the formulae in subsection \ref{sec:10} it should be possible to
estimate approximately, how our results would change for other
values of  $M_{Z'}$. In this context let us notice that in view
of moderate  NP effects found in $\model$ model
the modifications of rare
decay branching ratios due to NP will be governed by the interference of
SM and NP contributions and consequently will be inversely  proportional
to  $M_{Z'}$ as the corrections to the master functions.

\subsection{Simplified Analysis}
We first perform a simplified analysis of
$\Delta M_{d,s}$, $S_{\psi K_S}$
and $S_{\psi\phi}$
in order to identify oases in the space of four parameters in (\ref{newpar})
for which these four observables are consistent with experiment.
To this end we set all other input parameters at their central values
but in order to take partially hadronic
and experimental uncertainties into account we require the $\model$ model
to reproduce the data for $\Delta M_{s,d}$ within $\pm 5\%$ and the
data on $S_{\psi K_S}$ and $S_{\psi\phi}$ within experimental
$2\sigma$.

Specifically, our search is governed by the following allowed ranges:

\be
16.9/{\rm ps}\le \Delta M_s\le 18.7/{\rm ps},
\quad  -0.18\le S_{\psi\phi}\le 0.18, \label{oases23}
\ee
\be
0.48/{\rm ps}\le \Delta M_d\le 0.53/{\rm ps},\quad  0.64\le S_{\psi K_S}\le 0.72 . \label{oases13}
\ee

\begin{figure}[!tb]
\begin{center}
\includegraphics[width=0.45\textwidth] {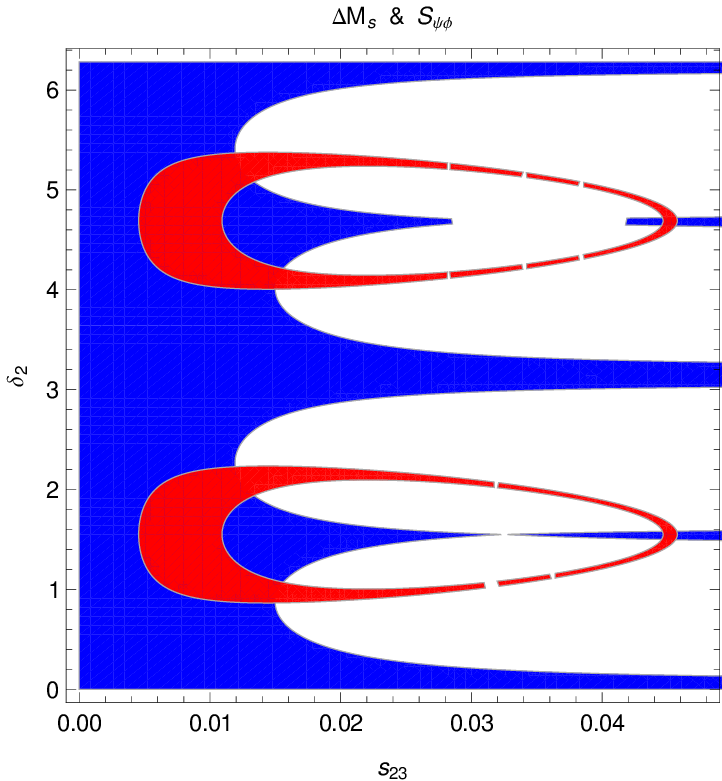}
  \includegraphics[width=0.45\textwidth] {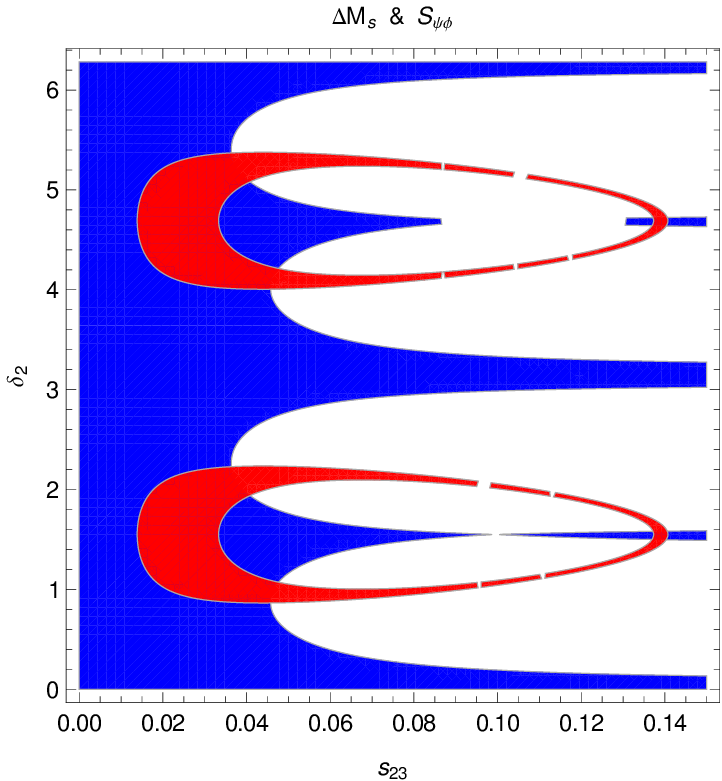}
\caption{\it  Ranges for $\Delta M_s$ (red region) and $S_{\psi \phi}$ (blue region) satisfying the bounds in eq. (\ref{oases23}).
 The plot on the left is obtained for $M_{Z^\prime}=1$ TeV, that on the right for  $M_{Z^\prime}=3$ TeV.
}\label{fig-oases23}~\\[-2mm]\hrule
\end{center}
\end{figure}

\begin{figure}[!tb]
\begin{center}
   \includegraphics[width=0.45\textwidth] {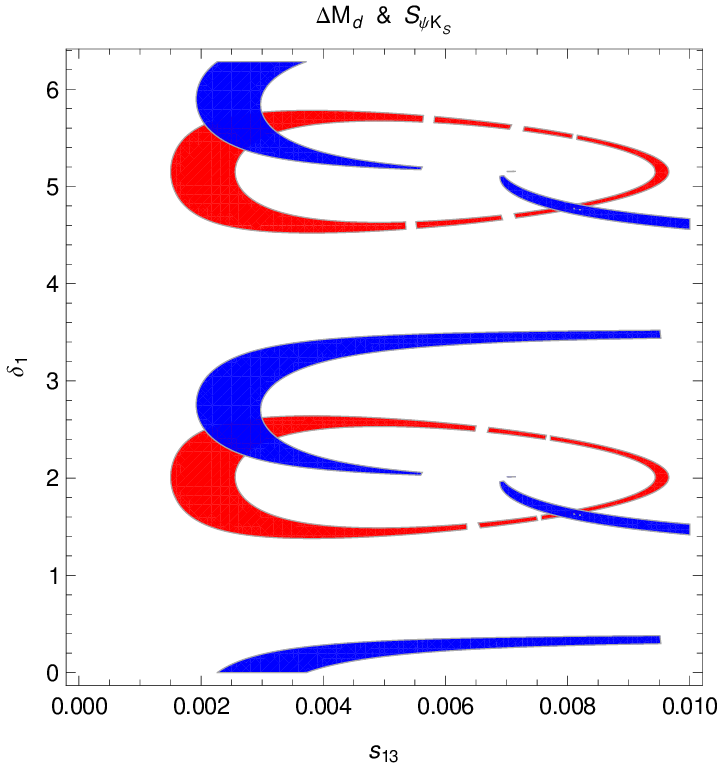}
   \includegraphics[width=0.45\textwidth] {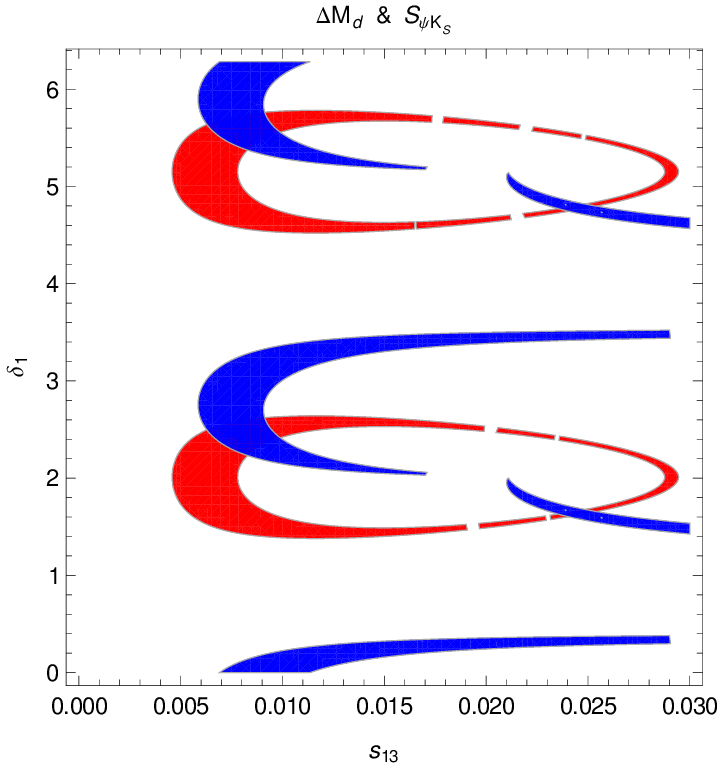}
\caption{\it
  Ranges for $\Delta M_d$ (red region) and $S_{\psi K_s}$ (blue region) satisfying the bounds in Eq.~(\ref{oases13}). The plot on
the left is obtained for $M_{Z^\prime}=1$ TeV, that on the right for  $M_{Z^\prime}=3$ TeV. }\label{fig-oases13}~\\[-2mm]\hrule
\end{center}
\end{figure}

The search for these oases is simplified by the fact that the pair
$(\Delta M_s,S_{\psi\phi})$ depends only on $(\tilde s_{23},\delta_2)$,
while the pair
$(\Delta M_d,S_{\psi K_S})$ only on $(\tilde s_{13},\delta_1)$.
The result of this search for $M_{Z'}=1\tev$ is shown on the left in Figs.~\ref{fig-oases23} and~\ref{fig-oases13}
for $(\tilde s_{23},\delta_2)$ and $(\tilde s_{13},\delta_1)$,
respectively.
The {\it red} regions correspond to the allowed ranges for $\Delta M_{d,s}$,
while the {\it blue} ones to the corresponding ranges for  $S_{\psi K_S}$ and $S_{\psi\phi}$. The overlap between red and blue regions identifies the
oases we were looking for. Analogous plots for $M_{Z'}=3\tev$ are shown on the right in the same
Figs.~\ref{fig-oases23} and~\ref{fig-oases13}.
%%%%%%%%%%%%%%%%%%%%%%%%%%%%%%%%%%%%%%%%%%%%%%%%%%%%%%%%
%%%%%%%%%%%%%%%%%%%%%%%%%%%%%2%%%%%%%%%%%%%%%%%%%%%%%%%%%
%%%%%%%%%%%%%%%%%%%%%%%%%%%%%%%%%%%%%%%%%%%%%%%%%%%%%%%%
From these plots we extract several oases that are collected in
Tables~\ref{s23oases} and \ref{s13oases}. We denote by
$A_i(M_{Z^\prime})$ and $B_i(M_{Z^\prime})$, the oases for $B_s$ and $B_d$
system, respectively. We observe the following pattern:
\begin{itemize}
\item
The increase of $M_{Z^\prime}$ by a factor of three, allows to increases $\tilde s_{13}$ and $\tilde s_{23}$ by the same factor. This
structure is evident from the formulae
for $\Delta S_q$. However, in view of the relation (\ref{master2R}) this
change will have impact on  rare $B_{s,d}$ decays, making the NP
effects in these decays for  $M_{Z^\prime}=3\tev$ smaller. The inspection of
the formulae for $\Delta X_K$ shows that when the constraints from B-physics
in (\ref{oases23}) and (\ref{oases13}) are imposed, the increase of
$M_{Z^\prime}$ is compensated in rare $K$ decays by the increase of
 $\tilde s_{13}$ and $\tilde s_{23}$ so that these decays practically
do not depend on $M_{Z^\prime}$. In the case of $\varepsilon_K$ the same
phenomenon even increases the room for NP effects with increasing $M_{Z^\prime}$
for values of several TeV, where all FCNC constraints can be satisfied.
\item
Simultaneously the ranges for $\delta_i$ remain unchanged.
\item
For each oasis with a given $\delta_i$ there is another oasis with $\delta_i$ shifted
by $180^\circ$.
\item
The oases with $i=2,4$ are very small.
\end{itemize}

In the rest of this subsection we will confine our
numerical analysis to these oases, investigating whether some of them can be
excluded by other constraints. In this context one should note that the oases
in the $(\tilde s_{23},\delta_2)$ and $(\tilde s_{13},\delta_1)$ spaces are without other
constraints independent of each other. Consequently when the full space of
four free variables is considered we deal really with 16 oases.  Fortunately,
as we will demonstrate at the end of this section, 8 oases involving
at least one of the oases $A_2,A_4$ can be eliminated by $\varepsilon_K$ and $\Delta M_s/\Delta M_d$ constraints
alone. Among the 8 remaining ones,  we will concentrate at first on the
following four oases:
\be\label{OASES}
(A_1,B_1), \qquad (A_1,B_3), \qquad (A_3,B_1), \qquad (A_3,B_3).
\ee
We will see that when all observables discussed by us will be measured we
will be able to identify {\it uniquely} the oasis that has the best chance
to describe the data correctly.

The case of the {\it small} oases involving one of the oases $B_2,B_4$ will be discussed in  Section \ref{small-oases}.

 As a final comment, we observe that the oases reported in Tables ~\ref{s23oases} and \ref{s13oases} actually describe squares in
the spaces $(\tilde s_{23},\delta_2)$ and $(\tilde s_{13},\delta_1)$, while the corresponding regions in Figs. \ref{fig-oases23}
and \ref{fig-oases13} have more complicated shapes. Indeed, in our numerical analysis of the various observables we have varied
the parameters in the {\it true} oases,  requiring that constraints (\ref{oases23}) and (\ref{oases13}) are satisfied.

\begin{table}[!tb]
\centering
\begin{tabular}{|c||c|c|c|c|c|}
\hline
 & $\tilde s_{23}$ & $\delta_2$   & $S_{\mu^+\mu^-}^s$ & $\Delta S_{\psi\phi}$ &
$\Delta\mathcal{B}(B_s\to\mu^+\mu^-)$  \\
\hline
\hline
  \parbox[0pt][1.6em][c]{0cm}{} $A_1(1)$ & $0.0044-0.0157$
& $49^\circ-129^\circ$ &  $-$ & $\pm$ & $\pm$ \\

 \parbox[0pt][1.6em][c]{0cm}{}$A_2(1)$&  $0.045-0.046$&$87^\circ-92^\circ$ & &&\\
\parbox[0pt][1.6em][c]{0cm}{} $A_3(1)$ & $0.0044-0.0157$
& $229^\circ-309^\circ$ &  $+$   & $\pm$&$\mp$ \\
 \parbox[0pt][1.6em][c]{0cm}{}$A_4(1)$&  $0.045-0.046$&$267^\circ-272^\circ$ & & & \\
\hline
\hline
  \parbox[0pt][1.6em][c]{0cm}{} $A_1(3)$ & $0.0140-0.0477$
& $49^\circ-129^\circ$ & $-$   & $\pm$ & $\pm$  \\

 \parbox[0pt][1.6em][c]{0cm}{}$A_2(3)$&  $0.138-0.140$&$87^\circ-92^\circ$ & && \\
\parbox[0pt][1.6em][c]{0cm}{} $A_3(3)$ & $0.0140-0.0477$
& $229^\circ-309^\circ$ & $+$    & $\pm$&$\mp$   \\
 \parbox[0pt][1.6em][c]{0cm}{}$A_4(3)$&  $0.138-0.140$&$267^\circ-272^\circ$ & && \\
\hline
\end{tabular}
\caption{\it Oases in the space $(\tilde s_{23},\delta_2)$ for $M_{Z^\prime}=1\tev$
and  $M_{Z^\prime}=3\tev$. The sign of $S_{\mu^+\mu^-}^s$ chooses the oasis
uniquely. The same applies to the pair $S_{\psi\phi}$ and
$\Delta\mathcal{B}(B_s\to\mu^+\mu^-)$ as discussed in the text.
}\label{s23oases}~\\[-2mm]\hrule
\end{table}

\begin{table}[!tb]
\centering
\begin{tabular}{|c||c|c|c|c|}
\hline
 & $\tilde s_{13}$ & $\delta_1$   & $\Delta\mathcal{B}(B_d\to\mu^+\mu^-)$ &
 $S_{\mu^+\mu^-}^d$ \\
\hline
\hline
  \parbox[0pt][1.6em][c]{0cm}{} $B_1(1)$ & $0.0020-0.0032$
& $128^\circ-150^\circ$ & $-$ & $+$ \\
 \parbox[0pt][1.6em][c]{0cm}{}$B_2(1)$&  $0.0078-0.0082$&$92^\circ-95^\circ$ & & $+$\\
\parbox[0pt][1.6em][c]{0cm}{} $B_3(1)$ & $0.0020-0.0032$
& $308^\circ-330^\circ$ & $+$ & $-$\\
\parbox[0pt][1.6em][c]{0cm}{}$B_4(1)$&  $0.0078-0.0082$&$272^\circ-275^\circ$ & &$-$\\
\hline
\hline
  \parbox[0pt][1.6em][c]{0cm}{} $B_1(3)$ & $0.0063-0.0099$
& $128^\circ-150^\circ$ & $-$ & $+$\\

 \parbox[0pt][1.6em][c]{0cm}{}$B_2(3)$&  $0.024-0.025$&$92^\circ-95^\circ$ & &$+$\\
\parbox[0pt][1.6em][c]{0cm}{} $B_3(3)$ & $0.0063-0.0099$
& $308^\circ-330^\circ$ & $+$ & $-$\\
\parbox[0pt][1.6em][c]{0cm}{}$B_4(3)$&  $0.024-0.025$&$272^\circ-275^\circ$ & &$-$\\
\hline
\end{tabular}
\caption{\it Oases in the space $(\tilde s_{13},\delta_1)$ for $M_{Z^\prime}=1\tev$
and  $M_{Z^\prime}=3\tev$. The enhancement or suppression of $\mathcal{B}(B_d\to\mu^+\mu^-)$ with respect to the SM value chooses the oasis
uniquely. The same
applies to the sign of $S_{\mu^+\mu^-}^d$.
}\label{s13oases}~\\[-2mm]\hrule
\end{table}

\subsection{The Search for the Optimal Oasis}

We will now demonstrate that it is possible to identify uniquely one of the
oases listed in (\ref{OASES}) by considering simultaneously a subset of
particular observables considered in our analysis. Once the optimal oasis has
been selected, the correlations between different observables in this oasis
will tell us whether  the $\overline{331}$ model is consistent with experiment.

Before entering the search, it should be emphasized that in view of the small
number of free parameters involved, the patterns of flavour violation
 and CP-violation
depends in the crucial manner on two conditions which led to the oases in
Tables~\ref{s23oases} and \ref{s13oases}:
\begin{itemize}
\item
The requirement that $S_{\psi K_S}$ has to be suppressed significantly below
its SM value to agree with data,
\item
Both $\Delta M_s$ and $\Delta M_d$ must also be suppressed below their
SM values.
\end{itemize}

While the first requirement is certain in view of small contributions to $\varepsilon_K$ and implied large value of $\vub$, the second
depends crucially on
the non-perturbative parameters  and it will be of interest to monitor how the
patterns of flavour violation presented below may change when the lattice
calculations will improve in precision.

Inspecting the expressions for various observables in different oases, we
have identified the fastest route to the optimal oasis. We describe this
route in the first step below. In the following steps we will demonstrate
how the correlations between various observables can give additional tests
once the analysis is confined to a particular oasis.
% {In all plots that follow the constraints in Eq.~(\ref{oases23}) and~(\ref{oases13}) are taken into account if not stated
% otherwise.}

{\bf Step 1:}

\begin{figure}[!tb]
\begin{center}
\includegraphics[width=0.45\textwidth] {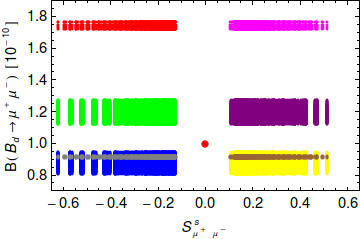}
  \includegraphics[width=0.45\textwidth] {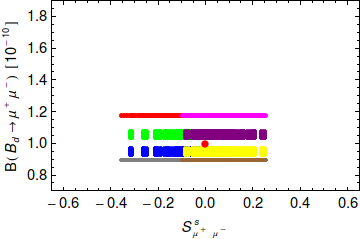}
\caption{\it
 $\mathcal{B}(B_d \to \mu^+ \mu^-)$ versus $S_{\mu^+ \mu^-}^s$. The plot on the left is obtained for $M_{Z^\prime}=1$ TeV, that on
the right for  $M_{Z^\prime}=3$ TeV. The colours indicate the oases to which the parameters belong: ($A_1$,$B_1$): blue,
$(A_1,B_3)$: green,
$(A_3,B_1)$: yellow, $(A_3,B_3)$: purple, $(A_1,B_2)$: red, $(A_1,B_4)$: gray, $(A_3,B_2)$: magenta,
$(A_3,B_4)$: brown. Red point: central
value of SM prediction.
 }\label{rainbow}~\\[-2mm]\hrule
\end{center}
\end{figure}

It turns out that the optimal oasis can be found by just measuring two
observables:
\be
\mathcal{B}(B_d\to\mu^+\mu^-), \qquad S_{\mu^+\mu^-}^s,
\ee
that can be considered as two coordinates. Already the sign of shifts of
them with respect to the SM values identifies the optimal oasis \footnote{Also
the sign of $S_{\mu^+\mu^-}^d$ distinguishes between big oases. But this asymmetry is much harder to measure than
$\mathcal{B}(B_d\to\mu^+\mu^-)$.}. This
we summarize schematically in Fig.~\ref{rainbow}.
%%%%%%%%%%%%%%%%%%%%%%%%%%%%%%%%%%%%%%%%%%%%%%%%%%%%%%%%
%%%%%%%%%%%%%%%%%%%%%%%%%%%%%%%%%%%%%%%%%%%%%%%%%%%%%%%%

We observe:
\begin{itemize}
\item
$(A_1,B_1)$ oasis ({\it blue}) is chosen when
\be
 S_{\mu^+\mu^-}^s<0, \qquad
\mathcal{B}(B_d\to\mu^+\mu^-)< \mathcal{B}(B_d\to\mu^+\mu^-)_{\rm SM}.
\ee
\item
$(A_1,B_3)$ oasis ({\it green}) is chosen when
\be
 S_{\mu^+\mu^-}^s<0, \qquad
\mathcal{B}(B_d\to\mu^+\mu^-)> \mathcal{B}(B_d\to\mu^+\mu^-)_{\rm SM}.
\ee
\item
$(A_3,B_1)$ oasis  ({\it yellow}) is chosen when
\be
 S_{\mu^+\mu^-}^s>0, \qquad
\mathcal{B}(B_d\to\mu^+\mu^-)< \mathcal{B}(B_d\to\mu^+\mu^-)_{\rm SM}.
\ee
\item
$(A_3,B_3)$ oasis ({\it purple}) is   chosen when
\be
 S_{\mu^+\mu^-}^s>0, \qquad
\mathcal{B}(B_d\to\mu^+\mu^-)> \mathcal{B}(B_d\to\mu^+\mu^-)_{\rm SM}.
\ee
\end{itemize}
In Fig.~\ref{rainbow} we also display the case in which $(\tilde s_{13},\delta_1)$ belong to one of the two small oases $B_2$,
$B_4$, but we discuss this case  below in Section \ref{small-oases}.

\begin{figure}[!tb]
 \centering
\includegraphics[width = 0.45\textwidth]{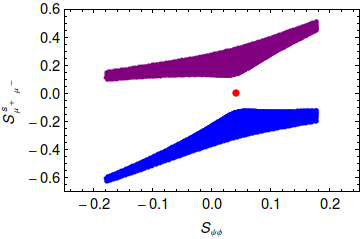}
\includegraphics[width = 0.45\textwidth]{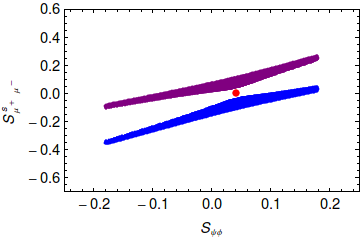}
\caption{\it $S_{\mu^+\mu^-}^s$ versus $S_{\psi\phi}$ for $M_{Z^\prime} = 1~$TeV (left) and $M_{Z^\prime} = 3~$TeV (right). Oasis
$A_1$: blue (lower region), $A_3$: purple (upper region). Red point: central
value of SM prediction.}\label{fig:SmuSphi}~\\[-2mm]\hrule
\end{figure}

In Fig.~\ref{fig:SmuSphi} we show  $ S_{\mu^+\mu^-}^s$  vs $S_{\psi\phi}$. Again the
requirement of suppression of $\Delta M_s$ requires  $ S_{\mu^+\mu^-}^s$  to
be non-zero. A {negative} value of  $ S_{\mu^+\mu^-}^s$  chooses scenario $A_1$ (blue),
while a {positive} one scenario $A_3$ (purple). Note that in both scenarios the sign
of $S_{\psi\phi}$ is not fixed yet but it will be fixed by invoking
 $\mathcal{B}(B_s\to\mu^+\mu^-)$ below. What is particularly remarkable
in this plot that for $M_{Z^\prime}=1\tev$, $|S_{\mu^+\mu^-}^s|$ can reach values as high as $0.5$ when  $|S_{\psi\phi}|\approx 0.2$.
Smaller values are found for
 $M_{Z^\prime}=3\tev$.

\begin{figure}[!tb]
 \centering
\includegraphics[width = 0.45\textwidth]{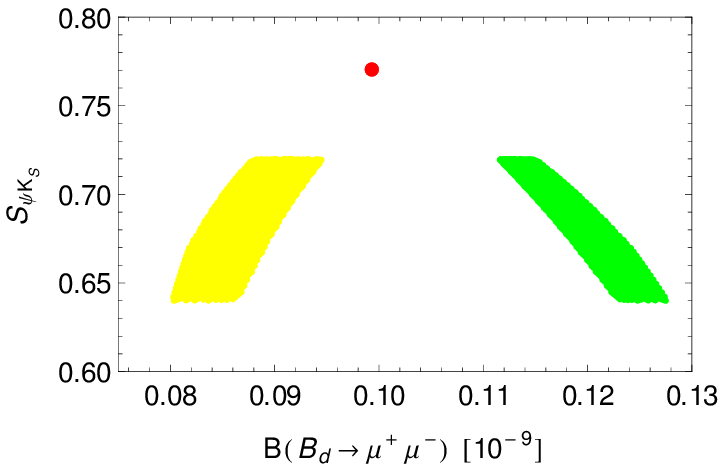}
\includegraphics[width = 0.45\textwidth]{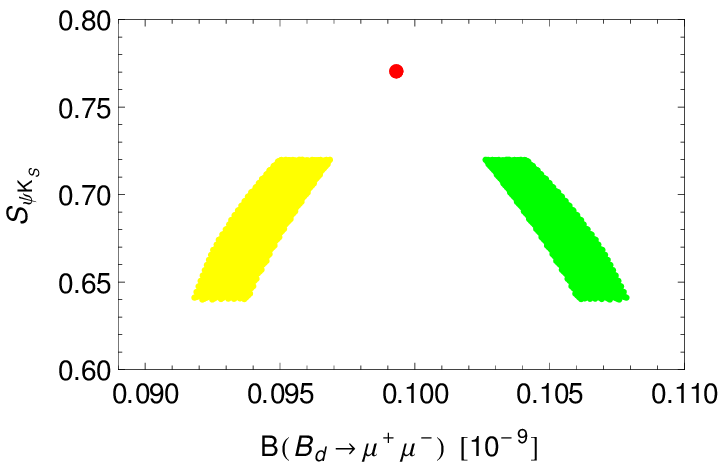}
\caption{\it  $S_{\psi K_S}$ versus $\mathcal{B}(B_d\to\mu^+\mu^-)$ for $M_{Z^\prime} = 1~$TeV (left) and $M_{Z^\prime} = 3~$TeV
(right). Oasis
$B_1$: yellow (left region), $B_3$: green (right region). Red point: central
value of SM prediction.}\label{fig:SKSBdmu}~\\[-2mm]\hrule
\end{figure}

In Fig.~\ref{fig:SKSBdmu} we show  $S_{\psi K_S}$ vs   $\mathcal{B}(B_d\to\mu^+\mu^-)$\footnote{ We should
remark that the central values for
$\mathcal{B}(B_{d}\to\mu^+\mu^-)^{\rm SM}=1.0\times 10^{-10}$ and
$\mathcal{B}(B_{s}\to\mu^+\mu^-)^{\rm SM}=3.1\times 10^{-9}$ shown in the plots
correspond to fixed CKM parameters chosen by us and differ from the ones
listed in~(\ref{LHCb2}) and (\ref{LHCb3}) but are fully consistent
with them.}.
The requirement on $S_{\psi K_S}$ and $\Delta M_d$ in (\ref{oases13})
forces
$\mathcal{B}(B_d\to\mu^+\mu^-)$ to differ from the SM value
but the sign of this
departure depends on the oasis considered. Here distinction is made
between $B_1$ (yellow) and $B_3$ (green) for which $\mathcal{B}(B_d\to\mu^+\mu^-)$ is
suppressed and enhanced with respect to the SM, respectively.
 These effects increase with decreasing $S_{\psi K_S}$.
These enhancements and suppressions amount to at most $\pm 25\%$ and
$\pm 10\%$ for   $M_{Z^\prime}=1\tev$ and  $M_{Z^\prime}=3\tev$, respectively.

The fact that  $ S_{\mu^+\mu^-}^s$  and $\mathcal{B}(B_d\to\mu^+\mu^-)$ are very
powerful in identifying the optimal oasis can be understood as follows.
 $S_{\mu^+\mu^-}^s$  is governed by the phase of the function $Y(B_s)$ that
originates in the $Z'$ contribution. It can distinguish between $A_1$ and
$A_3$ oasis because the new phase $\delta_2$  in these two oases differs by $180^\circ$
and consequently $\sin\delta_2$ relevant for this asymmetry differs by sign
in these two oases. Calculating the imaginary part of $Y(B_s)$ in
(\ref{YBq}) and taking into account that it is $\Delta_L^{sb}$ and not
$\Delta_L^{bs}$ that enters  $Y(B_s)$ one can convince oneself about the
definite sign of  $ S_{\mu^+\mu^-}^s$ in $A_1$ and $A_3$ oases as stated
above.

As far as  $\mathcal{B}(B_d\to\mu^+\mu^-)$
is concerned, it is correlated with $S_{\psi K_S}$ that is already
well determined. Therefore, the range of $\delta_1$ cannot be large.
 $\mathcal{B}(B_d\to\mu^+\mu^-)$  can then distinguish between $B_1$ and
$B_3$ oases because $\cos\delta_1$ differs by sign in these two oases.
We find then destructive interference of $Z'$ contribution with the
SM contribution in oasis $B_1$ and constructive one in oasis $B_3$ implying
the results quoted above.

It should be noted that on the basis of $\Delta F=2$ processes such a
distinction between these oases cannot be made because the
relevant amplitudes are governed by $2\delta_1$ and $2\delta_2$ which
differ by $360^\circ$.

\begin{figure}[!tb]
 \centering
\includegraphics[width = 0.45\textwidth]{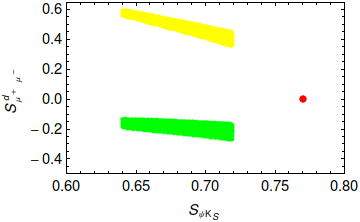}
\includegraphics[width = 0.45\textwidth]{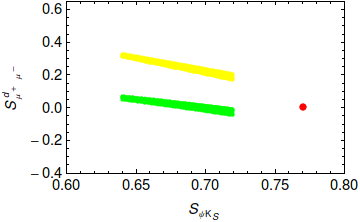}
\caption{\it $S_{\mu^+\mu^-}^d$ versus $S_{\psi K_S}$ for $M_{Z^\prime} = 1~$TeV (left) and $M_{Z^\prime} = 3~$TeV (right). Oasis
$B_1$: yellow (upper region), $B_3$: green (lower region). Red point: central
value of SM prediction.}\label{fig:SdmuSKS}~\\[-2mm]\hrule
\end{figure}

\begin{figure}[!tb]
 \centering
\includegraphics[width = 0.45\textwidth]{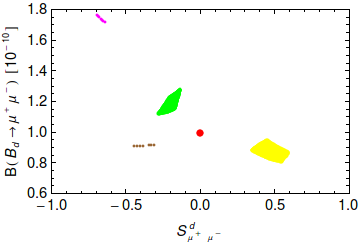}
\includegraphics[width = 0.45\textwidth]{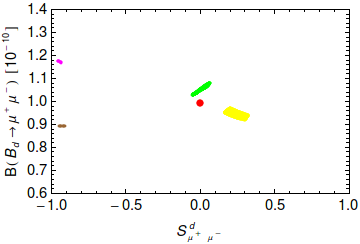}
\caption{\it $\mathcal{B}(B_d\to \mu^+ \mu^-)$ versus $S_{\mu^+\mu^-}^d$ for $M_{Z^\prime} = 1~$TeV (left)
and $M_{Z^\prime} = 3~$TeV (right). Oasis
$B_1$: yellow, $B_3$: green, $B_2$: magenta, $B_4$: brown. Red point: central
value of SM prediction.}\label{fig:BdmuSdmu}~\\[-2mm]\hrule
\end{figure}

We end this step by briefly discussing  $S_{\mu^+\mu^-}^d$ which as
seen in Table~\ref{s13oases} and  Fig.~\ref{fig:SdmuSKS} can also help
by means of its  sign to distinguish between different oases, {\bf although
it becomes difficult for $M_{Z^\prime} = 3~$TeV}. We note also
that the sign of  $S_{\mu^+\mu^-}^d$ is always opposite to the  sign
of the shift in the corresponding branching ratio, which can easily
be understood by inspecting the ranges of $\delta_1$. Moreover,
the predictions for $S_{\mu^+\mu^-}^d$ are rather precise. For instance
at $M_{Z^\prime} = 1~$TeV in oases $B_1$ and $B_3$, the allowed ranges
for  $S_{\mu^+\mu^-}^d$  amount to  $[0.27,0.44]$ and  $[-0.38,-0.24]$,
respectively. While being aware of the fact that this CP asymmetry will
be very hard to measure, this example shows that it can provide in
principle a very stringent test of the model in question and other
extensions of the SM.  We show this in Fig.~\ref{fig:BdmuSdmu}, where we have
also shown the results for small oases that we will discuss at the end
of this section.

{\bf Step 2:}

\begin{figure}[!tb]
 \centering
\includegraphics[width = 0.45\textwidth]{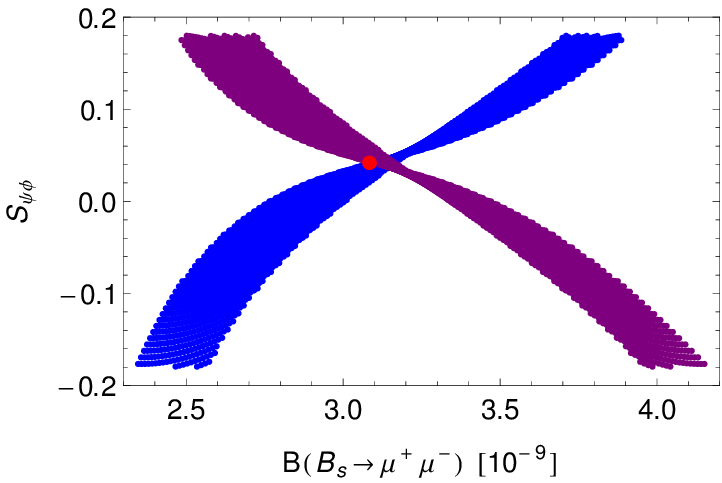}
\includegraphics[width = 0.45\textwidth]{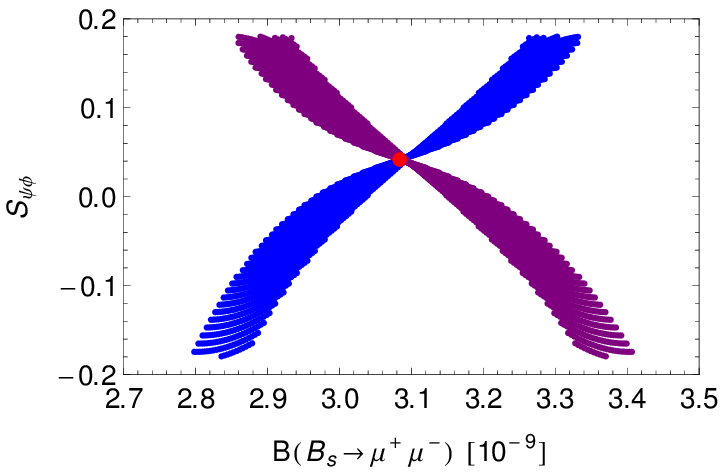}
\caption{\it  $S_{\psi\phi}$ versus $\mathcal{B}(B_s\to\mu^+\mu^-)$ for $M_{Z^\prime} = 1~$TeV (left) and $M_{Z^\prime} = 3~$TeV
(right).  Oasis
$A_1$: blue (down left to up right), $A_3$: purple (up left to down right). Red point: central
value of SM prediction.}\label{fig:SphiBsmu}~\\[-2mm]\hrule
\end{figure}

The reason why  $\mathcal{B}(B_s\to\mu^+\mu^-)$ cannot be presently
as powerful as  $\mathcal{B}(B_d\to\mu^+\mu^-)$ in the search for
oases is the significant experimental error  on $S_{\psi\phi}$ with which
this branching ratio is correlated. However, inspecting this correlation
in a given oasis
 constitutes
an important test of the model. We show this in Fig.~\ref{fig:SphiBsmu}. While in the
oasis $A_1$ (blue)
$S_{\psi\phi}$ increases (decreases) uniquely with increasing (decreasing)
$\mathcal{B}(B_s\to\mu^+\mu^-)$, in the oasis $A_3$ (purple), the increase of
$S_{\psi\phi}$ implies uniquely a decrease of $\mathcal{B}(B_s\to\mu^+\mu^-)$.
Therefore, while $\mathcal{B}(B_s\to\mu^+\mu^-)$ alone cannot uniquely
determine the optimal oasis, it can do it in collaboration with $S_{\psi\phi}$.
Therefore, finding both these observables above or below their SM expectations,
would select the oasis $A_1$, while finding one of them enhanced and the
other suppressed (opposite sign in the case of $S_{\psi\phi}$) would
select $A_3$ as the optimal oasis. We indicate this pattern in
Table~\ref{s23oases}.

If the favoured oasis will be found to  differ from the one found in step 1,
the
 $\overline{331}$ model will be in trouble. In terms of observables this
correlation between Step 1 and Step 2 could be as follows. Let us
assume that  $\mathcal{B}(B_s\to\mu^+\mu^-)$ will be found below its SM
value. Then the measurement of  $S_{\psi\phi}$ will uniquely tell us whether
$A_1$ or $A_3$ is the optimal scenario and consequently as seen in  Fig.~\ref{fig:SmuSphi} and
Table~\ref{s23oases}
we will be able to predict the sign of $S_{\mu^+\mu^-}^s$. Moreover, in the
case of $S_{\psi\phi}$ sufficiently different from zero, we will be able
to determine not only the sign but also the magnitude of  $S_{\mu^+\mu^-}^s$.
This discussion shows that we have a triple correlation
 $S_{\mu^+\mu^-}^s-S_{\psi\phi}-\mathcal{B}(B_s\to\mu^+\mu^-)$ in the
$\model$ model: once the sign of  $S_{\mu^+\mu^-}^s$ is known a unique
correlation $S_{\psi\phi}-\mathcal{B}(B_s\to\mu^+\mu^-)$ is found. If in
addition one of these three observables is precisely known the other
two can be strongly constrained.

Finally, we also note that  $\mathcal{B}(B_s\to\mu^+\mu^-)$ can for
 $M_{Z^\prime}=1\tev$ and  $M_{Z^\prime}=3\tev$
deviate from the SM value by
$\pm 30\%$ and $\pm 10\%$, respectively.

{\bf Step 3:}

\begin{figure}[!tb]
 \centering
  \includegraphics[width = 0.48\textwidth]{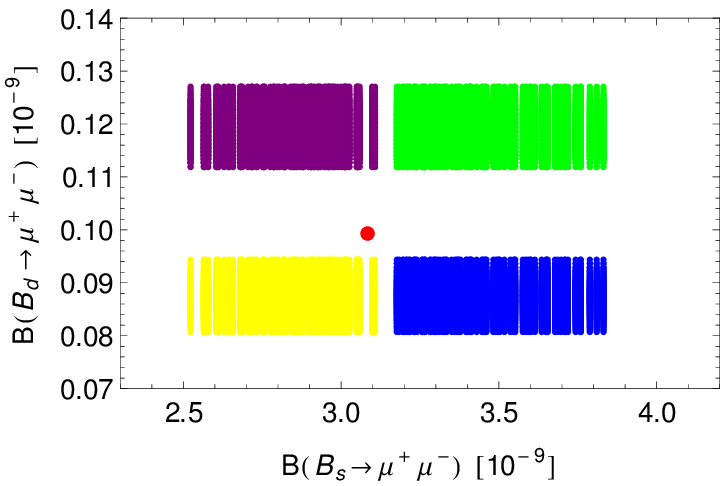}
  \includegraphics[width = 0.48\textwidth]{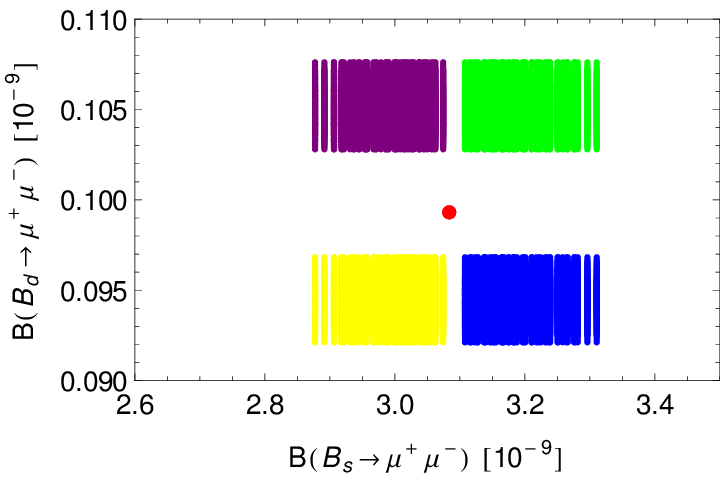}

 \includegraphics[width = 0.48\textwidth]{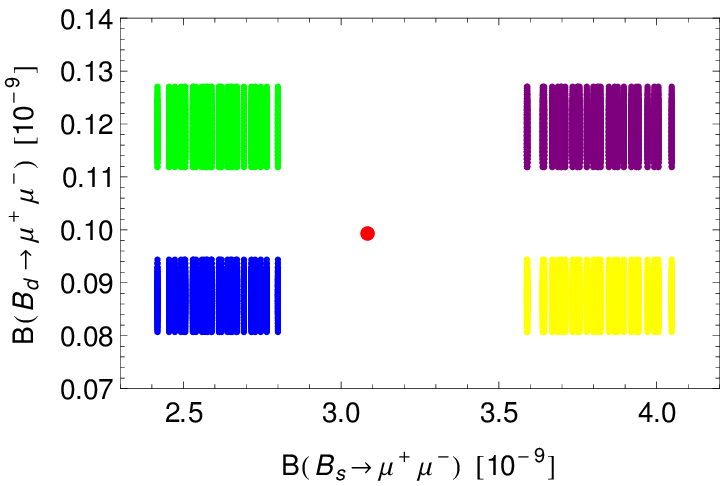}
 \includegraphics[width = 0.48\textwidth]{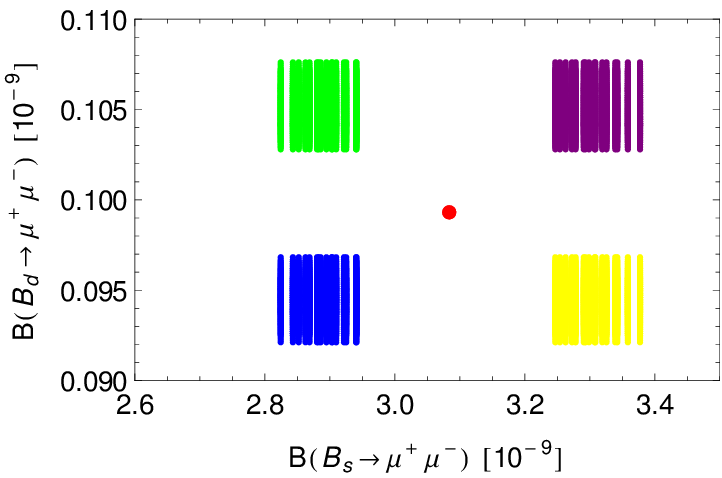}

\caption{\it $\mathcal{B}(B_d\to\mu^+\mu^-)$ versus $\mathcal{B}(B_s\to\mu^+\mu^-)$ for $M_{Z^\prime} = 1~$TeV (left) and $M_{Z^\prime} =
3~$TeV
(right).
In the upper panel: $S_{\psi\phi}>0.05 $, in the lower:  $S_{\psi\phi}<-0.05 $.
 Oases: $(A_1,B_1)$: blue,  $(A_1,B_3)$: green,
$(A_3,B_1)$: yellow,  $(A_3,B_3)$: purple. Red point: central
value of SM prediction. }\label{fig:BdBsmu-pos-neg}~\\[-2mm]\hrule
\end{figure}

In Fig.~\ref{fig:BdBsmu-pos-neg} we show
$\mathcal{B}(B_d\to\mu^+\mu^-)$ versus $\mathcal{B}(B_s\to\mu^+\mu^-)$
for $S_{\psi\phi}>0.05 $ (upper-panel) and $S_{\psi\phi}<-0.05 $ (lower-panel).
We observe four different areas corresponding to the
oases in question but the positions of these oases are different in these two plots.

%\begin{figure}[!tb]
% \centering
%\includegraphics[width = 0.45\textwidth]{pBdBsmu1.png}
%\includegraphics[width = 0.45\textwidth]{pBdBsmu3.png}
%\caption{\it $\mathcal{B}(B_d\to\mu^+\mu^-)$ versus $\mathcal{B}(B_s\to\mu^+\mu^-)$ for $M_{Z^\prime} = 1~$TeV (left) %and
%$M_{Z^\prime} = 3~$TeV
%(right). ($A_1$,$B_1$): blue,
%$(A_1,B_3)$: green,
%$(A_3,B_1)$: yellow, $(A_3,B_3)$: purple. Red point: central
%value of SM prediction.}\label{fig:BdBsmu}~\\[-2mm]\hrule
%\end{figure}

 {\bf Step 4:}

\begin{figure}[!tb]
 \centering
\includegraphics[width = 0.45\textwidth]{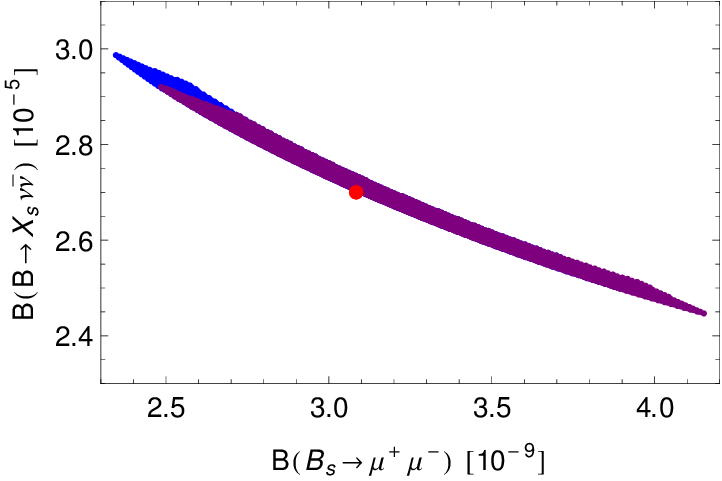}
\includegraphics[width = 0.45\textwidth]{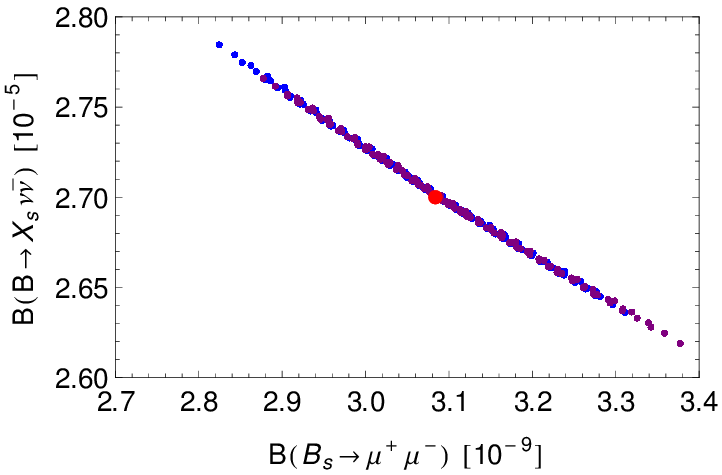}
\caption{ \it $\mathcal{B}(B\to X_s \nu\bar\nu)$ versus
$\mathcal{B}(B_s\to\mu^+\mu^-)$ for $M_{Z^\prime} = 1~$TeV (left) and
$M_{Z^\prime} = 3~$TeV
(right). The two oases $A_1$ (blue) and $A_3$ (purple) overlap in most parts. Red point: central
value of SM prediction.}\label{fig:BXsnuBsmu}~\\[-2mm]\hrule
\end{figure}

In Fig.~\ref{fig:BXsnuBsmu} we show
$\mathcal{B}(B\to X_s \nu\bar\nu)$ vs
$\mathcal{B}(B_s\to\mu^+\mu^-)$. This correlation is valid in any oasis due
to the general relation (\ref{master3R}) present in the $\model$ model.
As expected, NP effects are significantly larger in $\mathcal{B}(B_s\to\mu^+\mu^-)$ than $\mathcal{B}(B\to X_s \nu\bar\nu)$.

{\bf Step 5:}

\begin{figure}[!tb]
 \centering
\includegraphics[width = 0.45\textwidth]{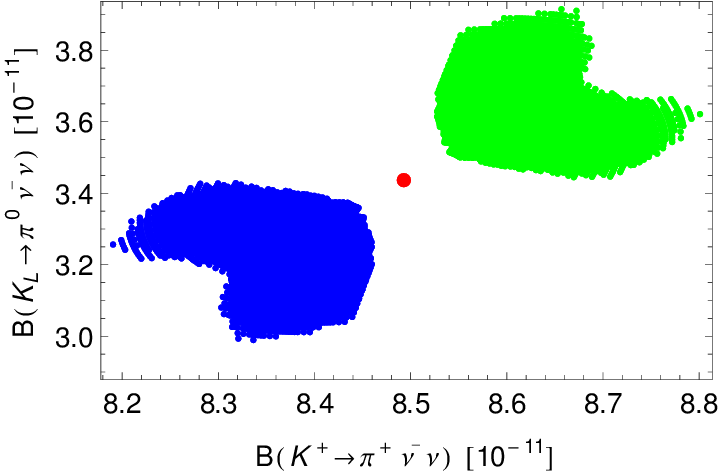}
\includegraphics[width = 0.45\textwidth]{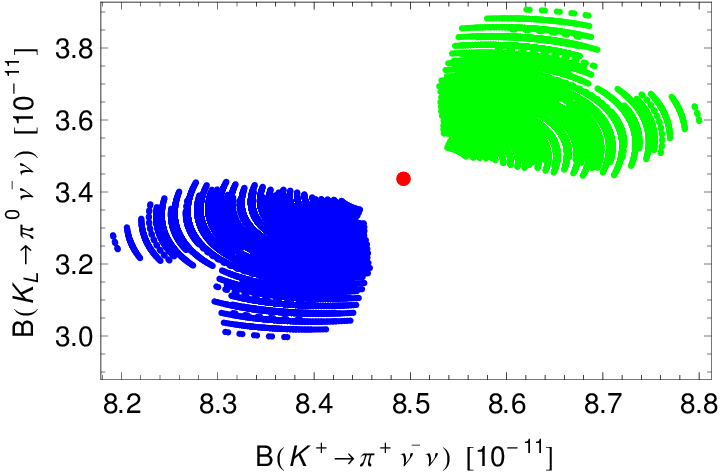}
\caption{\it  $\mathcal{B}(\klpn)$ versus  $\mathcal{B}(\kpn)$ for $M_{Z^\prime} = 1~$TeV (left) and
$M_{Z^\prime} = 3~$TeV
(right). ($A_1$,$B_1$) and $(A_3,B_3)$: blue,
$(A_1,B_3)$ and $(A_3,B_1)$: green. Red point: central
value of SM prediction.}\label{fig:KLKp}~\\[-2mm]\hrule
\end{figure}

We have calculated $\mathcal{B}(\klpn)$ vs  $\mathcal{B}(\kpn)$ (see Fig.~\ref{fig:KLKp}). It turns out
that in the oases $(A_1,B_1)$ and  $(A_3,B_3)$ (blue) both branching ratios
are suppressed with respect to the SM values but they are both enhanced
in scenarios $(A_1,B_3)$ and  $(A_3,B_1)$ (green). Combined with the remaining steps
the measurements of these branching ratios could in principle
 contstitute an important
test of the model in question. However, unfortunately, the deviations from
SM expectations are at most $5\%$ at the level of the branching ratios
so that these correlations cannot be tested in a foreseeable future.
On the other hand, finding experimentally
both branching ratios significantly different
from SM expectations would put the  $\model$ model in trouble. As
expected the NP effects basically do not depend on the mass of $Z'$.
Also NP contributions to
$\mathcal{B}(K_L\to\mu^+\mu^-)$, even if by a factor of two larger than
in $\kpn$, are
well within long distance uncertainties involved in this branching ratio.

\begin{figure}[!tb]
 \centering

\includegraphics[width = 0.48\textwidth]{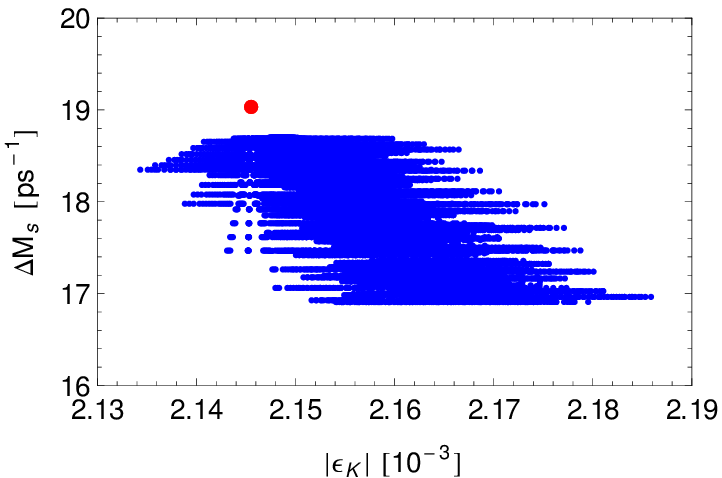}
\includegraphics[width = 0.48\textwidth]{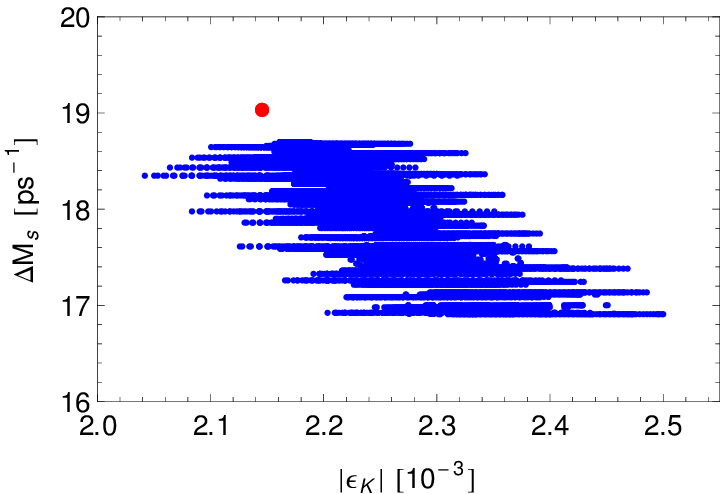}

\includegraphics[width = 0.48\textwidth]{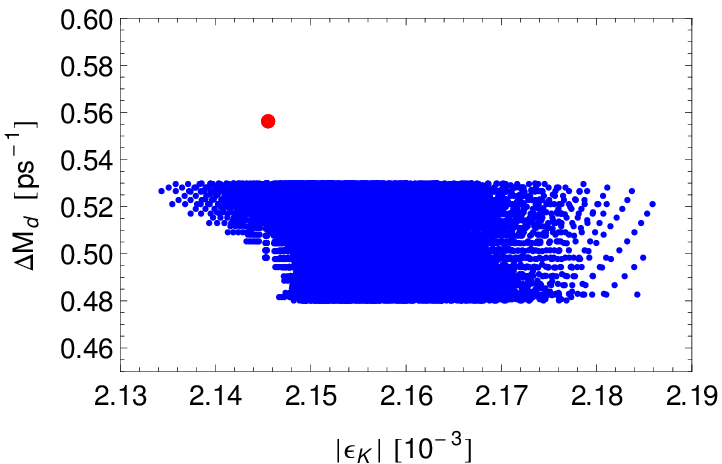}
\includegraphics[width = 0.48\textwidth]{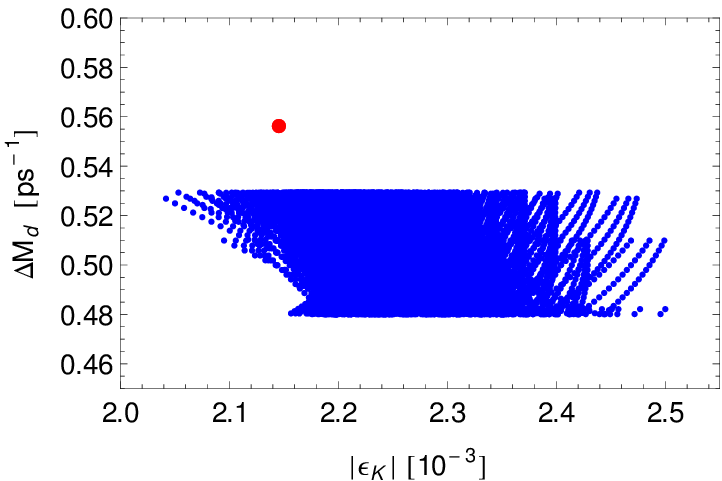}
\caption{\it  $\Delta M_s$ and $\Delta M_d$ versus $|\varepsilon_K|$ for $M_{Z^\prime} = 1~$TeV (left) and
$M_{Z^\prime} = 3~$TeV
(right). All oases ($A_1$,$B_1$), $(A_1,B_3)$, $(A_3,B_1)$, $(A_3,B_3)$ overlap. Red point:
central
value of SM prediction.}\label{fig:DMepsK}~\\[-2mm]\hrule
\end{figure}

In Fig.~\ref{fig:DMepsK}  we plot $\Delta M_s$ and $\Delta M_d$ as functions of
$|\varepsilon_K|$ in order to see how the correlation between these
three observables is modified within the $\overline{331}$ model.
These plots show that this model does not suffer from
$\Delta M_{s,d}$-$|\varepsilon_K|$ tension that has been identified
in CMFV models \cite{Buras:2012ts}.    Indeed, imposing, as we do, that $\Delta M_{s,d}$ satisfy the experimental values,
$\varepsilon_K$ is within a few $\%$
from the experimental central value.  As expected the room for NP
contributions in $\varepsilon_K$ is larger for larger $Z'$ mass.

\begin{figure}[!tb]
 \centering
\includegraphics[width = 0.48\textwidth]{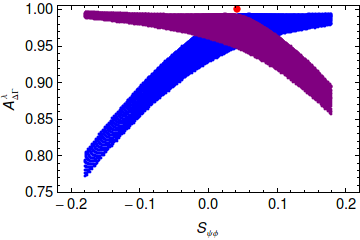}
\includegraphics[width = 0.48\textwidth]{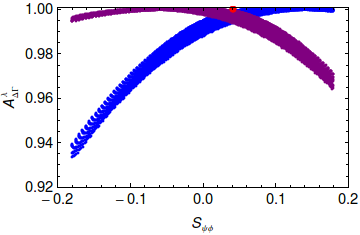}
\caption{\it  $\mathcal{A}^\lambda_{\Delta\Gamma}$ versus $S_{\psi\phi}$ for $M_{Z^\prime} = 1~$TeV (left) and
$M_{Z^\prime} = 3~$TeV
(right). Oasis
$A_1$: blue, $A_3$: purple.
}\label{fig:ASphi}~\\[-2mm]\hrule
\end{figure}

In Fig.~\ref{fig:ASphi} we plot $\mathcal{A}^\lambda_{\Delta\Gamma}$ vs  $S_{\psi\phi}$.
Only for  $M_{Z'}=1\tev$  and $S_{\psi\phi}$ significantly different from
zero, does $\mathcal{A}^\lambda_{\Delta\Gamma}$ differ significantly from
unity.

\subsection{A Brief Look at the Small Oases}\label{small-oases}
We will now briefly discuss the small oases. It turns out that they
correspond to NP contributions to $M_{12}^d$ and $M_{12}^s$ that are roughly
by a factor of two larger than their SM values but carry opposite signs. As
$\Delta M_{d,s}$ involve the absolute values of the mixing amplitudes these
oases cannot be eliminated on the basis of them. The situation is analogous
to the sign of the Wilson coefficient $C_{7\gamma}$ that cannot be fixed
by $B\to X_s\gamma$ decay. To this end other observables have to be invoked and
this is what we will do now in the case at hand.

We anticipated that among the 16 possible oases obtained when the two pairs of parameters $({\tilde s}_{23},\delta_2)$ and
$({\tilde s}_{13},\delta_1)$ belong to the regions denoted by $A_i$ and $B_i$ ($i=1,\dots 4$) in Tables \ref{s23oases} and
\ref{s13oases}, respectively, it possible to rule out the 8 ones obtained when the pair $({\tilde s}_{23},\delta_2)$
belongs either to $A_2$ or to $A_4$. This is a consequence of the requirement that $\varepsilon_K$ does not differ more than $5\%$
with respect to the experimental central value reported in Table \ref{tab:input},  an  uncertainty larger than the experimental
range reported in the same table. This request is violated when $(\tilde s_{23},\delta_2)$ vary in $A_2$ or in $A_4$, allowing us
to half the number of possible oases  provided the theoretical and parametric uncertainties in $\varepsilon_K$ could be lowered
down to $5\%$. This is clearly not the case at present,
primarly due to $\vcb^4$ dependence present in $\varepsilon_K$ but could
become realistic in the second half of this decade.

In the previous subsection we discussed the case of the {\it big} oases, i.e. that in which the parameters belong to $A_i$ and $B_i$ with
$i=1,3$. There are still other 4 possible regions obtained when the parameters $({\tilde s}_{13},\delta_1)$ vary in $B_2$ or $B_4$. The
resulting combinations of oases are therefore:
\be
(A_1,B_2), \hskip 1 cm (A_1,B_4), \hskip 1 cm (A_3,B_2), \hskip 1 cm (A_3,B_4). \hskip 1 cm
\label{small-o}
\ee
At present we cannot rule out these regions, but from Fig.~\ref{rainbow}
and \ref{fig:BdmuSdmu}
we can observe that
correlation between $B(B_d \to \mu^+ \mu^-)$ and $S_{\mu^+ \mu^-}^{s,d}$ can help distinguishing among these four as well as for
the {\it big} ones.
In particular:
\begin{itemize}
\item
In  $(A_1,B_4)$ ({\it gray})
\be
 S_{\mu^+\mu^-}^s<0, \qquad
\mathcal{B}(B_d\to\mu^+\mu^-)< \mathcal{B}(B_d\to\mu^+\mu^-)_{\rm SM},\qquad  S_{\mu^+\mu^-}^d<0.
\ee
\item
In  $(A_1,B_2)$ ({\it red})
\be
 S_{\mu^+\mu^-}^s<0, \qquad
\mathcal{B}(B_d\to\mu^+\mu^-)> \mathcal{B}(B_d\to\mu^+\mu^-)_{\rm SM},\qquad  S_{\mu^+\mu^-}^d>0.
\ee
\item
In  $(A_3,B_4)$ ({\it brown})
\be
 S_{\mu^+\mu^-}^s>0, \qquad
\mathcal{B}(B_d\to\mu^+\mu^-)< \mathcal{B}(B_d\to\mu^+\mu^-)_{\rm SM},\qquad  S_{\mu^+\mu^-}^d<0.
\ee
\item
In  $(A_3,B_2)$ ({\it magenta})
\be
 S_{\mu^+\mu^-}^s>0, \qquad
\mathcal{B}(B_d\to\mu^+\mu^-)> \mathcal{B}(B_d\to\mu^+\mu^-)_{\rm SM},\qquad  S_{\mu^+\mu^-}^d>0.
\ee
\end{itemize}

 However, compared to the big oases, these small oases are special since they give sharp predictions for the branching ratio of
$B_{d} \to \mu^+ \mu^-$ decay and this may one day help to rule them out or confirm.
We find:  $B(B_{d} \to \mu^+ \mu^-)=(1.74 \pm 0.02) \times 10^{-10}$ in $(A_1,B_2)$ and $(A_3,B_2)$ while $B(B_{d} \to \mu^+ \mu^-)=(0.911
\pm 0.005) \times 10^{-10}$ in $(A_1,B_4)$ and $(A_3,B_4)$, for $M_{Z^\prime}=1$ TeV, and $B(B_{d} \to \mu^+ \mu^-)=(1.172 \pm 0.005) \times
10^{-10}$ in $(A_1,B_2)$ and $(A_3,B_2)$ while $B(B_{d} \to \mu^+ \mu^-)=(0.892 \pm 0.001) \times 10^{-10}$ in $(A_1,B_4)$ and $(A_3,B_4)$,
for $M_{Z^\prime}=3$ TeV.
It can also be observed that, in the 3 TeV case there is no more overlap between the small and the big oases.

Finally, we note that the asymmetry $S_{\mu^+\mu^-}^d$ is very precisely
predicted in the small oases.  For instance
at $M_{Z^\prime} = 1~$TeV in oases $B_2$ and $B_4$, we find for
$S_{\mu^+ \mu^-}^d=-0.66 \pm 0.03 $ and  $ S_{\mu^+\mu^-}^d=-0.38 \pm 0.08$,
respectively.

\subsection{Deviations from CMFV Relations.}
The calculation of the various observables performed in the previous subsection  within the $\overline{331}$ model allows us to
discuss the departure from CMFV relations, i.e. the deviation of the functions in Eq.~(\ref{r1}) from 1.
We find that such functions vary in the following ranges when the parameters belong to the {\it big} oases:
\begin{itemize}
\item 1 TeV
\bea
r(\Delta M) &\in& [0.88,1.07]  \\
r(\nu \bar \nu) &\in& [0.84, 1.19]
\\
r(\mu^+ \mu^-) &\in& [0.60,1.62] \eea
\item 3 TeV
\bea
r(\Delta M) &\in& [0.88,1.07]  \\
r(\nu \bar \nu) &\in& [0.94, 1.06]
\\
r(\mu^+ \mu^-) &\in& [0.85,1.18] \,.\eea
\end{itemize}
We observe that deviation of $r(\Delta M)$ from 1 is approximately $10\%$, irrespective of the value of $M_{Z^\prime}$.
$r(\nu \bar \nu)$ can deviate by almost $20\%$ from 1 in the case $M_{Z^\prime}=1$ TeV, while the deviation reduces to $6\%$ for
$M_{Z^\prime}=3$ TeV.
$r(\mu^+ \mu^-)$ is by far the function that mostly deviates from CMFV prediction: its values can differ from 1 by $60\%$ for
$M_{Z^\prime}=1$ TeV, and almost $20\%$ for $M_{Z^\prime}=3$ TeV.

\begin{figure}[!tb]
 \centering
\includegraphics[width = 0.48\textwidth]{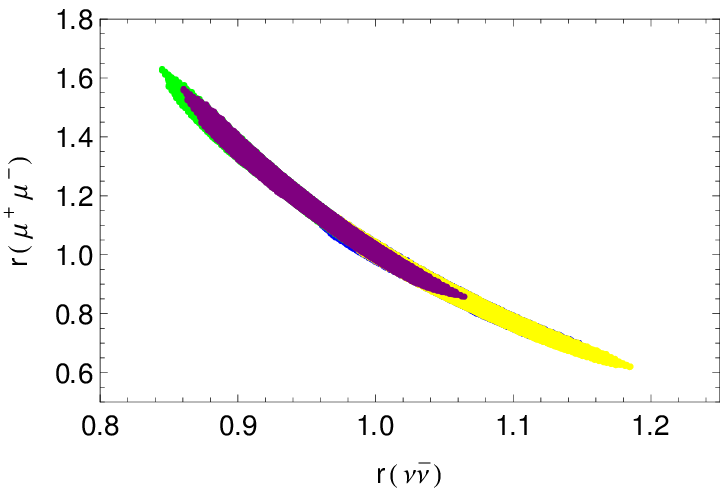}
\includegraphics[width = 0.48\textwidth]{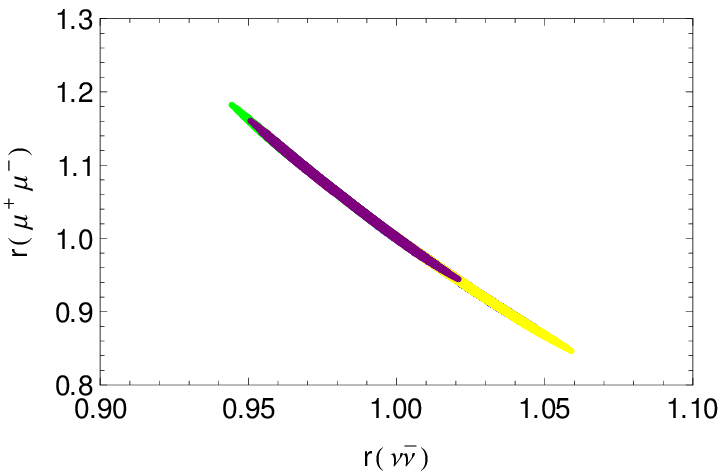}

\caption{\it Left: $M_{Z^\prime} = 1~$TeV, right: $M_{Z^\prime} = 3~$TeV. Oases: $(A_1,B_1)$: blue,  $(A_1,B_3)$: green,
$(A_3,B_1)$: yellow,  $(A_3,B_3)$: purple. }\label{fig:rplots}~\\[-2mm]\hrule
\end{figure}

The departure of these functions from 1 is anti-correlated in the case of $r(\mu^+ \mu^-)$ and $r(\nu \bar \nu)$, as can be seen
from Fig. \ref{fig:rplots} from which we argue that when $r(\nu \bar \nu)>1$ $(<1)$ it  implies $r(\mu^+ \mu^-)<1$ $(>1)$.
The largest possible values of $r(\nu \bar \nu)$ are obtained when $({\tilde s}_{13},\delta_1) \in B_1$, while
$r(\mu^+ \mu^-)$ is larger when $({\tilde s}_{13},\delta_1) \in B_3$.

The case in which the parameters belong to the {\it small} oases is not displayed in Fig. \ref{fig:rplots}.
However, we mention that in this case, for $M_{Z^\prime}=1$ TeV, the values of $r(\nu \bar \nu)$ can be larger, while $r(\mu^+ \mu^-)$ can
reach even the value $r(\mu^+ \mu^-)=2.2$ in the region $(A_1,B_2)$. For $M_{Z^\prime}=3$ TeV the contribution of the {\it small} oases is
almost indistinguishable from that of the {\it big} ones, except that $r(\mu^+ \mu^-)$ and $r(\nu \bar \nu)$ can be as large as 1.3 and 1.1,
respectively.

\begin{figure}[!tb]
 \centering
\includegraphics[width = 0.48\textwidth]{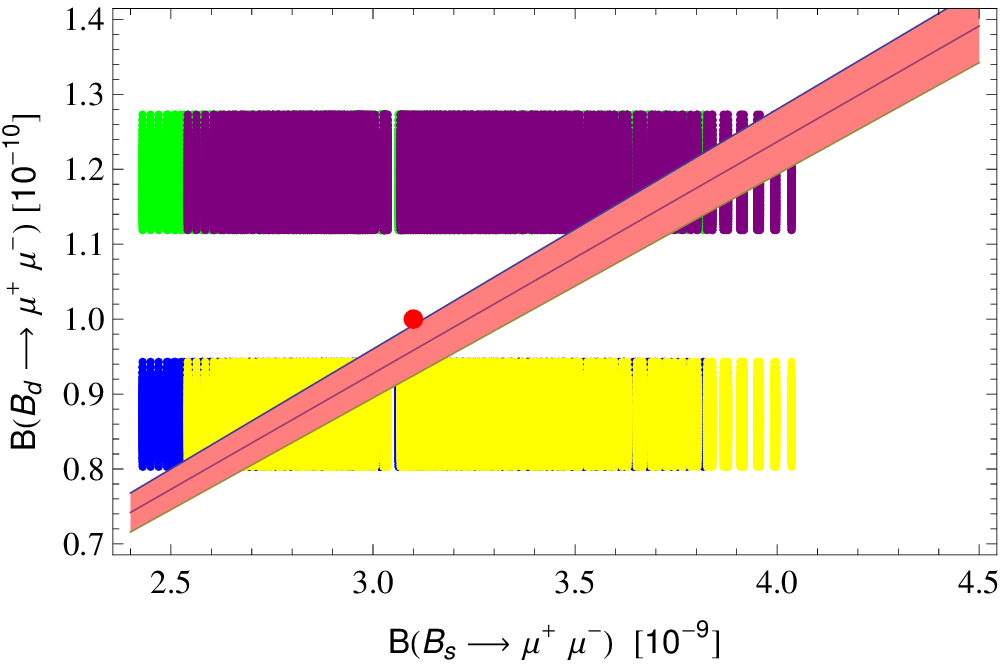}
\includegraphics[width = 0.48\textwidth]{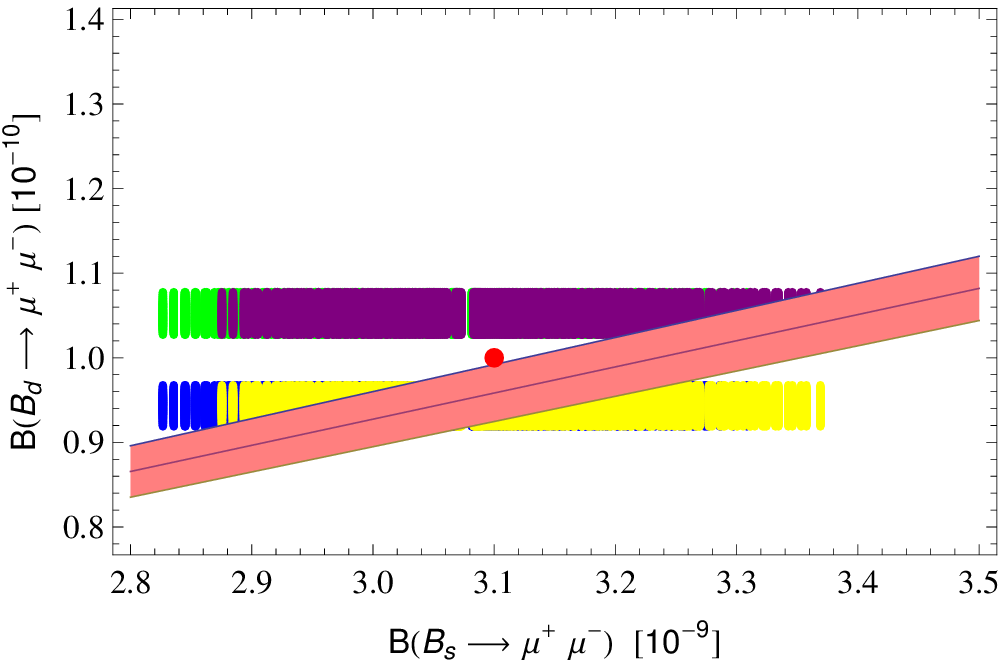}

\caption{\it Left: $M_{Z^\prime} = 1~$TeV, right: $M_{Z^\prime} = 3~$TeV. Oases: $(A_1,B_1)$: blue,  $(A_1,B_3)$: green,
$(A_3,B_1)$: yellow,  $(A_3,B_3)$: purple. Red point:
central
value of SM prediction. Pink band: CMFV relation from Eq. (\ref{R1}) with $r=1$. }\label{fig:pBdBsmu-cmfv}~\\[-2mm]\hrule
\end{figure}

The violation of CMFV relations in the $\overline{331}$ model can be discussed also comparing the actual results for the branching
ratios of $B_{d,s} \to \mu^+ \mu^-$ to what one would obtain from Eq.~(\ref{R1}) putting $r=1$. This is done in
Fig.~\ref{fig:pBdBsmu-cmfv}, where we plot $B(B_d \to \mu^+ \mu^-)$ vs $B(B_s \to \mu^+ \mu^-)$ and superimpose the band
corresponding to the relation (\ref{R1}) with $r=1$ and including the errors on the experimental determinations of $\Delta M_s$,
$\Delta M_d$, $\tau(B_s)$, $\tau(B_d)$. On the other hand, we used ${\hat B_s}/{\hat B_d}=1.05 \pm 0.03$, where the central
value coincides with the lattice determination reported in Table \ref{tab:input}, while we reduced the size of the uncertainty. An
improvement in this direction would indeed substantially help understanding if and how much CMFV relations are violated in the $\model$
model.

\subsection{What if $Z'$ is beyond the LHC Reach?}
Our numerical analysis assumed that $Z'$ will be found at the LHC,
that is $M_{Z^\prime}$ is somewhere in the range $1-3~$TeV. In this range
the B-physics constraints imply rather small values for $\tilde s_{13}$ and
$\tilde s_{23}$ and consequently for their product that enters $\varepsilon_K$.
NP effects in  $\varepsilon_K$ are then small and $\vub$ has to be sufficiently
large in order that $\varepsilon_K$ is consistent with the data.

But what if $Z'$ is beyond the LHC reach and its mass is as large as $10$~TeV?
We have analyzed this case with the following result:
\begin{itemize}
\item
The increase of  $M_{Z^\prime}$ to  $10$~TeV still allows to satisfy the
constraints in (\ref{oases23}) and  (\ref{oases13}) provided the values
of $\tilde s_{13}$ and $\tilde s_{23}$ are appropriately increased (see below).
\item
In turn, as already mentioned previously, NP effects in $\varepsilon_K$
can be larger so that for $M_{Z^\prime}=10$~TeV even the exclusive value for
$\vub$ allows to reproduce the data for $\varepsilon_K$, which is not possible
for the values of $M_{Z^\prime}$ in the few TeV range.
\item
NP effects in rare B-decays are strongly suppressed relatively to the effects
presented by us for $ 1\tev\le M_{Z'}\le 3\tev$ so that here similarly to rare $K$ decays everything is SM-like.
\end{itemize}

Thus the main effect of NP in this case in the $\model$ model is to remove
the anomalies in $\Delta F=2$ observables in the presence of exclusive value
of $\vub$ without basically any impact on rare $B$ and rare $K$ decays. This
rather uninteresting vision could become reality if eventually
$\mathcal{B}(B^+\to\tau^+\nu_\tau)$ will turn out to be SM-like, favouring
exclusive value of $\vub$.

We have investigated this case in more detail by setting
$\vub=3.1\cdot 10^{-3}$, $M_{Z^\prime}=10$~TeV and imposing additional constraint:
\be\label{C3}
0.75\le \frac{\Delta M_K}{(\Delta M_K)_{\rm SM}}\le 1.25,\qquad
2.0\times 10^{-3}\le |\varepsilon_K|\le 2.5 \times 10^{-3}.
\ee
This additional constraint is necessary as now NP effects in $\varepsilon_K$
for $\tilde s_{23}$ and $\tilde s_{13}$ allowed by $B$-physics $\Delta F=2$
observables in the presence of  $M_{Z^\prime}=10$~TeV can be large.
We  find then approximately:
\be
0.05\le \tilde s_{23}\le 0.12, \qquad 0.016\le \tilde s_{13}\le 0.030
\ee
and the ranges for $\delta_{1,2}$ slightly shifted to lower values. The
reason for some departure from the expected  $M_{Z^\prime}$ dependence
in the range discussed until now is the additional constraint in (\ref{C3})
which is immaterial for  $M_{Z^\prime}$ of a few TeV.

The fact that maximal values of $\varepsilon_K$ increase with  $M_{Z^\prime}$
when only B-physics constraints are taken into account is related to
the increase of $\tilde s_{23}$ and $\tilde s_{13}$ as already explained.
But when the $\Delta^{sd}(Z')$ reaches its maximal value, further
increase of  $M_{Z^\prime}$  will also decrease NP contributions to
 $\varepsilon_K$  so that for very large $M_{Z^\prime}$ SM results are
obtained as required by decoupling of NP.  However, this happens only for
$M_{Z^\prime}\ge 3000$~TeV. This result is just a confirmation of the known
fact that if the FCNC  $Z'$ couplings to quarks are $\ord(1)$,
$\varepsilon_K$ puts a very strong constraint on the scale of NP.

\subsection{Sensitivity to Non-Perturbative Parameters}
In the present paper we have used the most recent lattice input to find out
that the SM values for $\Delta M_s$ and $\Delta M_d$ were visibly
above the data, when the hadronic and CKM uncertainties were reduced
down to $\pm 5\%$. Here we would like to emphasize that some of the correlations  presented by us would change if $\Delta M_s$ and $\Delta
M_d$  would
both be found below the data. In order to illustrate this point we
have reduced the values of $\hat B_{B_q}$ by $20\%$ so that now
\be
(\Delta M_s)_{\rm SM}=15.2~{\rm ps}^{-1}, \quad (\Delta M_d)_{\rm SM}=0.45~{\rm ps}^{-1}.
\ee
This result would be welcomed by CMFV models in which $\Delta M_s$ and $\Delta M_d$ can only be enhanced keeping their ratio fixed. However,
the enhanced
value of $S_{\psi K_S}$ in the large $\vub$ scenario still would be problematic
for these models.

The $\model$ model faces this new situation as follows. It chooses the
phases $\delta_1$ and $\delta_2$, dependently on the oasis considered
close to $0^\circ$ or $180^\circ$.
 For $B_d$ system we find:
\be\label{B1B3}
 -12^\circ\le \delta_2\le 12^\circ\quad ({\rm green}), \qquad
 167^\circ\le \delta_2\le 192^\circ\quad ({\rm yellow})
\ee
with $0.0023\le\tilde s_{13}\le 0.0036$.
For $B_s$ system we find:
\be\label{A1A3}
 -30^\circ\le \delta_2\le 35^\circ\quad ({\rm purple}), \qquad
 150^\circ\le \delta_2\le 215^\circ\quad ({\rm blue})
\ee
with $0.01\le\tilde s_{23}\le 0.019$.

In both cases  $\Delta M_s$ and $\Delta M_d$
being sensitive to $2\delta_i$ are enhanced. In the case of $B_s^0-\bar B_s^0$ system, where the $S_{\psi\phi}$
asymmetry is small both in the SM and data, the apperance of a small
$2\delta_2$ is rather natural. In the case of $B_d^0-\bar B_d^0$ system
the addition of a new contribution with a new phase much smaller than
the SM phase lowers $S_{\psi K_S}$ through the interference of both
contributions, as desired.

\begin{figure}[!tb]
 \centering
\includegraphics[width = 0.45\textwidth]{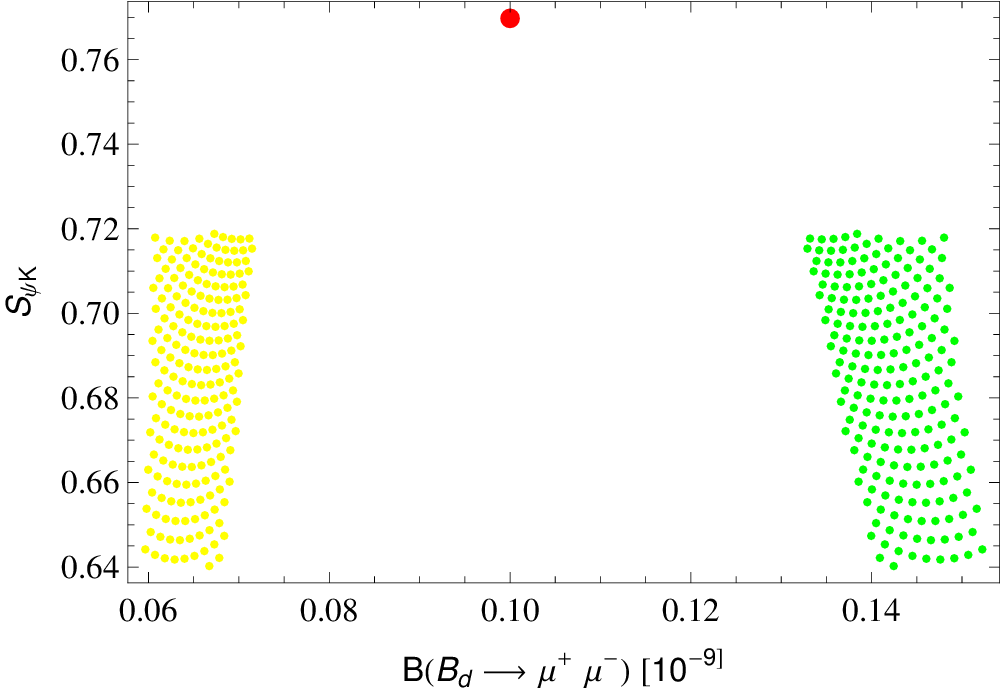}
\includegraphics[width = 0.45\textwidth]{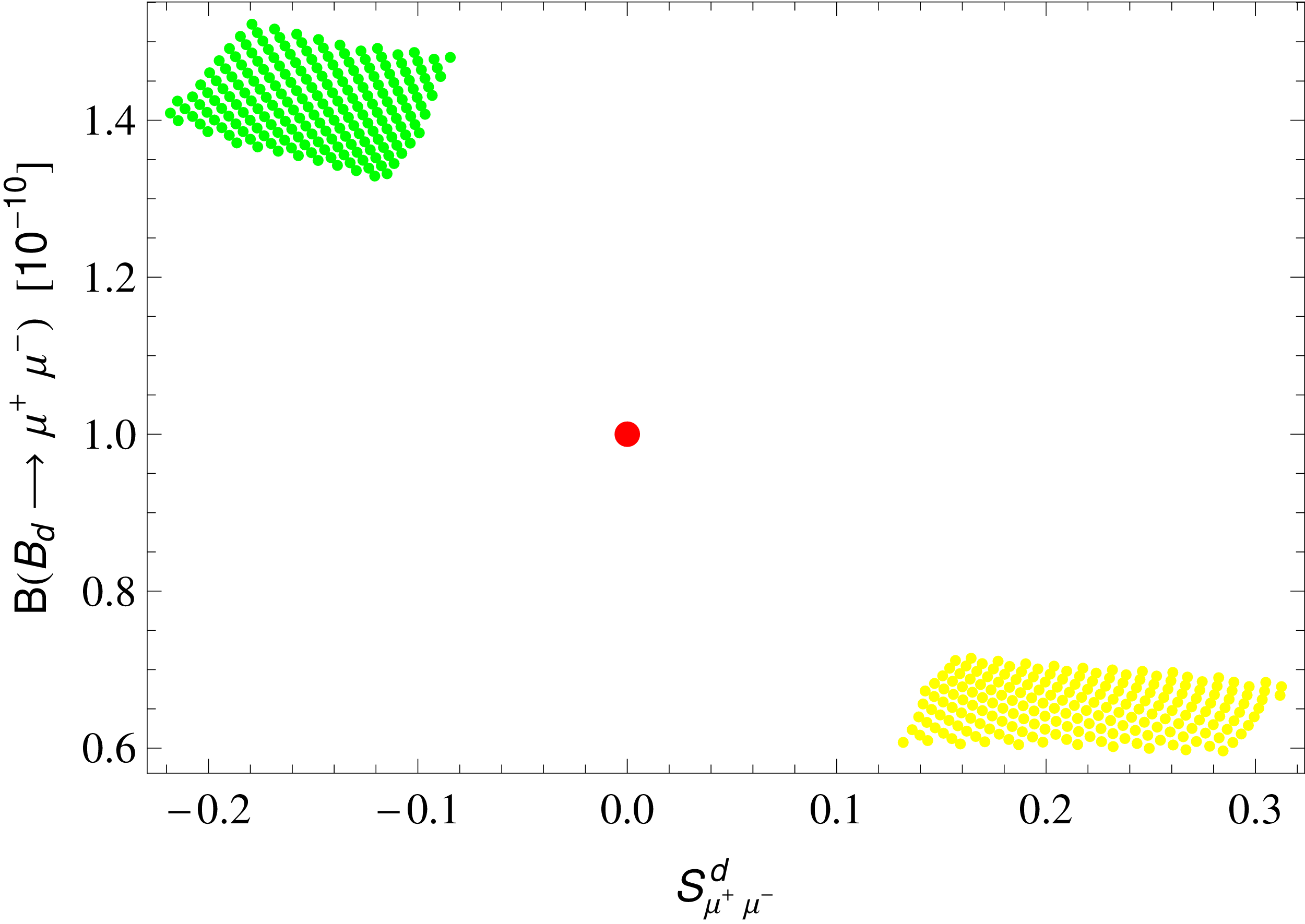}
\caption{\it  $S_{\psi K_S}$ versus $\mathcal{B}(B_d\to\mu^+\mu^-)$ (left) and $\mathcal{B}(B_d\to\mu^+\mu^-)$ versus $S^d_{\mu^+\mu^-}$
(right). Red point: central
value of SM prediction. For colour coding see (\ref{B1B3}).}\label{fig:SKSBdmuNEW}~\\[-2mm]\hrule
\end{figure}

\begin{figure}[!tb]
 \centering
\includegraphics[width = 0.45\textwidth]{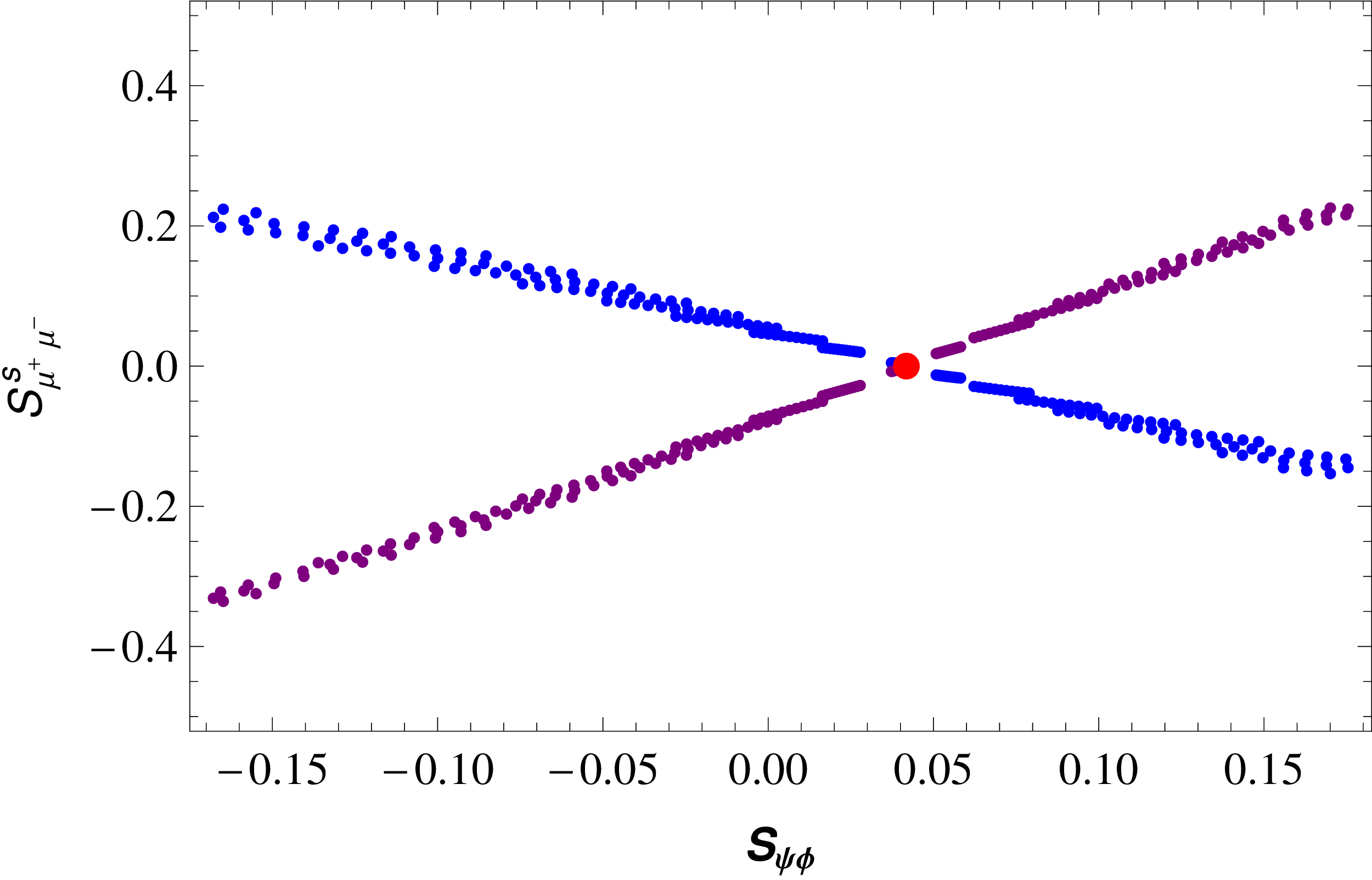}
\includegraphics[width = 0.45\textwidth]{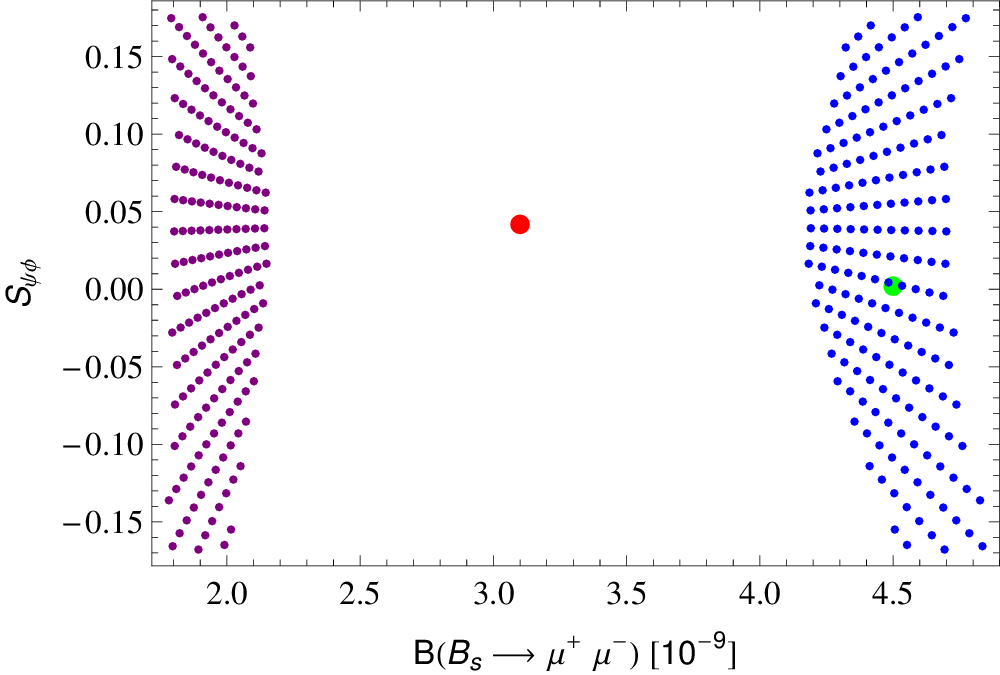}
\caption{\it $S_{\mu^+\mu^-}^s$ versus $S_{\psi\phi}$ (left) and $S_{\psi\phi}$ versus $\mathcal{B}(B_d\to\mu^+\mu^-)$  (right).
Red point: central
value of SM prediction.  For colour coding see (\ref{A1A3}).                 }\label{fig:SmuSphiNEW}~\\[-2mm]\hrule
\end{figure}

The question then arises what impact does this new structure have on correlations presented above.
Setting $M_{Z'}=1\tev$ we answer this question in Figs.~\ref{fig:SKSBdmuNEW}
and \ref{fig:SmuSphiNEW} for most
interesting correlations. We observe:
\begin{itemize}
\item
As seen in Fig.~\ref{fig:SKSBdmuNEW} in the $B_d$ meson system, the general structure of
correlations is unchanged but the NP effects this time in $B_d\to\mu^+\mu^-$
are found to be much larger than in $S^d_{\mu^+\mu^-}$.
\item
As seen in Fig.~\ref{fig:SmuSphiNEW} the impact on the $B_s$ system is much larger. Now the
roles of $B_s\to\mu^+\mu^-$
 and $S^s_{\mu^+\mu^-}$ in the search for optimal oasis are interchanged.
The suppressions or enhancements of $\mathcal{B}(B_s\to\mu^+\mu^-)$ relatively
to the SM uniquely identify the optimal oasis. Moreover,  $\mathcal{B}(B_s\to\mu^+\mu^-)$ is bound to be different from the SM values.
$S^s_{\mu^+\mu^-}$ can
now be vanishing and can only identify the optimal oasis in combination with
$S_{\psi\phi}$, moreover it is now significantly smaller than
previously.
\end{itemize}

This analysis demonstrates clearly that it is crucial the have precise
determinations of flavour observables within the SM in order to be able
to search indirectly for NP.

\section{Summary}\label{sec:8}
In this paper we have reconsidered the flavour structure of the 331 models
concentrating the phenomenology on a particular model,
the  $\overline{331}$ model. In this model the dominant NP contributions
come from $Z'$ tree-level exchanges. For $M_{Z'}$ of order of few TeV,
visible departures from the SM expectations are possible. As a detailed
analysis of phenomenological implications and related correlations have
been presented in the previous section we only list few highlights:
\begin{itemize}
\item
Favouring the inclusive value of $\vub$,
for $1\tev \le M_{Z'}\le 3\tev$, the  { $\overline{331}$}   model  agrees
well with the most recent
$B^+\to\tau^+\nu_\tau$ data,
removes
the $\varepsilon_K-S_{\psi K_S}$ tension present in the SM and
the $\varepsilon_K-\Delta M_{s,d}$ tension present in CMFV models.
Simultaneously, while differing from the SM, it is consistent with the present data on $B_{s,d}\to\mu^+\mu^-$ and $S_{\psi\phi}$.
\item
We identify four {\it big} oases in the parameter space that can be uniquely
distinguished by the values of $\mathcal{B}(B_d\to\mu^+\mu^-)$ and $S_{\mu^+\mu^-}^s$.  Also the asymmetry $S_{\mu^+\mu^-}^d$ is
helpful in this respect.
\item
In each oasis there is a definite correlation between
$\mathcal{B}(B_s\to\mu^+\mu^-)$ and $S_{\psi\phi}$ so that also these two
variables, when considered simultaneously can give valuable information
about the model.
\item
 There are also four {\small} oases that cannot be ruled out at present but
as predictions for various observables are very sharp in these oases,
it is likely that they will not survive more precise data.
\item
A characteristic feature of the model is the anti-correlation between
$\mathcal{B}(B_s\to\mu^+\mu^-)$ and $\mathcal{B}(B\to X_s \nu\bar\nu)$ .
\item
$Z'$ contributions to $b\to s\nu\bar\nu$
transitions, $\kpn$ and $\klpn$ are found typically below $5-10\%$.
\item
The violation of CMFV relations are clearly visible with the largest
one related to $B_{s,d}\to\mu^+\mu^-$ decays.
\item
We have investigated the case of $M_{Z^\prime}=10$~TeV, demonstrating that
for such high values of $Z'$ mass the pattern of departures from the
SM changes,  allowing  to remove all anomalies even for exclusive values
of $\vub$. But then NP effects in rare $B$ and $K$ decays are very small.
\end{itemize}

 We are aware of the fact that some of the correlations presented
by us would be washed out if we included all existing uncertainties. Yet, our
simplified numerical analysis had as the main goal to illustrate how the
decrease of theoretical, parametric and experimental uncertainties in the
coming years might allow to exhibit certain features of NP, even if
the deviations from the SM will be only moderate. In fact within the
coming years the size of the assumed uncetainties in our analysis
 could likely become
reality.

The correlations between various observables found in this paper
 will allow to distinguish it from other simple secenarios.
In particular we have demonstrated that sizable violation of CMFV relations
are still possible in this model. Here we would like to make a brief
comparision of the $\model$ model with the ${\rm 2HDM_{\overline{MFV}}}$ and
$MU(2)^3$ models. The most striking differences are as follows:
\begin{itemize}
\item
${\rm 2HDM_{\overline{MFV}}}$ \cite{Buras:2010mh} and $\model$ model
share the property that
the NP contributions to $\varepsilon_K$ are small favouring thereby inclusive
value
of $\vub$. In ${\rm 2HDM_{\overline{MFV}}}$, the data on $S_{\psi K_S}$
imply then automatically a positive value of $S_{\psi\phi}\ge 0.15$ \cite{Buras:2012ts}. In
 $\model$ model this is not required and any value within the LHCb range
is still possible. Concerning $\Delta F=1$ transitions let us just mention
that in  ${\rm 2HDM_{\overline{MFV}}}$ there is some tendency that with
increasing $S_{\psi\phi}$ also the lower bound on
$\mathcal{B}(B_s\to\mu^+\mu^-)$ increases. In $\model$ model, as seen
in Fig.~\ref{fig:SphiBsmu}, the correlation
$S_{\psi\phi}-\mathcal{B}(B_s\to\mu^+\mu^-)$ is much more transparent.
Moreover, while in the oasis $A_1$ these two observables are correlated,
they are anti-correlated in the oasis $A_3$. In both models
$\mathcal{B}(B_s\to\mu^+\mu^-)$ can
be smaller and larger than the SM value, but in  ${\rm 2HDM_{\overline{MFV}}}$
the deviations can be larger, a characteristic property of scalar currents.
\item
In the case of $MU(2)^3$ models, the most striking differences from
the $\model$ model are the correlation
$S_{\psi K_S}-S_{\psi\phi}-\vub$ \cite{Buras:2012sd} that is absent in the $\model$ model and
the fact that  $MU(2)^3$ models can be made consistent with the FCNC data
for exclusive value of $\vub$, while this is rather
difficult in the $\model$ model
for $M_{Z'}\le 5 \tev$.
Finding in the future that exclusive value for $\vub$ is chosen by nature
and $M_{Z'}\le 5 \tev$  would
favour $MU(2)^3$ models over $\model$ model. On the other hand if the
inclusive value of $\vub$ is the right one and $S_{\psi\phi}$ is found to
be  negative, this would put  $MU(2)^3$ models into difficulties while
the $\model$ model would face such data without any problems.
\end{itemize}

In view of these definite findings
we are looking forward to improved experimental data and improved lattice
calculations. The plots presented by us should facilitate monitoring
the future confrontations of  the  $\overline{331}$ model  with the data
and to find out whether this simple model can satisfactorily describe
the observables considered  by us.

The generalization of our analysis to $Z^\prime$ models with both right-handed and left-handed couplings
and sizable NP contributions to $\varepsilon_K$ and rare $K$ decays is presented in \cite{Buras:2012jb}.

\section*{Acknowledgments}
We would like to thank Pietro Colangelo and  Robert Fleischer for discussions.
This research was done and financed in the context of the ERC Advanced Grant project ``FLAVOUR''(267104) and was partially supported by the DFG cluster
of excellence ``Origin and Structure of the Universe''.

\appendix
\section{Appendix}\label{app:Deltas}
We collect the formulae for the couplings of $Z^\prime$ to quarks and leptons especially for $\beta = \sqrt{3}$ and
$\frac{1}{\sqrt{3}}$ that
enter various expressions in our paper.

{\bf Case 1: arbitrary $\beta$}
{\allowdisplaybreaks
\begin{subequations}
\begin{align}
 \Delta_L^{ij}(Z') &=\frac{g}{\sqrt{3}c_W\sqrt{1-(1+\beta^2)s_W^2}}c_W^2 v_{3i}^*v_{3j}\,,\\
 \Delta_L^{ji}(Z') &=\left[ \Delta_L^{ij}(Z')\right]^\star\,,\\
\Delta_L^{\nu\bar\nu}(Z') & = \frac{g~\left[1-(1+\sqrt{3}\beta)s_W^2\right]}{2 \sqrt{3}c_W\sqrt{1-(1+\beta^2)s_W^2}}\,,\\
\Delta_L^{\mu\bar\mu}(Z')&=\Delta_L^{\nu\bar\nu}(Z')\,,\\
\Delta_R^{\mu\bar\mu}(Z')&=\left\{\begin{array}{cc}
                                   \frac{-g~\beta~ s_W^2}{c_W\sqrt{1-(1+\beta^2)s_W^2}} \,,&\quad \beta\neq\sqrt{3}\\
\frac{g\sqrt{1-4s_W^2}}{ \sqrt{3}c_W}\,, & \quad \beta=\sqrt{3}\\
                                  \end{array}\right.
\end{align}
\end{subequations}}%

{\bf Case 2: $\beta = \sqrt{3}$}
{\allowdisplaybreaks
\begin{subequations}
\begin{align}
 \Delta_L^{sd}(Z')&=\frac{g}{\sqrt{3}c_W\sqrt{1-4s_W^2}}c_W^2 v_{32}^*v_{31}
=1.84~g \tilde s_{13}\tilde s_{23}\tilde c_{13} e^{i(\delta_2-\delta_1)}\,,\\
\Delta_L^{bd}(Z')&=\frac{g}{\sqrt{3}c_W\sqrt{1-4s_W^2}}c_W^2 v_{33}^*v_{31}
=-1.84~g  \tilde s_{13}\tilde c_{13}\tilde c_{23} e^{-i\delta_1}\,,\\
\Delta_L^{bs}(Z')&=\frac{g}{\sqrt{3}c_W\sqrt{1-4s_W^2}}c_W^2 v_{33}^*v_{32}
= -1.84~g \tilde s_{23}\tilde c^2_{13}\tilde c_{23} e^{-i\delta_2}\,,\\
\Delta_L^{\nu\bar\nu}(Z')&=\frac{g\sqrt{1-4s_W^2}}{2 \sqrt{3}c_W}=0.090~g\,,\\
\label{eq:relationcouplings}
\Delta_L^{\mu\bar\mu}(Z')&=\frac{1}{2}\Delta_R^{\mu\bar\mu}(Z')=\Delta_L^{\nu\bar\nu}(Z')=\Delta_A^{\mu\bar\mu}(Z')\,,\\
\Delta_V^{\mu\bar\mu}(Z')&=0.271~g\,.
\end{align}
\end{subequations}}%

{\bf Case 3: $\beta = \frac{1}{\sqrt{3}}$}
{\allowdisplaybreaks
\begin{subequations}
\begin{align}
\Delta_L^{sd}(Z')&=\frac{g}{\sqrt{3}c_W\sqrt{1-\frac{4}{3}s_W^2}}c_W^2 v_{32}^*v_{31}
=0.61~g \tilde s_{13}\tilde s_{23}\tilde c_{13} e^{i(\delta_2-\delta_1)}\,,\\
\Delta_L^{bd}(Z')&=\frac{g}{\sqrt{3}c_W\sqrt{1-\frac{4}{3}s_W^2}}c_W^2 v_{33}^*v_{31}
=-0.61~g  \tilde s_{13}\tilde c_{13}\tilde c_{23} e^{-i\delta_1}\,,\\
\Delta_L^{bs}(Z')&=\frac{g}{\sqrt{3}c_W\sqrt{1-\frac{4}{3}s_W^2}}c_W^2 v_{33}^*v_{32}
= -0.61~g \tilde s_{23}\tilde c^2_{13}\tilde c_{23} e^{-i\delta_2}\,,\\
\Delta_L^{\nu\bar\nu}(Z')&=\frac{g\,\left(1-2s_W^2\right)}{2 \sqrt{3}c_W\sqrt{1-\frac{4}{3}s_W^2}}=0.213~g\,,\\
\Delta_L^{\mu\bar\mu}(Z')&=\Delta_L^{\nu\bar\nu}(Z')\,,\\
\Delta_R^{\mu\bar\mu}(Z')&=-\frac{g \, s_W^2}{\sqrt{3} c_W \sqrt{1-\frac{4}{3}s_W^2}} = \frac{2
s_W^2}{1-2s_W^2}\Delta_L^{\mu\bar\mu}(Z') =- 0.183~g\,\\
\Delta_A^{\mu\bar\mu}(Z')&=-0.396~g\,,\\
\Delta_V^{\mu\bar\mu}(Z')&=0.030~g\,.
\end{align}
\end{subequations}}%

\bibliographystyle{JHEP}
\bibliography{allrefs}
\end{document}